\documentclass[prb,twocolumn,superscriptaddress,showpacs,amssymb]{revtex4-1}

\usepackage{psfrag,graphicx}
\usepackage{tikz}
\usepackage{bbm}
\usepackage{pgfplots}
\usepackage{dcolumn}
\usepackage{bm}
\usepackage{amsfonts,amssymb,amsmath}
\usepackage{xcolor}
\usepackage{tikz}
\usepackage{times}
\usetikzlibrary{decorations.markings}
\usetikzlibrary{arrows}
\usepackage{epstopdf}
\usepackage{soul}
\usepackage{wasysym}

\definecolor{jens}{rgb}{.2,0.7,.9}

\usepackage{mathbbol}

\newcommand{\be}{\begin{equation}}
\newcommand{\ee}{\end{equation}}
\newcommand{\bq}{\begin{eqnarray}}
\newcommand{\eq}{\end{eqnarray}}

\newcommand{\rf}[1]{(\ref{#1})}

\newcommand{\tr}{\mathrm{tr}}
\newcommand{\ket}[1]{\left |#1 \right\rangle}

\newcommand{\Z}{\mathbbm{Z}}

\begin{document}

\title{Anatomy of Fermionic Entanglement and Criticality in Kitaev Spin Liquids}

\author{K.\ Meichanetzidis}\email{mmkm@leeds.ac.uk}
\affiliation{School of Physics and Astronomy, University of Leeds, Leeds LS2 9JT, United Kingdom}
\author{M.\ Cirio}
\affiliation{  Interdisciplinary Theoretical Science Research Group (iTHES), RIKEN, Wako-shi, Saitama 351-0198, Japan }	
\author{ J.\ K.\ Pachos}
\affiliation{School of Physics and Astronomy, University of Leeds, Leeds LS2 9JT, United Kingdom}
\author{V.\ Lahtinen}
\affiliation{Dahlem Center for Complex Quantum Systems, Freie Universit{\"a}t Berlin, 14195 Berlin, Germany}	

\date{\today}

\begin{abstract}
We analyse in detail the effect of non-trivial band topology on the area law behaviour of the entanglement entropy in Kitaev's honeycomb model. By mapping the  translationally invariant 2D spin model into 1D fermionic subsystems, we identify those subsystems responsible for universal entanglement contributions in the gapped phases and those responsible for critical entanglement scaling in the gapless phases. For the gapped phases we analytically show how the topological edge states contribute to the entanglement entropy and provide a universal lower bound for it. For the gapless semi-metallic phases and topological phase transitions the identification of the critical subsystems shows that they fall always into the Ising or the XY universality classes. As our study concerns the fermionic degrees of freedom in the honeycomb model, qualitatively similar results are expected to apply also to generic topological insulators and superconductors. 
\end{abstract}

\maketitle

\section{Introduction}\label{Introduction}
Fundamental characteristics of quantum many-body systems at zero temperature are manifested in their entanglement properties. A common measure of quantum correlations between two complementary regions of a system is the entanglement entropy. While it in general obeys an area law, and is hence widely considered non-universal,\cite{Area} it can also successfully capture some universal features of quantum states. The most famous ones include central charges describing entanglement scaling in critical 1D states~\cite{Vidal03, Cardy} and subleading corrections, such as topological entanglement entropy, in gapped long-range entangled topologically ordered states with anyonic excitations~\cite{Hamma,KitaevPreskill, LevinWen}. Progress has also been made in understanding corner entanglement that may characterize critical states in higher dimensions.\cite{Bueno15} 

Due to recent experimental progress, short-range entangled gapped topological phases, more commonly known as topological insulators and topological superconductors, have been the subject of much interest.\cite{Hasan10} Depending on the symmetries of the state, their band structures are characterised by non-trivial topological indices,~\cite{Schnyder08} whose physical manifestation is the appearance of protected gapless states at the physical boundary of the system. These are intimately related to the entanglement in such states.~\cite{Li08,Qi} When the system without a boundary is partitioned into two halves, topologically protected \emph{virtual} edge states localized on the partition boundary appear also in the single-particle entanglement spectrum~\cite{Fidkowski10, MongShivamoggi}. These describe the short-range entanglement across the boundary and provide the dominant area law contributions to the entanglement entropy.\cite{Area} As such edge states cannot be adiabatically removed,~\cite{Turner10,Eisert} their correlations impose a lower bound to the entropy~\cite{Klich,Fidkowski10}. The area law part of entanglement entropy can thus also contain universal features that can serve as an entropic criterion of short-range entangled topological phases in both free and interacting fermionic models.~\cite{Mei15} 

% A similar behaviour is also expected in the gapless critical points between gapped topological phases. The virtual edge states in this case contribute maximally to the entropy which again follows an area law. However, it is the critical contributions that dominate the entropy since they logarithmically diverge due to the diverging correlation length.
%

In this work we revisit in detail these entropic properties in the context of Kitaev's honeycomb model.\cite{KitaevHoneycomb} As an exactly solvable model for a spin liquid -- a topological state of spins with anyonic excitations -- it has been subject of numerous studies. Recently, it has gained direct experimental relevance due to the confirmation\cite{Chun15,Banerjee16} that certain magnetic materials may host such states.\cite{Jackeli09,Singh12} With these developments in mind, we focus on the exactly solvable original model that hosts a rich phase diagram containing both gapless semi-metallic states as well as gapped phase characterized by various Chern numbers.\cite{Pachos07, Lahtinen08, Lahtinen10, Lahtinen} While the original spin model is long-range entangled, as required for anyons to emerge, it has the curious property that the topological entanglement entropy takes the same value in all topological phases.\cite{KitaevHoneycomb} It does not depend on the emergent fermionic degrees of freedom and contains no information about a particular topological phase of the model.\cite{Yao10} Here we focus on these short-range entangled degrees of freedom to identify the universal entropic properties that characterize the different phases of the honeycomb model. We explicitly study how these correlations, originating from the topological structure of the fermionic bands, manifest themselves in the area law behavior. Thus while we focus on a particular model that in principle has long-range entanglement, qualitatively similar results should apply also to generic short-range entangled topological insulators and superconductors. 

To analyze the anatomy of the fermionic entanglement, we place the model on a torus and study the correlations across a cylindrical partition. Translational invariance along the partition allows us to apply a partial Fourier transformation that reduces the 2D problem to a set of 1D models. We identify the effective 1D models that in the gapped phases encode the topological nature of the 2D state and give rise to the universal entropy lower bound, and in the gapless phases give rise to critical entropy scaling. By analyzing this scaling and the corresponding 1D models, we show that the gapless phases and topological phase transitions of the full 2D model are all in the universality classes of 1D Ising or XY models. Furthermore, we show that also gapless states follow the area law that in general comprises of critical, quasi-critical and edge state contributions.

The paper is structured as follows. We start by reviewing entanglement in translationally invariant fermion system in Section \ref{Entanglement spectrum and entropy}. Section \ref{KitaevHoneyModel} introduces our case study -- Kitaev's honeycomb model -- and reviews its phase diagram. 
In Section \ref{Entanglement Entropy of Topological Phases} we study the entanglement in its gapped phases characterized by different Chern numbers. We solve analytically for the edge states and show quantitatively how they dominate the entanglement entropy and give rise to a lower bound. In Section \ref{gapless phases tomography} we focus on both gapless phases and critical points between gapped topological phases. We identify the dominant contributions from critical and quasi-critical 1D subsystems and identify the universality classes of the transitions.

In Appendix \ref{EntropyCorrMat} we present in detail the form of the entanglement entropy in terms of the correlation matrix and the derivation of the entropic lower bound. In Appendix \ref{KitaevModelApp} we review in detail Kitaev's honeycomb model, its analytic solution and the decomposition in terms if 1D wires. Details of the derivation of the edge-mode dispersion using the generating function method are presented in Appendix \ref{GenFunMethod}. 

\section{Entanglement in free fermion systems}
\label{Entanglement spectrum and entropy}

We analyze the fermionic entanglement in Kitaev's exactly solvable model for a 2D spin liquid by mapping it to free Majorana fermions.\cite{KitaevHoneycomb} Thus qualitatively similar analysis applies also to generic topological insulators and superconductors. We begin by reviewing the key concepts in order to study entanglement properties in such systems.

\begin{figure}[t]
\includegraphics[scale=0.48]{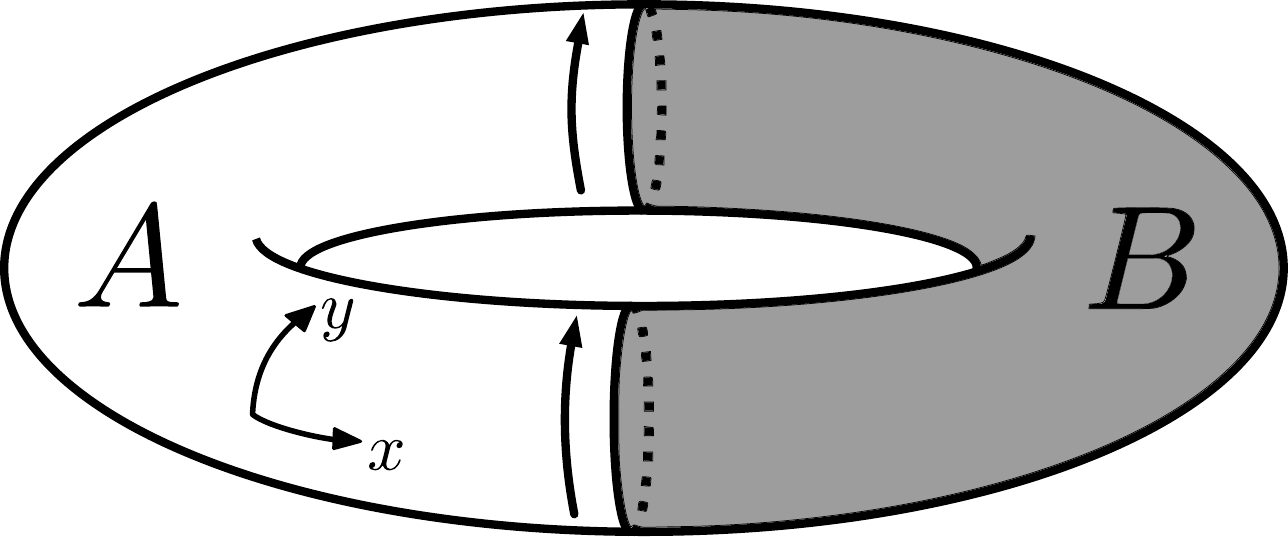}
\caption{\label{fig:Torus} A 2D system defined on a torus of size $L_x\times L_y$ is bi-partitioned in regions $A$ and $B$ such that translational symmetry along $L_y$ is preserved. If the band structure is characterized by a topological invariant, both the edge energy spectrum and the entanglement spectrum exhibit localized states along the cut $\partial A$.}
\end{figure}

Any system of non-interacting fermions $f_{i}$ can be written in the basis of Majorana fermions by writing ${f_i= (\gamma_{2i-1}+i \gamma_{2i})/\sqrt{2}}$, i.e. by introducing two Majorana sites for each physical site, where the Majorana fermion operators satisfy the reality condition ${ {\gamma_{i}}^\dagger=\gamma_{i} }$ and the anti-commutation relation ${\{ \gamma_{i},{\gamma_{j}}\}=2\delta_{i,j}}$.  The most general Hamiltonian takes then the form
\be
\label{eq:quadraticHam}
H=\sum_{i,j \in \Lambda} i H_{i,j}  \gamma_{i} \gamma_{j},
\ee
where $\Lambda$ denotes the lattice for the Majorana fermions. The skew-symmetric and real matrix $H_{i,j}$ encodes all the information about the tunneling and possible pairing amplitudes, as well as about on-site terms for the fermions $f_i$. We take the Majorana lattice to be of size $|\Lambda | =L_x \times L_y$ and define it on a torus, as shown in Fig.~\ref{fig:Torus}. Assuming full translational invariance along the periodic $y$-direction, a Fourier transformation brings the Hamiltonian then into a block-diagonal form
\be
\label{eq:chainsHam}
H=\sum_{p=0}^{\pi} \sum_{ i,j\in \Lambda_x} i {H_{i,j}}(p)  {\gamma^\dagger_{i,p}} \gamma_{j,p},
\ee
where the $L_y$ blocks $H(p)$ describe decoupled chains of length $L_x$ on the 1D lattice $\Lambda_x$. Due to the Fourier transform the Majorana operators behave as complex fermions subject to the constraint $\gamma_{i,p}^\dagger = \gamma_{i,-p}$. Thus the Brillouin zone is halved in order to avoid double counting of the chains.

Due to Wick's theorem for non-interacting systems, all the information about the ground state is contained in the correlation matrix~\cite{Peschel} ${C_{i,j}(p)=\langle {\gamma_{i,p}}^\dagger \gamma_{j,p} \rangle}$, which is also block-diagonal in momentum $p$. To analyze the entanglement entropy of the ground state, $\Lambda$ is partitioned into regions $A$ and $B$, such that translation invariance in the $y$-direction is preserved, as shown in Fig.~\ref{fig:Torus}. Similar to the reduced density matrix, the reduced correlation matrix is defined by restricting to the submatrix $C^A_{i,j}(p)$, whose entries $i,j$ have only support in $A$. The entanglement entropy of each chain is then obtained from the \emph{entanglement spectrum}, the spectrum  $\{\lambda(p) \in[0,1] \}$ of $C^A(p)$ as\cite{Peschel, Vidal03}
\bq
\label{eq:entropyFromProbabilities}
S(p)=-\frac{1}{2} \sum_j &&\Big[\lambda_{j}(p)\log\lambda_{j}(p)\nonumber\\
&&+\big(1-\lambda_{j}(p)\big)\log \big(1-\lambda_{j}(p)\big)\Big],
\eq
whose derivation we review in Appendix \ref{EntropyCorrMat}. This form immediately shows how different entanglement eigenvalues contribute to the entanglement entropy. Eigenvalues close to $\lambda=1/2$ contribute maximally to the total entropy, while the contribution of the eigenvalues close to $\lambda=0$ or $1$ is small. 

One physical origin of such different contributions has been shown to be related to the topology of the band structure. For non-interacting topological insulators, there is an adiabatic correspondence between the entanglement spectrum $\{\lambda\}$ defined on a torus partitioned to $A$ and $B$ by the cut $\partial A$ and the energy spectrum of the Hamiltonian $H_A$ defined on the cylindrical region $A$ with open boundaries $\partial A$.~\cite{Fidkowski10,Mei15} In the latter case the topological band structure manifests itself as the existence of edge states localized along $\partial A$ (we refer to these as physical edge states), while in the entanglement spectrum their counterpart is the existence of entanglement states crossing from $\lambda=0$ to $\lambda=1$ (we refer to these as virtual edge states). Thus the existence of physical edge states necessarily implies maximally entangled modes that reflect strong short-range correlations across the cut $\partial A$.

Due to the additivity of the entropy for block-diagonal Hamiltonians, the entropy $S$ of the 2D system \rf{eq:chainsHam} is given by
\be
\label{eq:Ssumchains}
S=\sum_{p=0}^{\pi} S(p).
\ee
This opens up the possibility to microscopically trace the behavior of the full 2D model back to a few number of chains that dominate the entanglement entropy. Our strategy is to isolate chains that are either in a topological phase or exhibit quantum criticality, and thereby analyze the anatomy of entanglement entropy in Kitaev spin liquids.

%In this work we analyze in detail the relatation between entanglement and physical edge states and explicitly show how they give rise to the celebrated area law $S=\alpha |\partial A|$ in a 2D system. The key method is to decompose the 2D system into 1D chains \rf{eq:entropyFromProbabilities}, which implies that also the entanglement entropy is additive in individual chain contributions

\section{Kitaev's honeycomb lattice Model}
\label{KitaevHoneyModel}

Kitaev's honeycomb model describes spin-$1/2$ particles on the vertices of a honeycomb lattice $\Lambda$ subject to anisotropic nearest-neighbour exchange interactions $J_r \sigma_i^r \sigma_j^r$.~\cite{KitaevHoneycomb}. The  magnitude $J_r$ and the type of interaction depend on the three distinct orientations $r=x,y,z$ of the links of the honeycomb lattice. The beauty of the model lies on the property that time-reversal symmetry can be broken by introducing a three-spin term of magnitude $K$, enabling chiral topological phases characterized by Chern numbers to emerge, while still preserving the exact solvability of the model. 

This solution is reviewed in Appendix \ref{KitaevModelApp}. There it is shown that the interacting spin Hamiltonian can be mapped to the Hamiltonian 
\be
\label{H_honey_Maj_main}
	H = \frac{i}{2}\sum_{\langle i,j \rangle\in\Lambda} u_{i,j} J_r \gamma_{i} \gamma_{j} + \frac{i}{2} K \sum_{\langle \langle i,j\rangle\rangle\in\Lambda} {u}_{i,m} {u}_{m, j} \gamma_{i} \gamma_{j},
\ee
describing free Majorana fermions $\gamma_i$ living on the vertices of the lattice, coupled to a $\mathbb{Z}_2$ gauge field $u_{i,j}$, that lives on the links. While the latter fluctuate, the model can be analyzed in a fixed gauge by specifying their configuration $\{ u_{i,j}=\pm 1 \}$. Fixing the gauge specifies a physical sector of the model through $\Z_2$ valued Wilson loop operators,
\be
{W}_{\hexagon}=\prod_{\langle i,j \rangle\in {\hexagon}} {u}_{i,j},
\ee
that correspond to gauge-invariant products of the gauge fields around each hexagonal plaquette $\hexagon$. Even there are many gauges $\{u_{ij} \}$ giving rise to the same vortex sector, the physics depends only on  $\{W_{\hexagon}\}$. The eigenvalues $W_{\hexagon}=-1$ are interpreted as the presence of a $\pi$-flux \emph{vortex} on $\hexagon$, with the pattern of Wilson loop values $\{W_{\hexagon}\}$ being referred to as a \emph{vortex sector}. The ground state of the model lies in the vortex-free sector for which $W_{\hexagon}=+1$ on all plaquettes.

Restricting to a particular vortex sector by fixing the gauge brings the Hamiltonian to the quadratic form \rf{eq:quadraticHam}, that in the case of the honeycomb model always describes a $p$-wave type topological superconductor. Depending on the vortex sector it has been shown to support both gapless phases as well as topological phases characterized by distinct Chern numbers.\cite{Pachos07,Lahtinen10,Lahtinen} Hence, it provides a wide variety of regimes with diverse entanglement properties. 

\subsection{The phase diagram}

To restrict to a part of the phase diagram over all vortex sectors that supports topological phases characterized by Chern numbers $\nu=0,\pm1$ and $\pm2$, we set $J_x=J_y=J$ and $J_z=1$, while keeping $K>0$ as a free parameter. To interpolate between vortex sectors, we adopt a perspective that the gauge variables $u_{ij}$ can be absorbed into the signs of the corresponding tunneling terms. This enables to smoothly tune  between topological phases that appear in different vortex sectors by allowing for staggered configurations of couplings. Restricting to considering only vortex-free $\{ W_{\hexagon}=+1\}$ and full-vortex $\{ W_{\hexagon}=-1\}$ configurations,  the honeycomb model can be brought into a block-diagonal form in chains \rf{eq:chainsHam}. Each chain is described by the 1D Hamiltonian
\bq
\label{eq_HforChains_main}
H(p)&&=\frac{i}{2}\sum_{j}\Big[J e^{-i p}a^\dagger_j b_j+ J a^\dagger_j b_{j-1} +\Theta_j a^\dagger_j b_j\nonumber\\
&&+K\Big(  e^{-i p}  a^\dagger_j a_{j+1} + \Theta_j e^{i p}  a^\dagger_j a_{j} + \Theta_{j+1}  a^\dagger_j a_{j-1}   \nonumber\\
&&+ e^{i p}b^\dagger_j b_{j-1}+\Theta_j e^{-ip} b^\dagger_j b_{j}+\Theta_{j+1} b^\dagger_j b_{j+1} \Big)\Big]+\text{h.c.}.\nonumber\\
\eq
where we introduced $\Theta_j=(1-\theta)+\theta (-1)^j$ to stagger the couplings such that one can smoothly interpolate between the vortex-free ($\theta=0$) and the full-vortex ($\theta=1$) sectors. For brevity, we have suppressed here the momentum index $p \in [0,\pi]$ of the Majorana operators $a_i \equiv a_{i,p}$ and $b_i \equiv b_{i,p}$ living on the two distinct triangular sublattices of the honeycomb lattice, respectively. One should keep in mind though that apart from $p=0,\pi$, at which $a_i$ and $b_i$ become two distinct species of Majorana operators, each chain describes complex fermions.

\begin{figure}[t]
\includegraphics[scale=0.362]{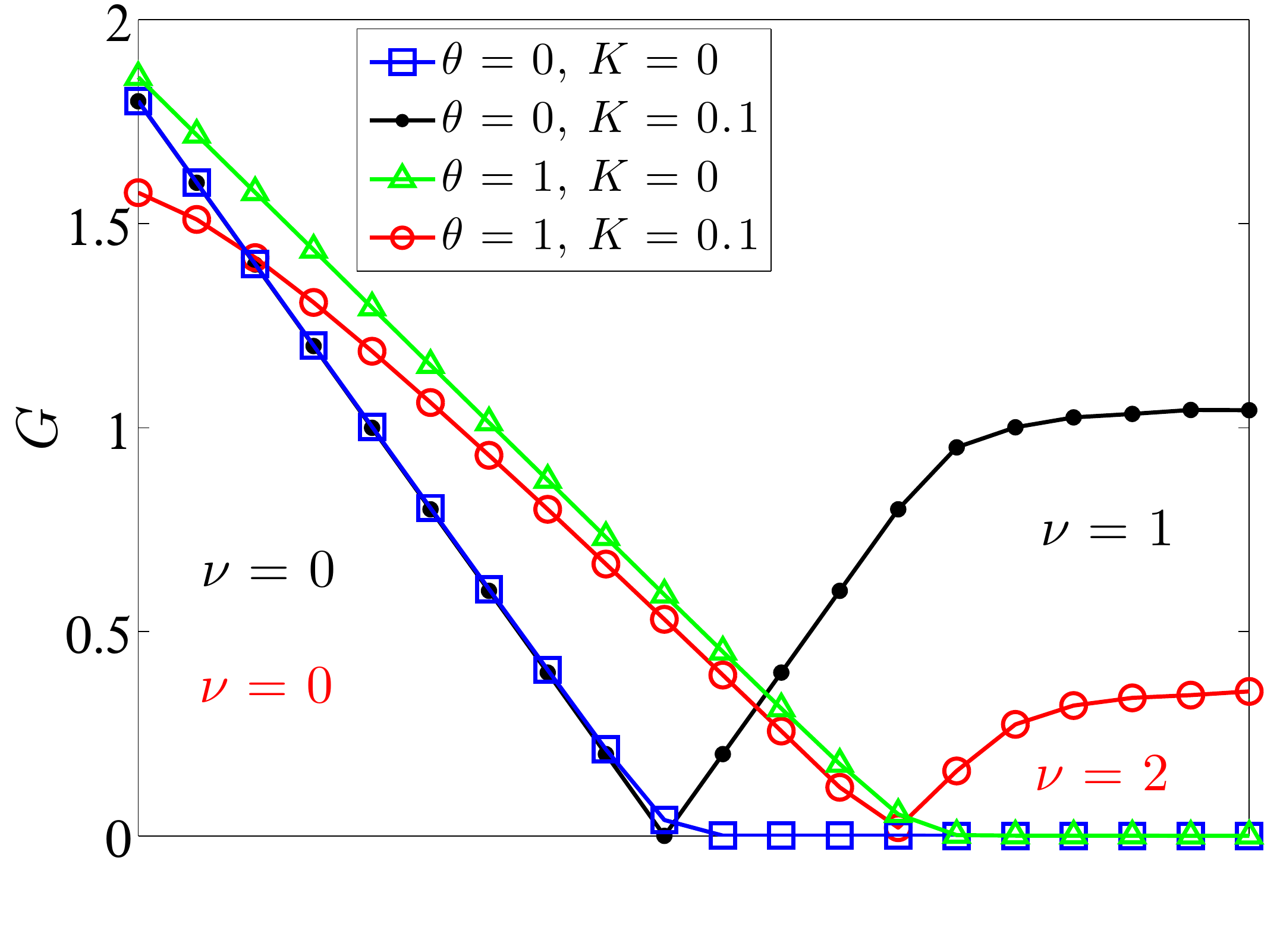}

\includegraphics[scale=0.36]{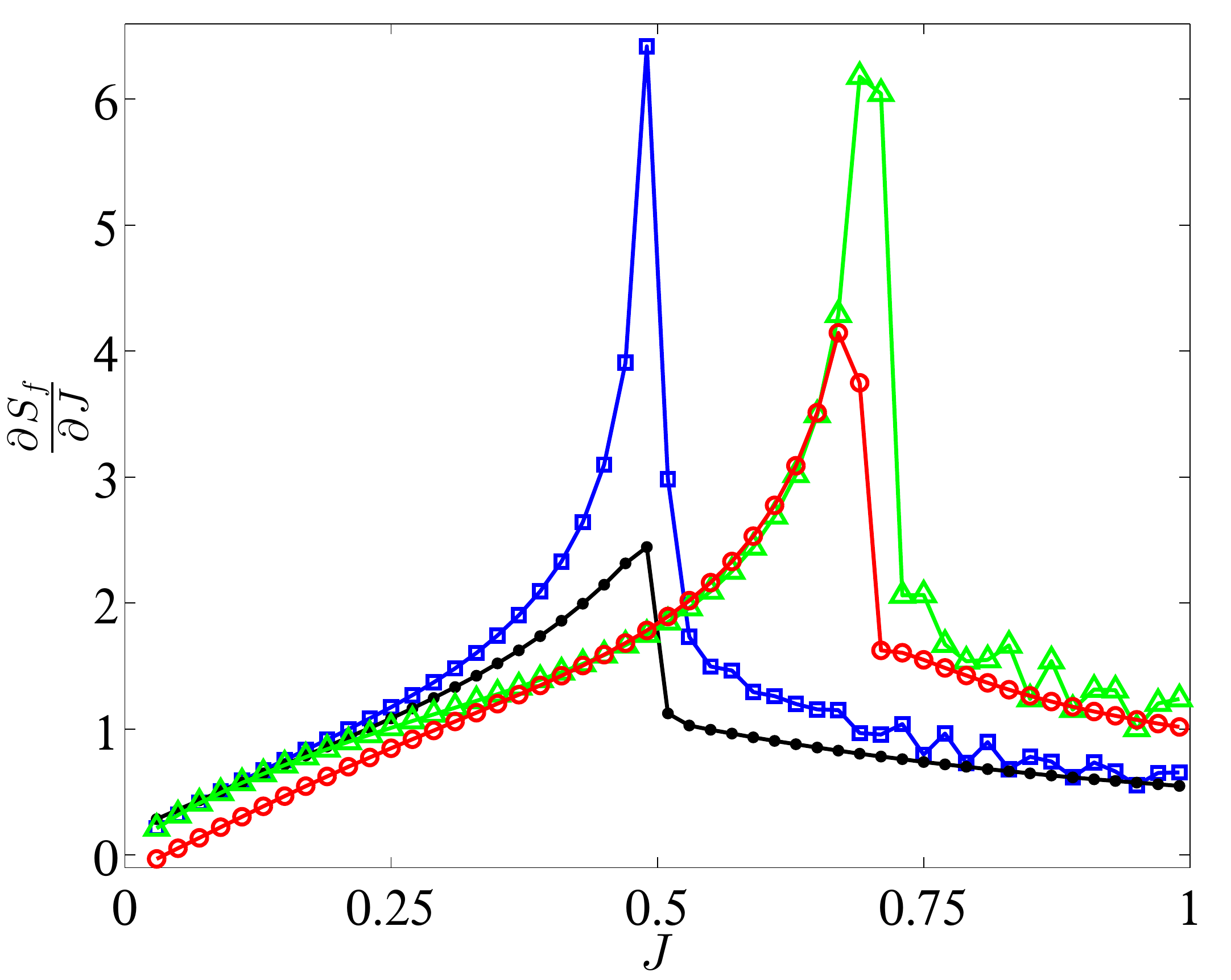}

\begin{picture}(0,0)
\put(23.2,287){\includegraphics[scale=0.238]{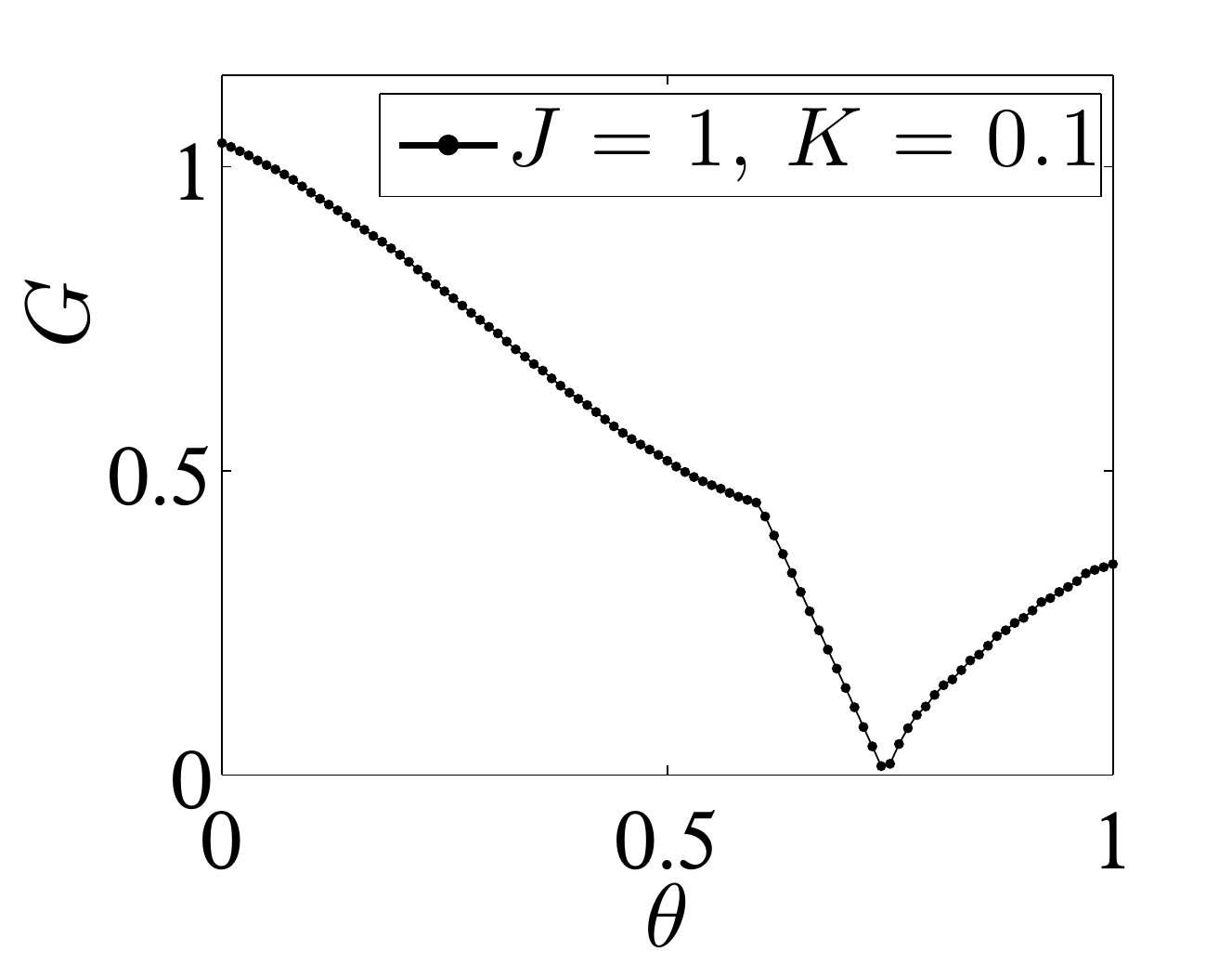}}
\put(-88,98){\includegraphics[scale=0.27]{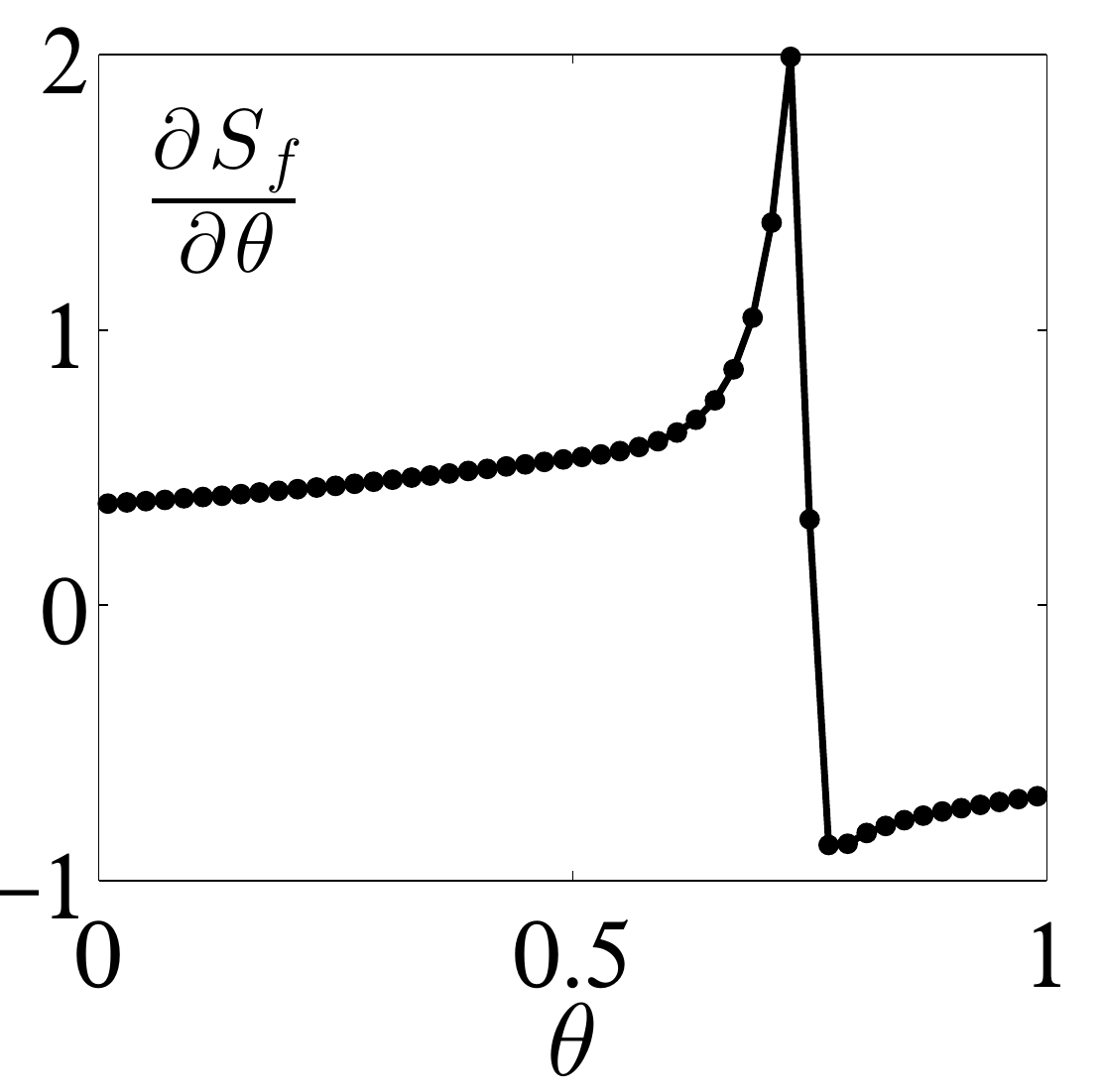}}
\end{picture}
\caption{\label{fig:Gaps_PTs} (Top) The energy gap $G$ in the vortex-free ($\theta=0$) and full-vortex ($\theta=1$) sectors. In the vortex-free sector the phase transitions between the $\nu=0$ and the gapless phase for $K=0$ or the $\nu=1$ phase for $K>0$ occurs at $J_c=1/2$, while at the full-vortex sector the transition to the $\nu=2$ phase or the corresponding gapless phase takes place at $J_c=\sqrt{\frac{1-K^2}{2}}$. (Inset) The transition between $\nu=1$ and $\nu=2$ topological phases occurs for staggered couplings corresponding to $\theta_c=\frac{3+K^2}{4}$.
(Bottom) Divergence of the rate of change in the entropy correlates with the quantum phase transitions. Here $L_x=L_y=144$ in the main figures and $L_x=L_y=100$ in the insets.
}
\end{figure}

The phase diagram over the vortex-free and full vortex sectors has been studied in Refs \onlinecite{Pachos07,Lahtinen10,Lahtinen}. In the vortex-free sector ($\theta=0$) the system is in a non-chiral Abelian phase characterized by $\nu=0$ for $J<1/2$, while for $J>1/2$ the system is in a gapped topological non-Abelian phase with $\nu=\textrm{sign}(K)$. In the non-Abelian phase, vortices bind localised Majorana modes and behave thus as non-Abelian anyons.\cite{KitaevHoneycomb,Lahtinen08}  In the full-vortex sector ($\theta=1$), the toric code Abelian phase occurs for $J<\sqrt{\frac{1-K^2}{2}}$, while for $J>\sqrt{\frac{1-K^2}{2}}$ the system is in a chiral Abelian phase with $\nu=2\textrm{sign}(K)$. One can also tune directly from the $\nu=\pm 1$ phase to the $\nu= \pm 2$ phase by changing $\theta$, in which case for $J=1$ the phase transition occurs at $\theta_\text{c}=\frac{3+K^2}{4}$.\cite{Lahtinen10} Fig.~\ref{fig:Gaps_PTs} shows that the bulk energy gap $G$ indeed vanishes at all these critical points, which is also witnessed as the diverging rate of change in the entropy. As the chiral topological phase with $\nu \neq 0$ exist only for $K \neq 0$, for $K=0$ time-reversal symmetry is restored and the regimes of the chiral states become gapless semi-metallic states. These can be viewed as critical points between chiral phases characterized by $\nu$ and $-\nu$.

The key property relating to entanglement in the distinct topological phases is the existence of $|\nu|$ Majorana edge states per physical edge. In the first part of our study, we explicitly show how these give the dominant contributions to the entanglement entropy and provide a universal lower bound to it. In the second part we focus on the gapless phases and critical points and identify their universality classes via critical entanglement scaling.

%\begin{figure}[t]
%\includegraphics[scale=0.5]{PDfiniteK}
%\includegraphics[scale=0.5]{PDChern}
%\caption{\label{fig:PD} Phase diagram showing the energy gap $G$ (Top)
%and the corresponding Chern number $\nu$, for $K>0$.
%\kon{Between $\nu=0$ and $\nu=2$ for $\theta>1/2$: hofstadter model with staggered strips, has $\\nu=1$.}
%\kon{Chern $2$ between the vf and fv TC's for $J\approx < 0.2$ ?: Running more data points to see what is there.} }
%\end{figure}

\subsection{Long-range vs. short-range entanglement in the honeycomb model}

Before proceeding, we review briefly the long-range entanglement that arises in the full honeycomb model when not restricted to a specific vortex sector. The entanglement entropy of the full spin model has been studied in Ref \onlinecite{Yao10}. There it is shown that it splits into two independent contributions
\be
  S={S}_{\rm{f}}+{S}_{\rm{g}},
\ee
where $S_{\rm{f}}$ is the entanglement of the fermionic degrees of freedom, while $S_{\rm{g}}$ depends only on the fluctuating gauge degrees of freedom $\{ u_{ij} \}$. In particular, it was shown that $S_{\rm{g}}$ contains the universal topological entanglement entropy~\cite{KitaevPreskill} and that it takes always the value of $\gamma=-\log{2}$ regardless of the microscopic paramaters $J_r$, $K$ and $\theta$. Hence, the honeycomb model exhibits intrinsic long-range entanglement akin to fractional quantum Hall states. Nevertheless, the extraction of $\gamma$ gives no information about the particular phase of the model. Indeed, all the different anyon models that appear for different Chern numbers have total quantum dimension of $D=2$ consistent with $\gamma=\log{D}$.\cite{KitaevHoneycomb}

Thus area law part of entanglement entropy due to $S_{\rm{f}}$, that is usually considered to be non-universal, must contain all the information about the excitations and the edge spectrum. The main result of our work is to uncover these universal signatures related to the topological phases hiding in the area law. As we will always be working on a fixed vortex sector, we neglect below the constant $S_{\rm{g}}$. When referring to the entanglement entropy $S$, it is understood that we refer to the fermionic entropy $S_{\rm{f}}$ and we omit the subscript for clarity.

\section{Anatomy of Entanglement in the Gapped Phases}
\label{Entanglement Entropy of Topological Phases}

In this section we study in detail the structure of the entanglement entropy in the gapped topological phases of the honeycomb model.
In such a phase, entanglement for a given bipartition is always short-ranged and concentrated along the cut, thus giving rise to the area law~\cite{Area}.
We split the fermionic entropy into $S_\text{edge}$ due to edge states and to $S_\text{bulk}$ due to bulk contributions
\be
	 S=S_\text{edge}+S_\text{bulk}.
\ee
In the following, we focus on $S_{\text{edge}}$ to study how it encodes the dominant short-ranged entropy along the cut and the universal signatures of the topological phases.
Our motivation to focus on $S_\text{edge}$ is to explicitly study how the entanglement due to edge states as it encodes universal signatures of topological phases.
By analytically solving for the dispersion of the edge states, we obtain an approximate expression for the $S_\text{edge}$ that shows how it depends on their velocity. This result reproduces the known lower bound of entanglement entropy\cite{Fidkowski10,Klich,Mei15} due to the presence of topological edge states. We elaborate on the origin of this lower bound by showing that it arises from 1D subsystems of the full 2D model having completely decoupled edge states. Moreover, we construct a simple effective model that accurately approximates the total entropy due to their hybridization.

\begin{figure}[t]
$\begin{array}{cc}
\includegraphics[scale=0.222]{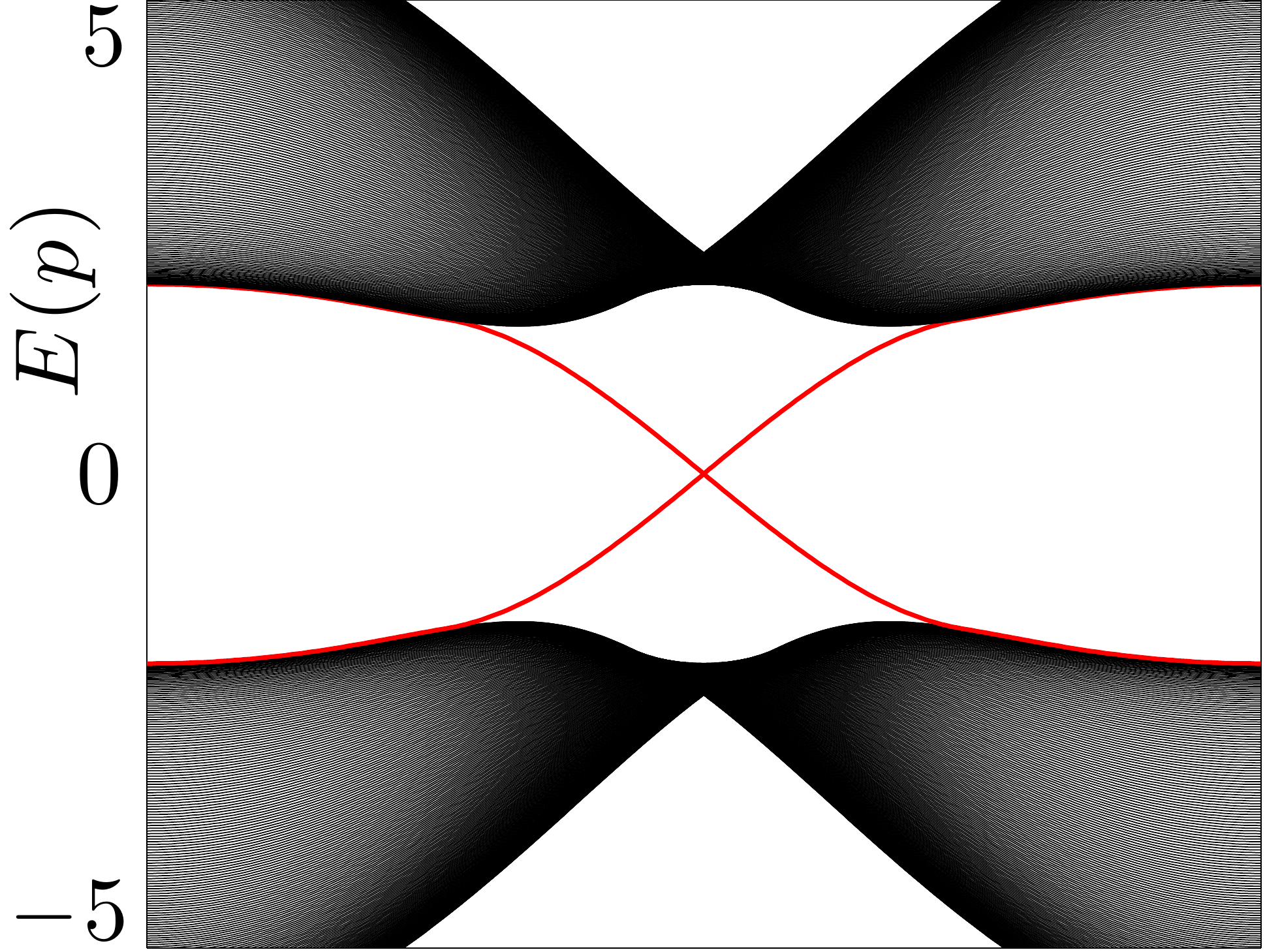}&\includegraphics[scale=0.222]{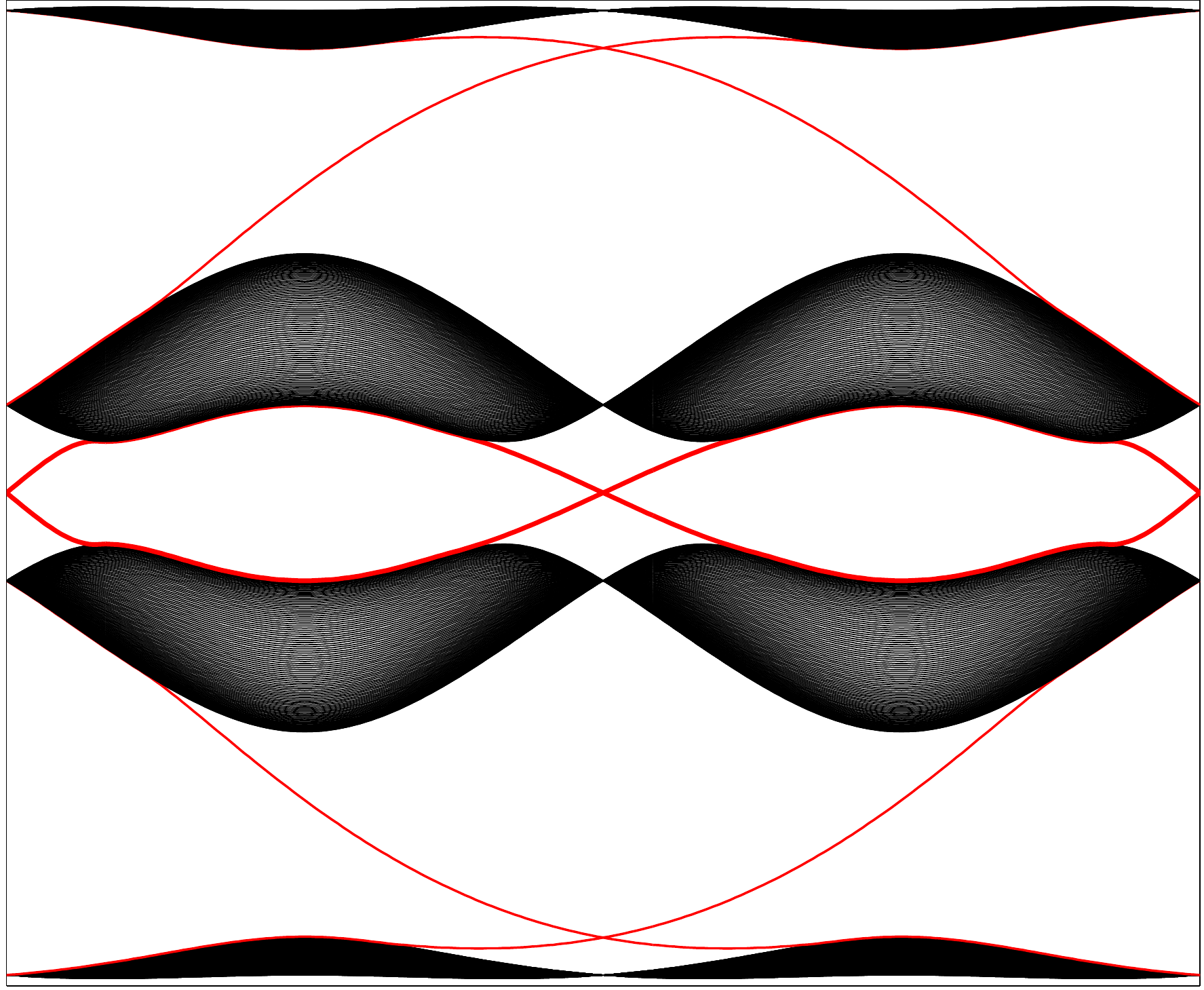}\\
\includegraphics[scale=0.222]{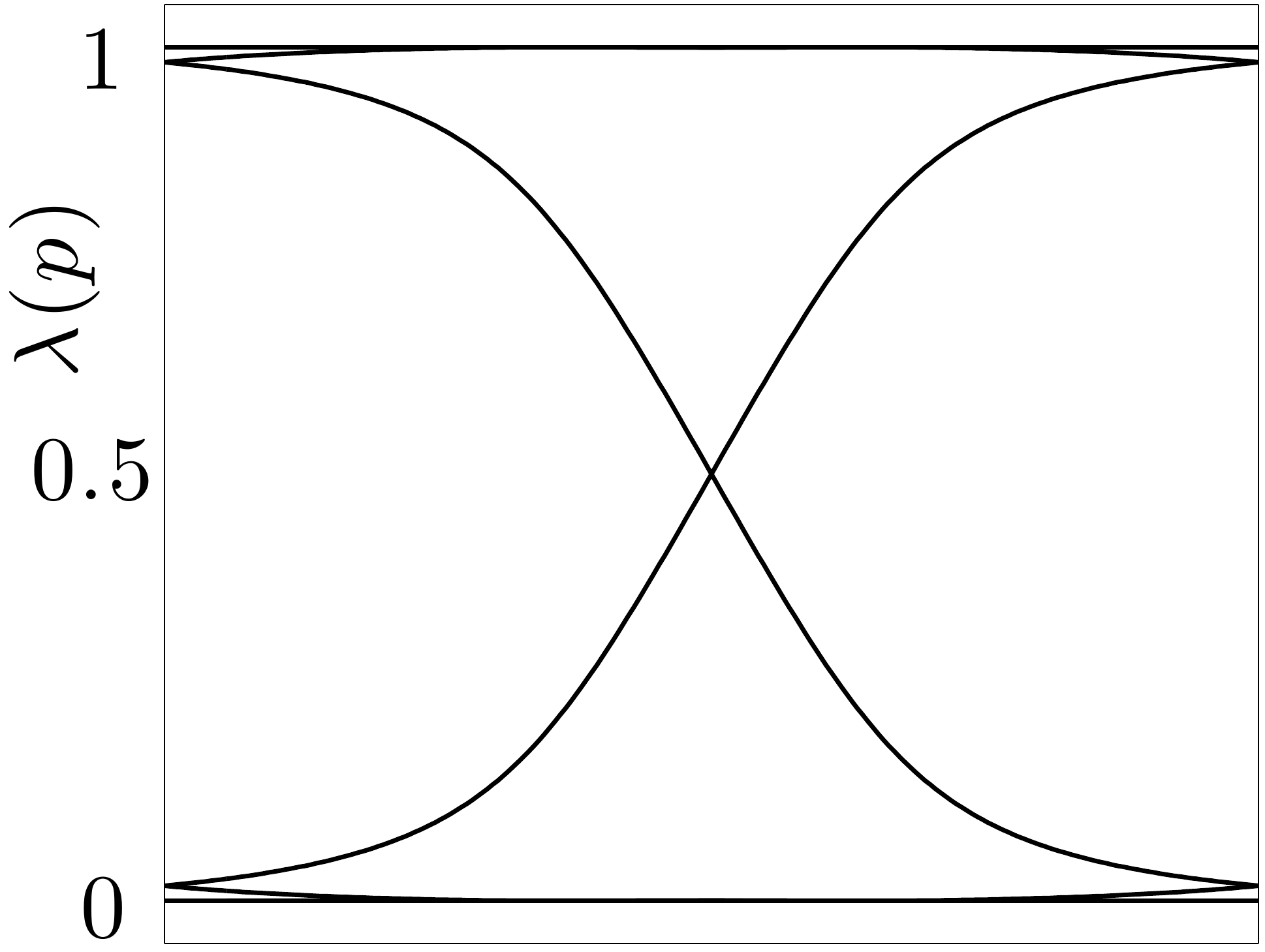}&\includegraphics[scale=0.222]{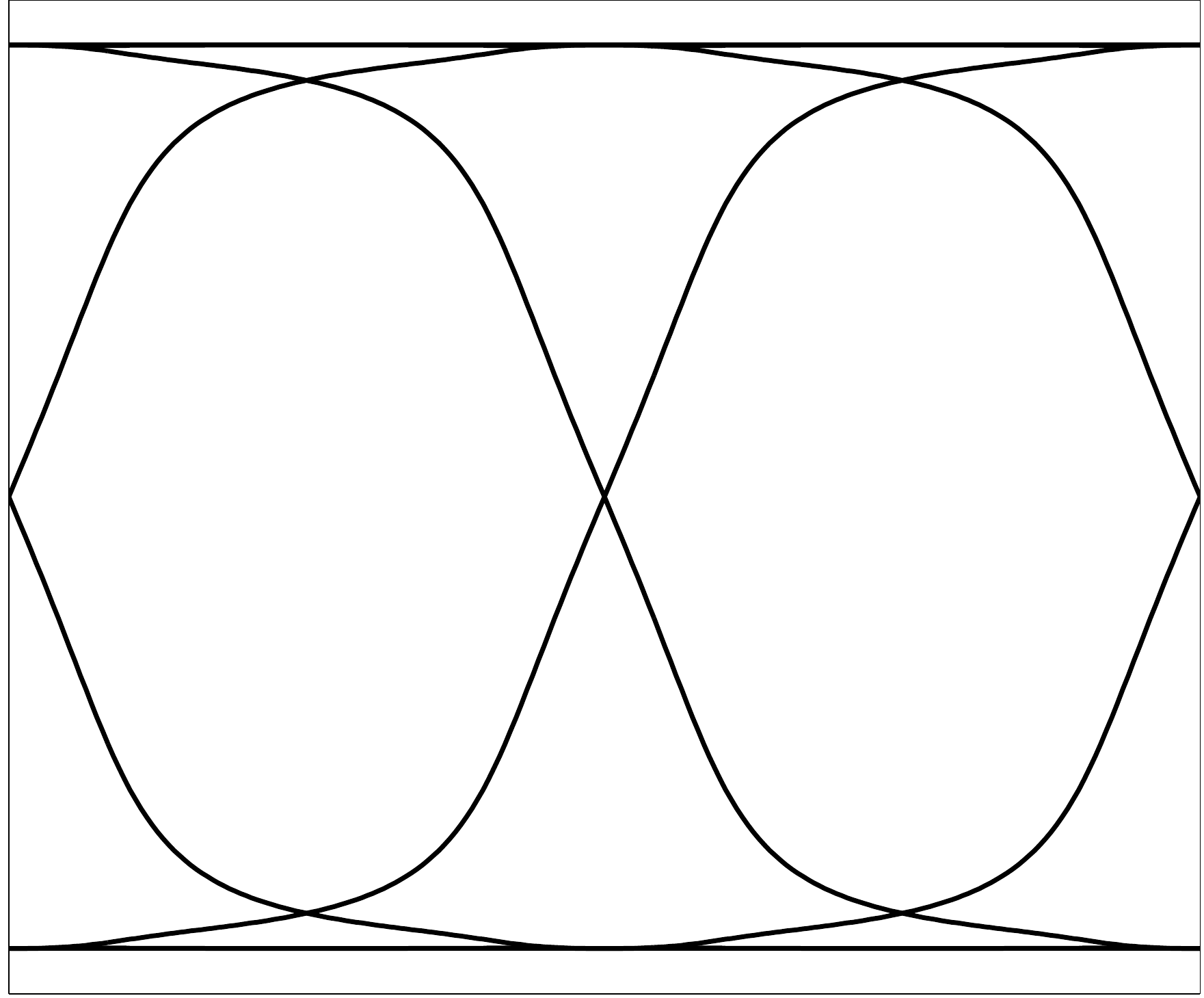}\\
\includegraphics[scale=0.222]{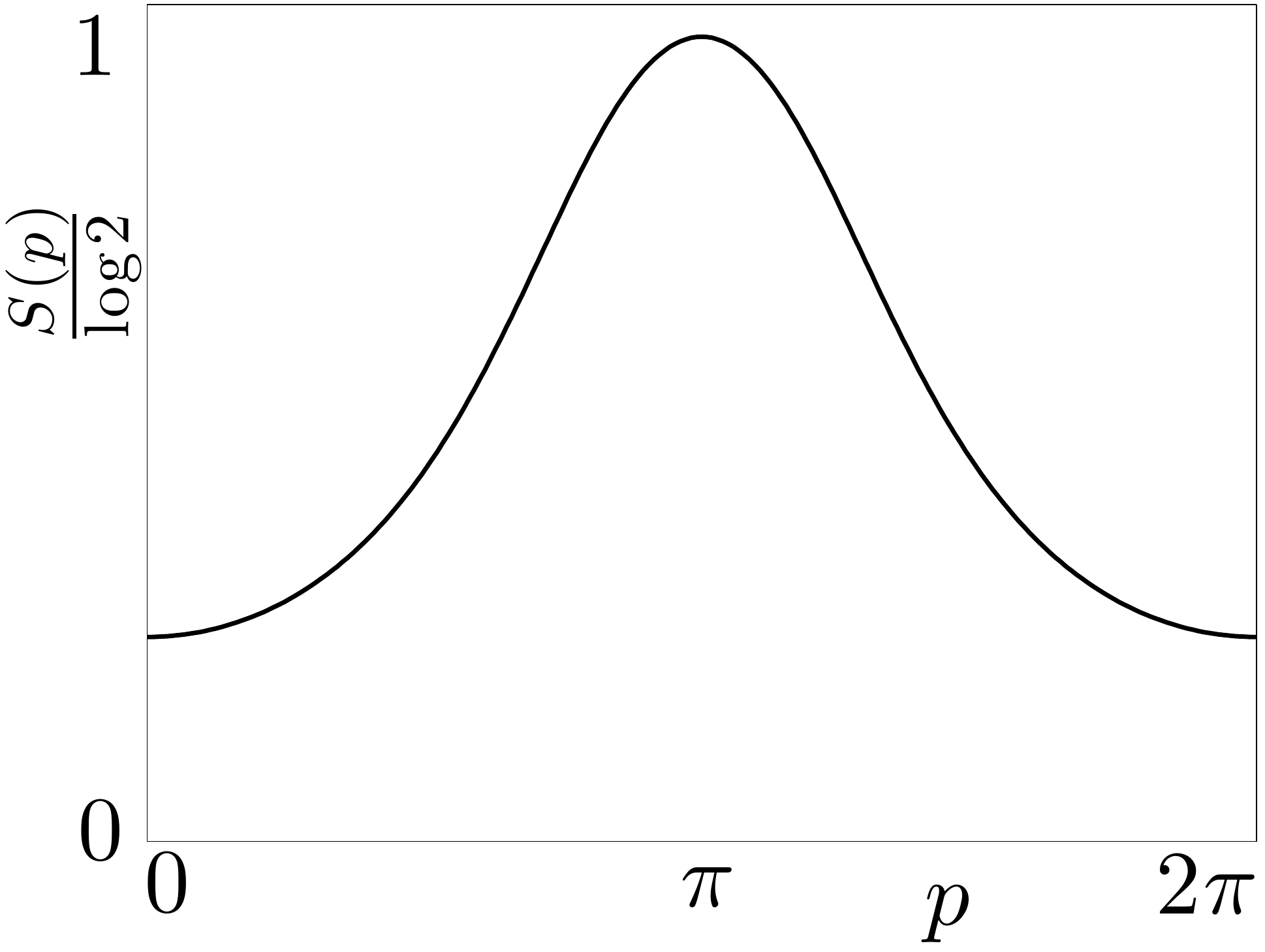}&\includegraphics[scale=0.222]{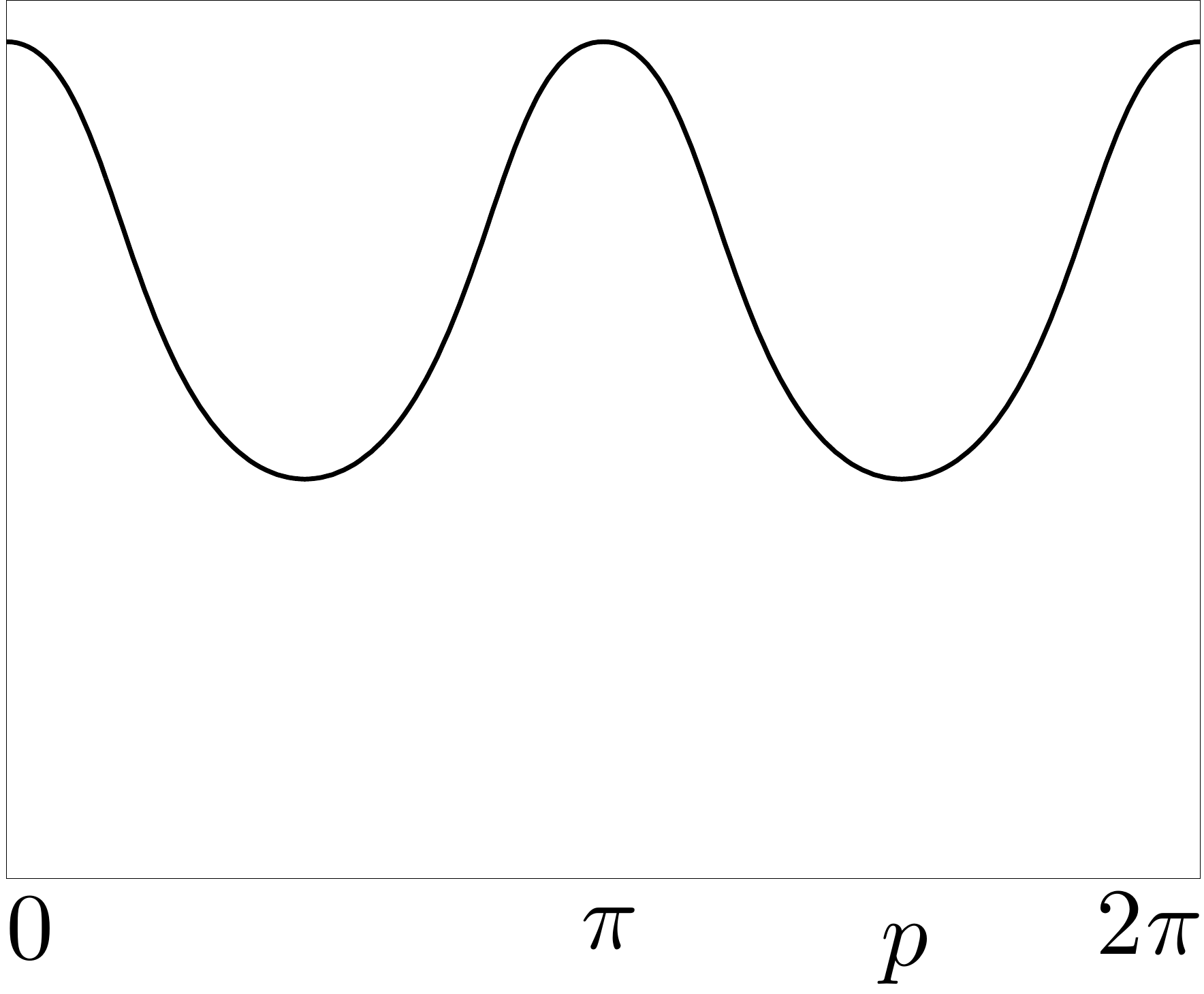}
\end{array}$
\caption{\label{fig:lambdaGapped}
Energy spectrum $E(p)$ on a cylinder and the corresponding the entanglement spectrum $\lambda(p)$ and the entanglement dispersion $S(p)$ for the $\nu=1$ (Left) and $\nu=2$ (Right) phases. The real space edge states are indicated in red. Here $J=1$ and $K=0.15$.}
\end{figure}

\subsection{Edge-Mode Dispersion}\label{edgedisp}

To explicitly show how the edge states give rise to the dominant area law behavior, we focus on the vortex-free sector that supports a topological phase characterized by $\nu=1$, implying a single Majorana edge state per edge. To connect these edge states to the entanglement entropy, we analytically evaluate their real space energy dispersion for generic values of the microscopic parameters $J$ and $K$~\cite{Pershoguba, Sticlet, Sticlet2, MongShivamoggi}. Here we present the main steps of the calculation and refer to Appendix \ref{GenFunMethod} for the details.

In the vortex-free sector ($\theta=0$) each chain in the Hamiltonian \rf{eq_HforChains_main} can be written as
\bq
H(p)=\sum_{j} \Big( &&\chi_j^\dagger \Gamma_1\chi_j+\chi_j^\dagger \Gamma_2\chi_{j-1}+\chi_j^\dagger \Gamma_2^\dagger\chi_{j+1}\nonumber\\
&&+\chi_1^\dagger \Gamma_1\chi_1+\chi_1^\dagger \Gamma_2^\dagger\chi_2\Big),
\eq
respectively, where $\chi_j={\left(\begin{array}{cc}a_j&b_j\end{array}\right)}^T$, $\Gamma_1= \left(\begin{array}{cc}\zeta&\omega\\{\omega}^*&-\zeta\end{array}\right)$ and $\Gamma_2=\left(\begin{array}{cc}\gamma&J\\0&-\gamma\end{array}\right)$
with
$\zeta=- K \sin{p}$, $\omega=i(J e^{i p}+1)$, and $\gamma=-\frac{i}{2}K(e^{-i p}-1)$. Imposing open boundary conditions in one direction, i.e. placing the system on a cylinder, the Schr\"odinger equation $H(p)\Psi=E\Psi$ leads to the following recursive equation
\be
\Gamma_2\Psi_{j-1}+(\Gamma_1-E)\Psi_j+\Gamma_2^\dagger\Psi_{j+1}=0,
\label{eqn:req}
\ee
and boundary condition
\be 
(\Gamma_1-E)\Psi_1+\Gamma_2^\dagger\Psi_{2}=0,
\label{eqn:bound}
\ee
where $\Psi=\sum_{j=1}^\infty \chi_j^\dagger\Psi_j\ket{0}$ and $\Psi_j=(\Psi_j^1~\Psi_j^2)^T$. Multiplying \rf{eqn:req} by $z^j\in\mathbb{C}$, summing over $j$ and using the boundary condition \rf{eqn:bound} we obtain
\be\label{eq:GenFunction}
\big[z^2 \Gamma_2+z(\Gamma_1-E)+\Gamma_2^\dagger\big]G(z)=\Gamma_2^\dagger\Psi_1.
\ee
The generating function $G(z)=\sum_{j=1}^\infty z^{j-1}\Psi_j$ can be explicitly solved and it is given by
\be
G(z)=\big[z^2 \Gamma_2+z(\Gamma_1-E)+\Gamma_2^\dagger\big]^{-1}\Gamma_2^\dagger \Psi_1.
\label{eq:GGG}
\ee
The edge states and their energy dispersion can be obtained from the generating function by analyzing its poles: If $G(z)$ has a pole at $z$ with $|z|>1$, then the corresponding state $\Psi_j$ has energy $E$ and it is exponentially localised at the boundary.~\cite{Pershoguba} 

In the vortex-free sector the zero energy states are found to appear at $p^\star=\pi$. Linearizing around it, we obtain the edge state dispersions
\be
\label{eq:EnergyBoundaryModesLinear_main}
E_\pm(p) = \pm v_E p + \ldots \, ,
\ee
where the velocity is given by
\be\label{eq:velEnergyBoundaryModesLinear_main}
	v_E = \left(2J+1\right)\cos\left(\arctan{\frac{J}{4K}}\right).
\ee
This analytic expression for the velocity is valid in the momentum range $\Delta p$ that is given by the existence of poles in the generating function $G(z)$ (see Appendix C). While it can not in general be obtained analytically, one finds that $\Delta p$ coincides accurately with the distance between the Fermi momenta between which the edge states exist. Thus this momentum interval contains the dominanant and universal entanglement properties of the topological phase.

\begin{figure}[t]
\includegraphics[scale=0.35]{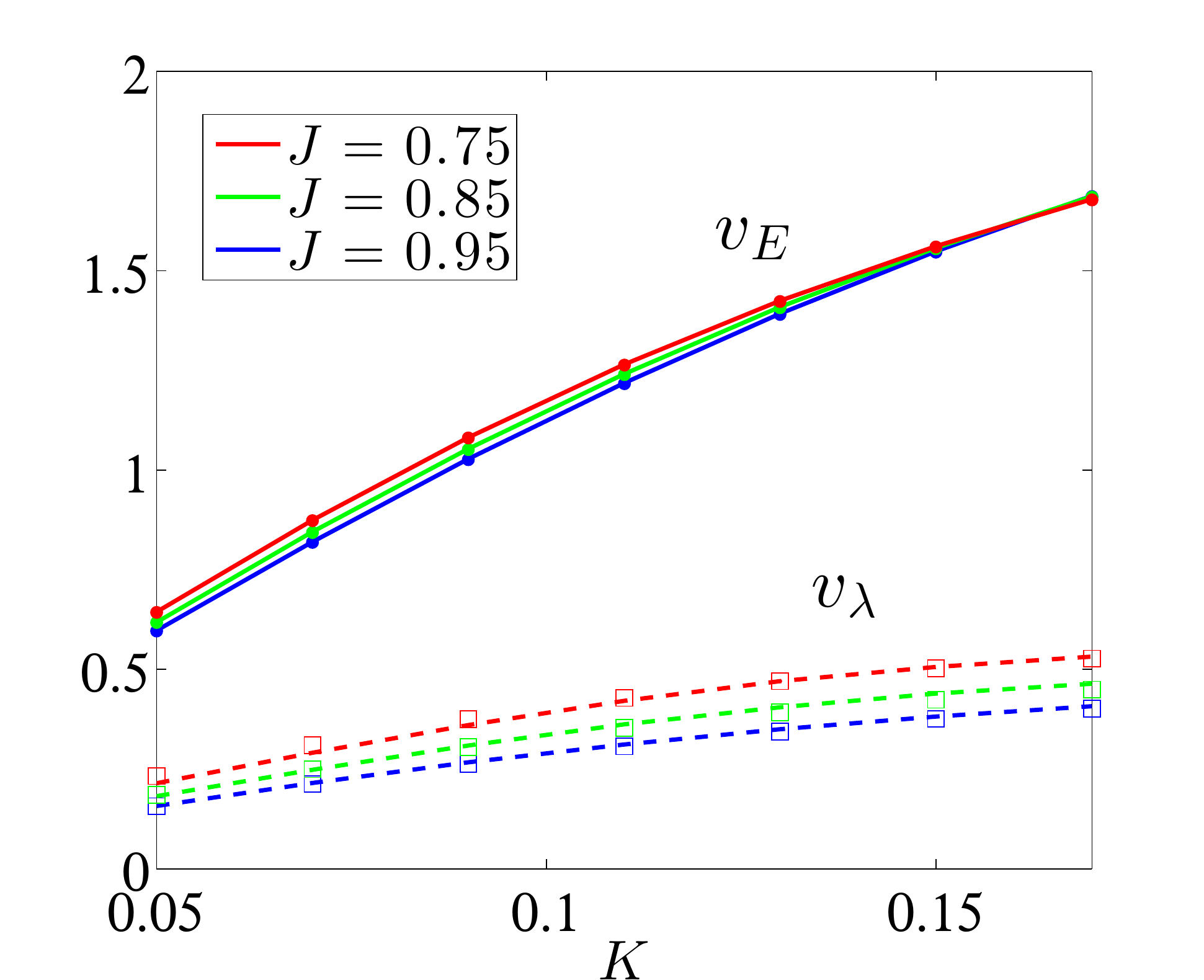}
\caption{\label{fig:velocitieskappa}
The correspondence between the physical velocities $v_E$ and the virtual velocities $v_\lambda$.
The analytic expression (solid line) given by (\ref{eq:velEnergyBoundaryModesLinear_main}) for the physical velocities agrees with the numerical values (dots). Via the ansatz \rf{eq:velocities} (dashed line) it reproduces the numerically obtained virtual velocities (squares). %Here $L_x=20, L_y=60$.
}
\end{figure}

To employ the physical edge states to approximate the entanglement entropy, the physical velocity $v_E$ needs to be related to the virtual velocity $v_\lambda$ of the virtual edge states in the entanglement spectrum.
We recall that the energy spectrum with open boundaries and the entanglement spectrum for periodic boundaries are adiabatically connected~\cite{Fidkowski10}, i.e. the latter is a band-flattened version of the former. Due to this fact, it is both physically and intuitively motivated to assume that the velocities are also proportional to each other and we write,
\be\label{eq:velocities}
v_\lambda\approx{\kappa}v_E,
\ee
since a rescaling of the slope of the edge states is the simplest deformation to consider.
To obtain an ansatz for $\kappa$, we first assume that the energy spectrum is flattened around its half-bandwidth energy at $p^\star$ and the energy of this band is rescaled to unity. Subsequently, we center the entanglement spectrum around zero and rescale it as $\lambda\rightarrow 2(\lambda-\frac{1}{2})$ such that every entanglement level takes values between $\pm 1$. This leads to the ansatz ${\kappa = (2G(p^\star)+W(p^\star))^{-1}}$, where $G(p^\star)$ and $W(p^\star)$ are the energy gap and the bandwidth at $p^\star$, respectively.
Indeed, Fig.~\ref{fig:velocitieskappa} confirms this ansatz to provide an accurate approximation of $v_\lambda$ for a wide range of couplings $J$ and $K$.
Even thought this ansatz is phenomenological, it concerns the velocity-dependent part of the entanglement spectrum which contributes most of the edge entanglement, and thus is should be general for topological phases exhibiting similar entanglement structure.

\begin{figure}[t]
\includegraphics[scale=0.2839]{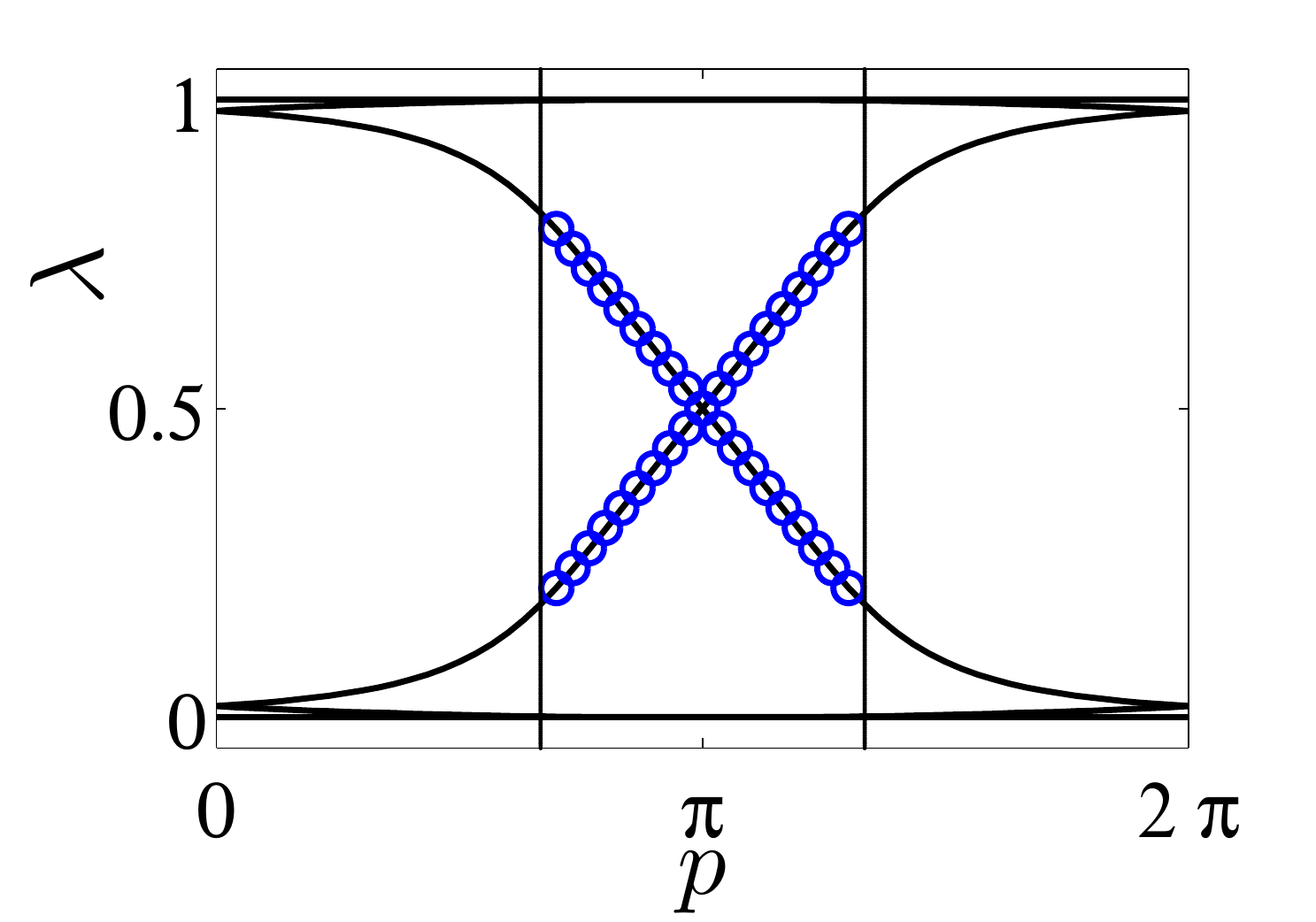}
\includegraphics[scale=0.2839]{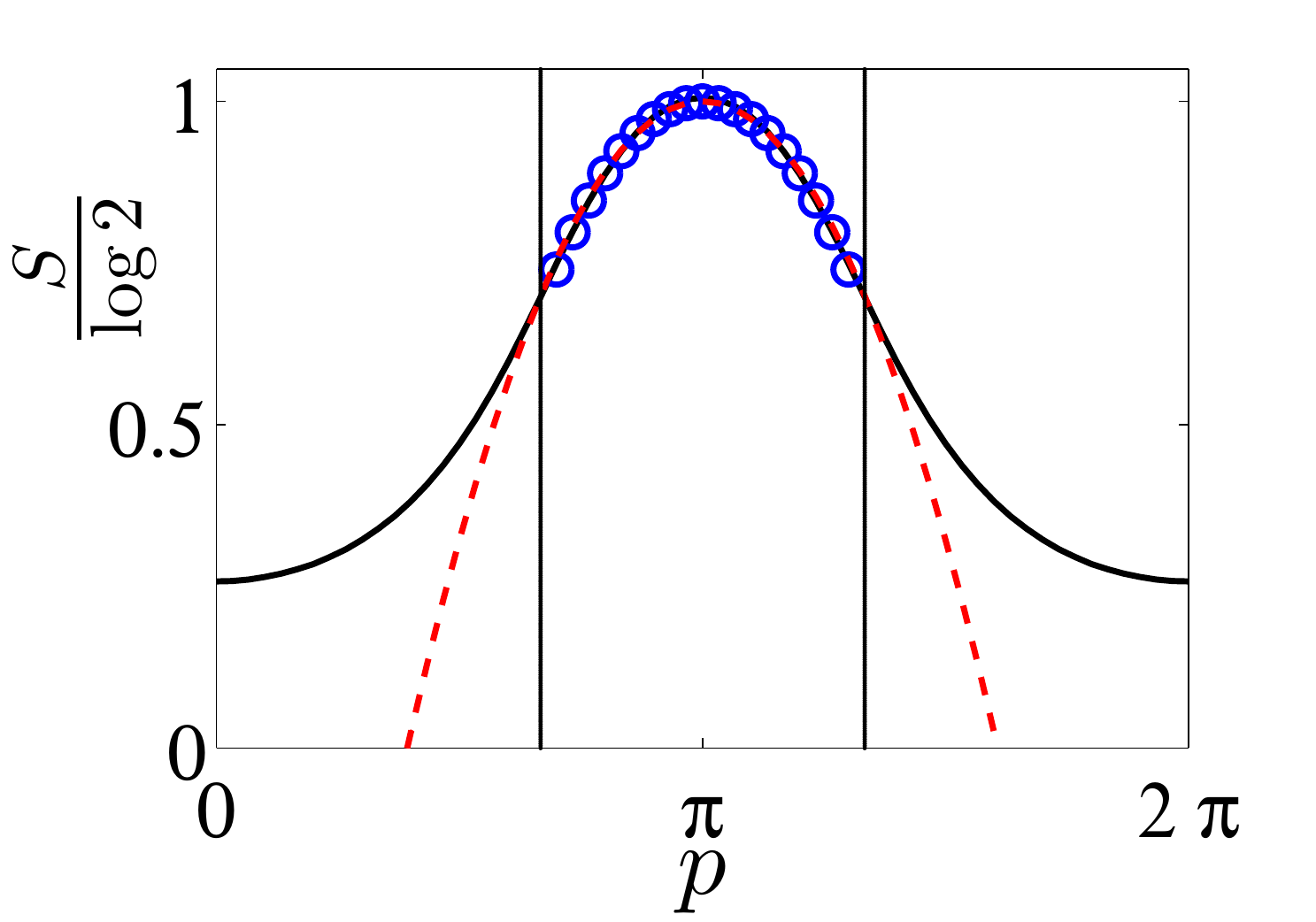}
\caption{\label{fig:dominance} 
Entanglement spectrum (Left) and entropy (Right) in the momentum interval $\Delta p$ (between vertical lines) in the vortex-free ($\theta=0$) sector with $J=1$ and $K=0.12$.
A linear approximation $\lambda^\pm$ with velocity $v_\lambda$ obtained form \rf{eq:velocities} contributes almost all of the entropy inside $\Delta p$ (blue circles). The analytic expression \rf{eq:Sedgelinq} (red dashed) as a quadratic approximation to $S_\text{edge}(p)$ reproduces the entropy within accuracy of $0.1\%$ inside $\Delta p$.
}
\end{figure}

The entanglement spectrum can then be approximated by the linearized expression
\be\label{eq:lambdalin}
	\lambda^\pm(p)=\pm {v}_\lambda p+\frac{1}{2},~~~p\in\Delta p.
\ee
Inserting this into \rf{eq:entropyFromProbabilities} and expanding around $p^\star$, the dominant entropy contributions can then be approximated by
\be\label{eq:Sedgelinq}
S_\text{edge}(p)\approx \log{2} - 2 v_\lambda^2 p^2 + \mathcal{O}(p^4),
\ee
where we have retained only the lowest order terms in $p$. Fig.~\ref{fig:dominance} shows that this expression provides an excellent approximation of the entropy within the momentum interval $\Delta p$ where the virtual velocity is directly obtained from the physical velocity of the edge states.

\begin{figure}[t]
\centering
\includegraphics[scale=0.38]{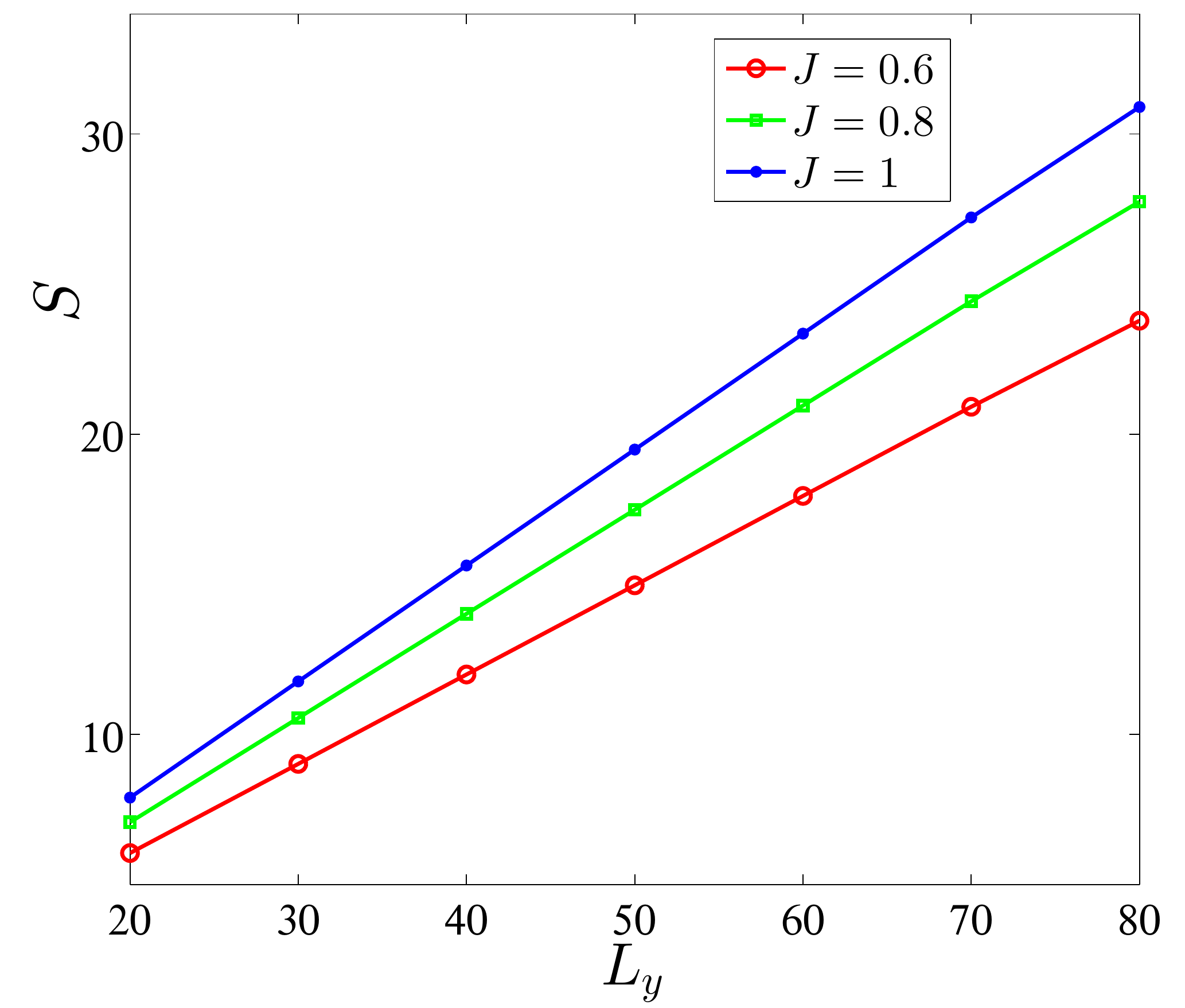}
\begin{picture}(0,0)
\put(-206,103){\includegraphics[scale=0.21]{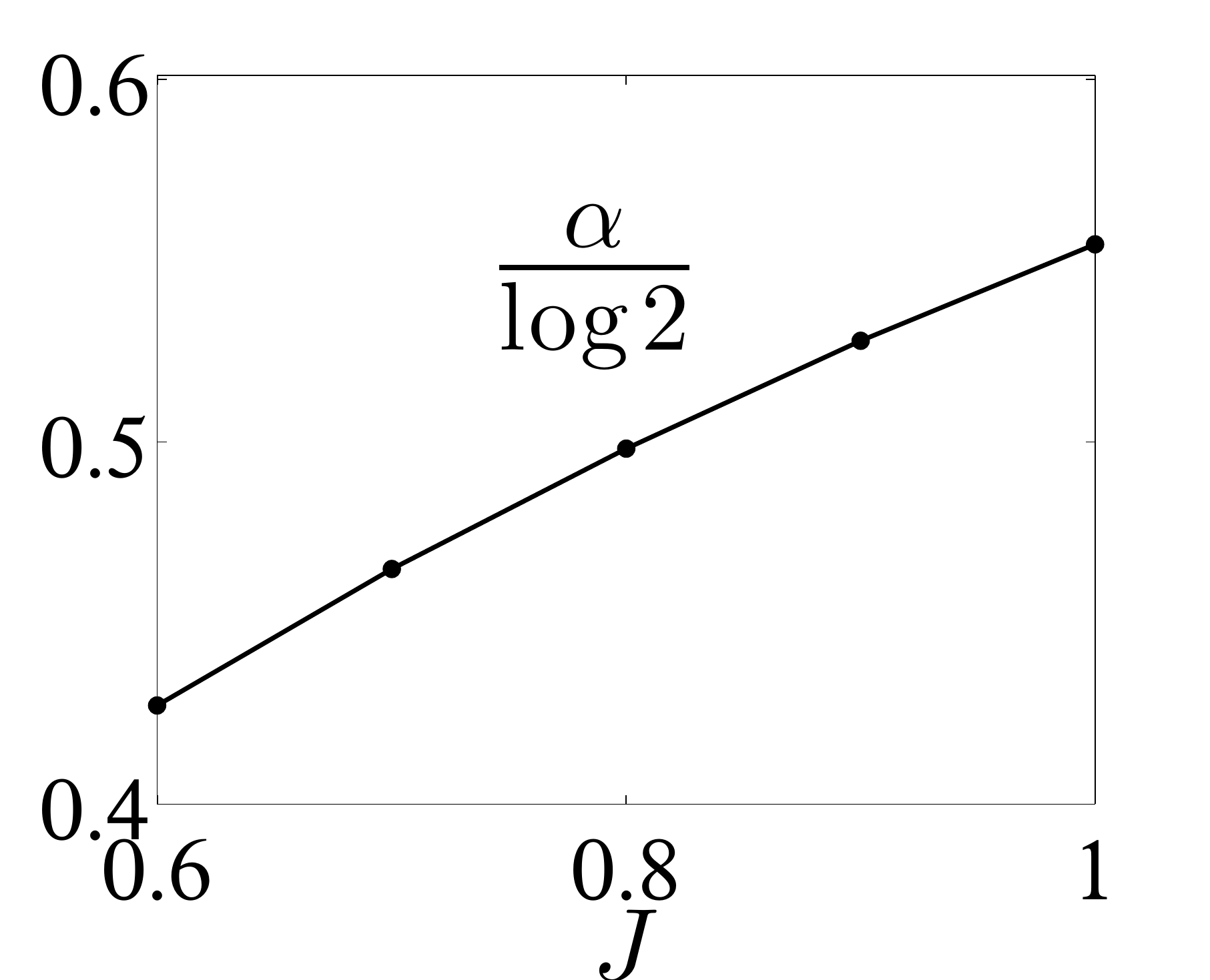}}
\put(-90,25){\includegraphics[scale=0.34]{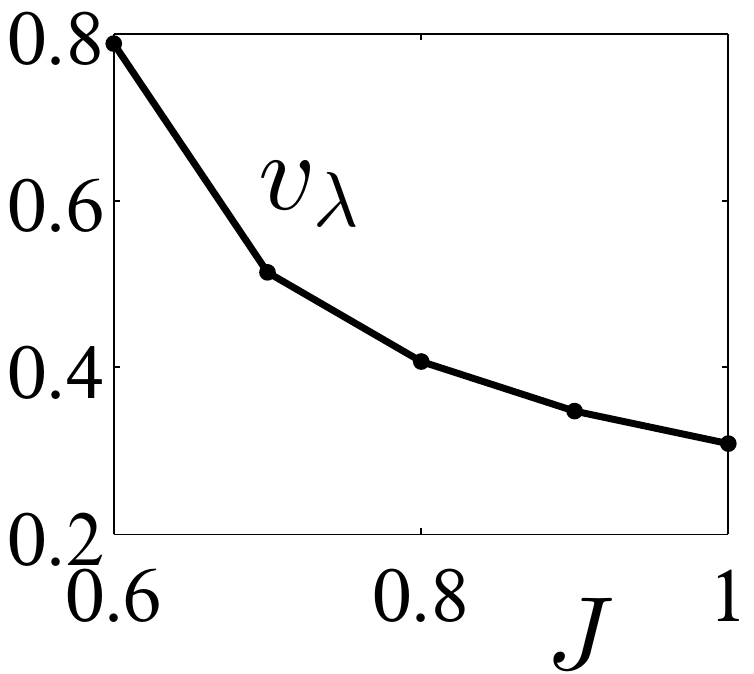}}
\end{picture}
\caption{\label{fig:alphaupperbound} 
Area law $S=\alpha L_y$ obeyed by total entropy. The area law coefficient $\alpha$, theoretically upper bounded by $\log{2}$, increases as the virtual velocity $v_\lambda$ decreases as a function of $J$ (insets). Here $K=0.12$.
}
\end{figure}

This expression shows two general properties of the entanglement entropy due to the presence of topological edge states. First, the contribution by each momentum $p$ is upper bounded by $\log 2$ with the contribution decreasing with increasing velocity $v_\lambda$. As the perimeter of the cut $\partial A$ is directly proportional to the momentum range $\Delta p$, this means that the area law contribution $S \sim \alpha |\partial A|$ should be smaller when the velocity increases. Indeed, Fig. \ref{fig:alphaupperbound} shows that as the microscopic parameters are tuned to give smaller velocities, the entanglement entropy grows faster with increasing of the cut length $|\partial A|$, i.e. the area law coefficient $\alpha$ increases. Second, as the velocity can in principle take arbitrary values, the upper bound for each $S(p)$ implies that the total entropy has a theoretical lower bound of $\log 2$, consistent with earlier studies.\cite{Fidkowski10,Klich,Mei15} It can in principle be recovered via adiabatically tuning to the limit $v_\lambda \to 0$. Physically, this is artificial though as it corresponds to a Hamiltonian with infinite range couplings.\cite{Eisert}

\subsection{Microscopics of the Lower Bound}

We now turn to analyze the microscopic origin of the lower bound due to edge states~\cite{Fidkowski10}. Previously, the lower bound was considered from the perspective of particle number fluctuations~\cite{Klich} and edge state entanglement~\cite{Mei15}.
Here we provide a complimentary perspective for the lower bound in terms of the end states of the microscopic 1D chains at the momenta $p^\star$.
To this end we focus first on the 1D wire corresponding to the only crossing point of the edge states at $p^\star=\pi$ in the vortex free sector, as shown in Fig.~\ref{fig:lambdaGapped}. For $\theta=0$ the general Hamiltonian \rf{eq_HforChains_main} takes the simple form 
\be\label{eq:Hp=pi}
H(\pi)=i\sum_{j\in\Lambda_{\pi}}\Big[(1-J) a_j b_j+ J a_j b_{j-1}+K (a_j a_{j-1}- b_j b_{j-1})\Big],
\ee
where the operators $a_j$ and $b_j$ are now again Majorana fermions. As explicitly shown in Appendix \ref{KitaevModelApp}, this can be recognized as the Hamiltonian of Kitaev's Majorana chain~\cite{KitaevWire} with $\mu=2(1-J)$, $t=\text{Re}(\Delta e^{i\phi})=2 J$ and $\text{Im}(\Delta e^{i\phi})=-4 K$. The Kitaev chain has two phases, a topologically trivial and a non-trivial one. The non-trivial phase occurs for $|t|>|\mu|$, which corresponds to $J>1/2$, in agreement with this regime in the vortex-free sector being in the topological phase characterized by $\nu=1$. In fact, the chain at $p^\star =\pi$ is special as the phase diagram does not depend on $K$ ($\Delta_I$ can be gauged away). Thus this wire exhibits also a topological phase with Majorana edge states for $K=0$ that corresponds to the gapless phase of the honeycomb model, consistent with Fig. \ref{fig:lambdaGapless} showing a flat band of edge states between the Fermi points. 

Majorana wires of the form \rf{eq:Hp=pi} have exact topologically protected zero energy edge modes throughout the topological phase~\cite{FendleyChain}. When the wire with periodic boundary conditions is partitioned into $A$ and $B$, two Majorana end modes $a_L$ and $a_R$ appear at the boundary $\partial A$ that are entangled with the counterpart edge modes $b_L$ and $b_R$ in $B$. Thus when tracing out $B$ to calculate the entanglement entropy, each end mode in $A$ contributes entropy of $\log \sqrt{2}$ corresponding to the half a fermion they encode in $A$. These are precisely the maximally entangled $\lambda=1/2$ eigenvalues in the entanglement spectrum. Since their origin is topological and hence can not be adiabatically removed,\cite{Turner10} the total entropy in the topological $\nu=1$ phase, and the corresponding gapless phase for $K=0$, must satisfy $S\geq 2\log{\sqrt{2}}$.
Due to the fact that the existence of edge states in \rf{eq:Hp=pi} does not depend on $K$, this lower bound of fermionic entanglement entropy applies both to the gapped $\nu=1$ topological phase as well as to the gapless semi-metallic phase of the honeycomb model.

Similar analysis can be carried out in the full-vortex sector ($\theta =1$). As shown in Fig.~\ref{fig:lambdaGapped}, exact zero energy edge states correspond now to two chains at $p^\star=0,\pi$. In the limit $K=0$, one finds that at $p^\star=0$ the Hamiltonian \rf{eq_HforChains_main} has three decoupled Majorana sites at the edges (that are coupled between each other), while at $p^\star=\pi$ there is only a single decoupled Majorana site. Thus both momenta support a single zero energy Majorana end mode, while $p^\star=0$ supports also a single complex fermion mode at finite energy, consistent with Fig. \ref{fig:lambdaGapped}. In line with our analysis above, the doubled number of Majorana edge modes in $A$ entangled with their counterparts in $B$ gives then rise to a lower bound of the entanglement entropy that is twice that of the vortex-free sector $S \geq 4\log{\sqrt {2}}$.

These lower bounds give joint information about the number and dimensionality of the edge states. While neither can be accessed independent of each other by purely entropic means, they can serve as valuable complementary information to identify the nature of topological phases in both free and interacting systems.\cite{Mei15}

\subsection{Effective Model for Edge Entropy}

\begin{figure}[t]
\centering
\includegraphics[scale=0.35]{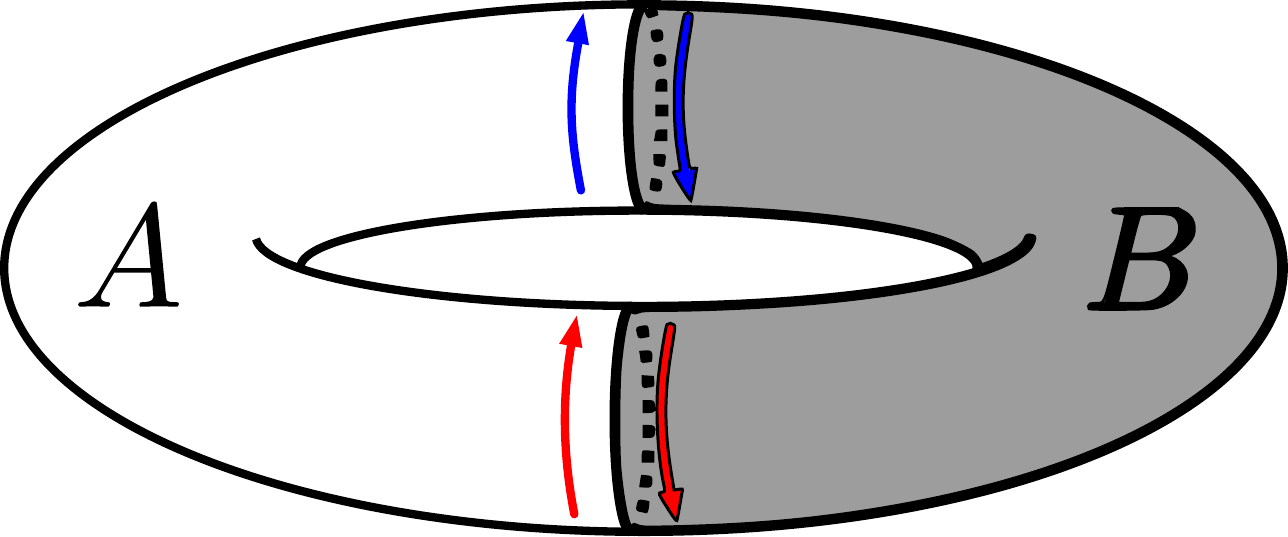}~~~\includegraphics[scale=0.33]{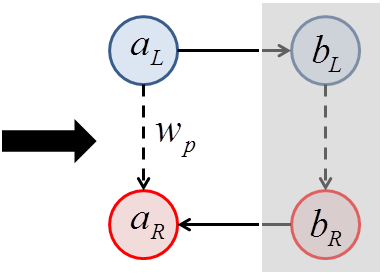}
\includegraphics[scale=0.34]{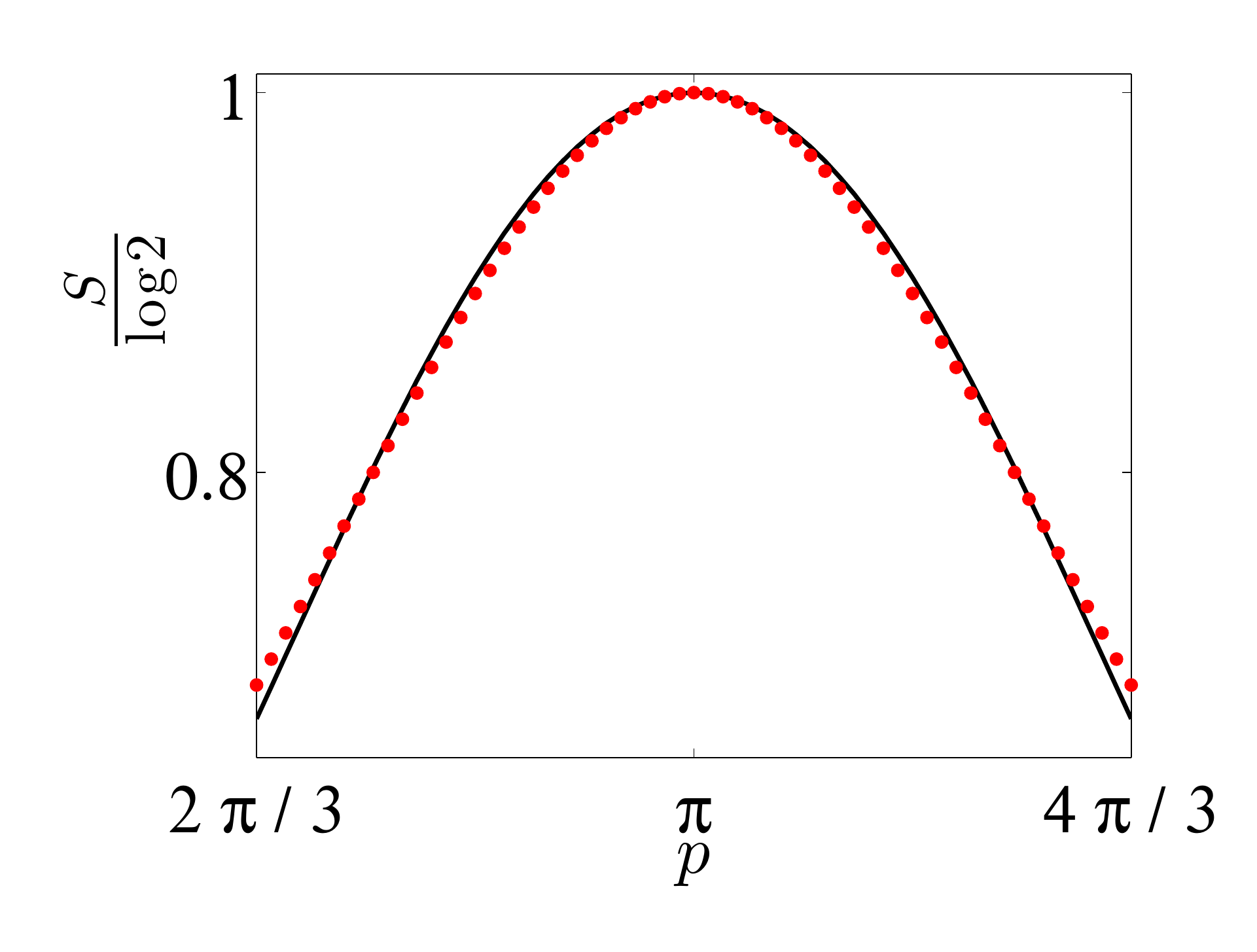}%
\begin{picture}(0,0)
\put(-128,30){\includegraphics[scale=0.28]{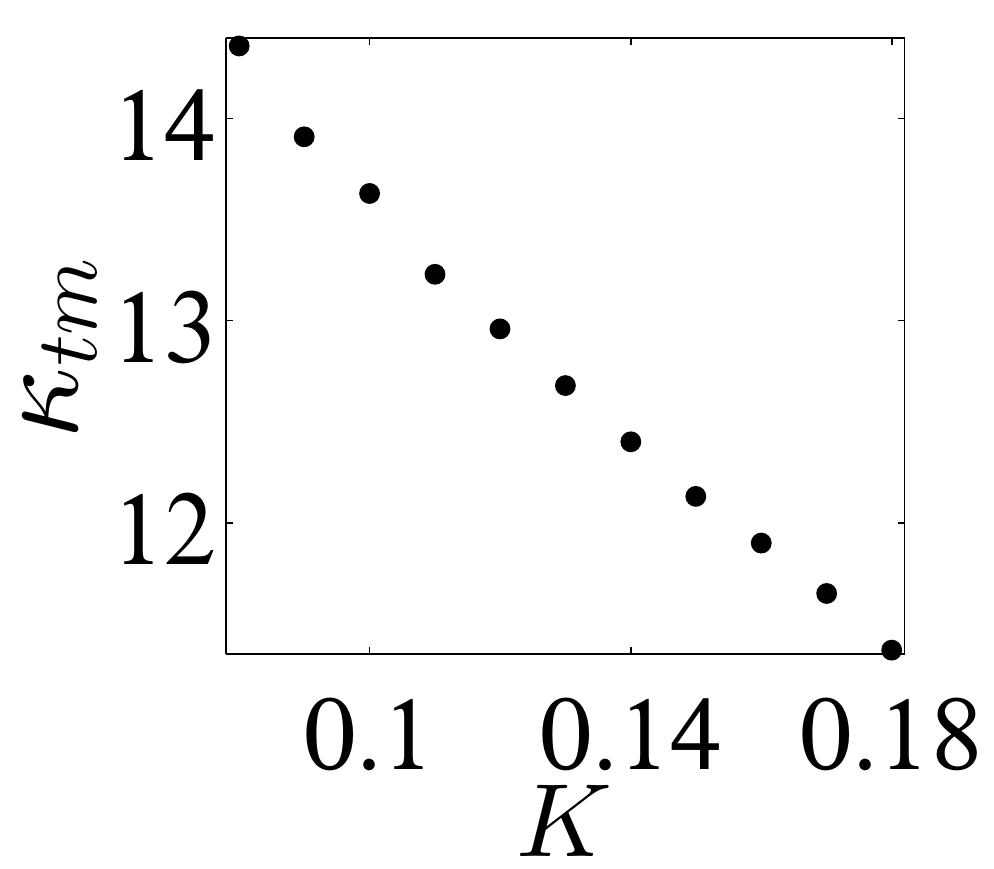}}
\end{picture}
\caption{\label{fig:TM}
The effective model for the edge entropy $S_\text{edge}$. (Top) Partitioning the system on a torus to $A$ and $B$ gives rise to two edges with counter-propagating Majorana edge states in both partitions. We take the edges to correspond to the sites of a four-site tight-binding model, with Majorana modes $a_L$ and $a_R$ living on the edges of $A$ and $b_L$ and $b_R$ living on the edges of $B$. These hybridize according to the Hamiltonian \rf{eq:effectiveHam}, where the solid (dashed) arrows correspond to strong (weak) tunneling of magnitude $1$ ($w_p=\kappa_{\text{tm}} |E_\pm(p)|$). The sign of the tunneling is positive along the arrows, which gives $\pi$-flux on the single plaquette. (Bottom) Entropy $S_{\text{tm}}$ (red dots) between the Fermi points obtained from the effective model reproduces acccurately the entropy $S$ (black line) of the $\nu=1$ phase of the honeycomb model. The data is for $L_x=24, L_y=60$, $K=0.125$ and the value of the optimized proportionality constant $\kappa_{\text{tm}}$ as the function of $K$ is shown in the inset. }
\end{figure}

We have shown how the protected edge states dominate the entropy and give rise to a lower bound to it. Now we turn this perspective around and ask how much of the total entropy can be understood in terms of the edge states only?
Discarding all bulk entanglement, we model the entanglement due to edge states with a simple four-site tight-binding model, as illustrated in Fig.~\ref{fig:TM}. Each site corresponds to one of the four edges of the two cylinders that result from partitioning the system on a torus to regions $A$ and $B$. Denoting the Majorana edge modes on the left and right edges of $A$ by $a_L$ and $a_R$ (and similarly for $B$), the Hamiltonian is given by
\be\label{eq:effectiveHam}
H=i (a_L b_L - a_R b_R ) + i w_p ( a_L a_R + b_L b_R),
\ee
where the signs are chosen such that there is $\pi$-flux going through the chain (fermions on a square lattice have a ground state in the $\pi$-flux sector). In this toy model we have set the inter-partition coupling to unity, while the intra-partition coupling is denoted by $w_p$. 

To connect this simple model to the honeycomb model, we define $w_p=\kappa_{\text{tm}} |E_\pm(p)|$, where $\kappa_{\text{tm}}$ is a proportionality constant which is needed due to arbitrary normalization of the tunneling across the cut.
%and the lack of an obvious relation of these tunneling amplitudes with the full honeycomb model.
We set the tunneling through the bulk to be directly proportional to the linearized edge state dispersion.
This is motivated by the physical edge states being completely decoupled from the bulk at $p^\star$ where $E_\pm(p^\star)=0$, while away from this special momenta they penetrate into the bulk. We interprete the finite energy $|E_\pm(p)|$ as reflecting an effective coupling between the edge states belonging to the same partition due to their hybridization. Hence, at $p^\star$ the weak tunneling $w_p$ vanishes and one immediately recovers the lower bound since the model describes two decoupled pairs of Majoranas, with one from each pair belonging to $A$ and $B$. As one moves away from $p^\star$, i.e. as the edge states penetrate deeper into the bulk, all the Majorana modes hybridize. This results in general in lower entanglement since in the limit $w \gg 1$ the two subsystems, $A$ and $B$, completely decouple.

To verify this, we compute the entanglement entropy $S_{\text{tm}}(w_p)$ by tracing out the Majoranas $b_L$ and $b_R$ in $B$, as shown in Fig.~\ref{fig:TM}. The von Neumann entropy of the ground state is given by
\be
 S_{\text{tm}}(w_p)=\frac{1}{1+s^2} (-2 s^2 \log{s} + (1+s^2) \log({1+s^2})),
\ee
where $s=-\frac{w_p-\sqrt{w_p^2+4}}{2}$. To compare this expression with $S_\text{edge}$ obtained from the full model, we choose the proportionality constant $\kappa_{\text{tm}}$ such that $\frac{|S-S_{\text{tm}}|}{S}<1\%$ in the range $\Delta p$. With this free parameter fixed, Fig.~\ref{fig:TM} shows that $S_{\text{tm}}$ indeed accurately approximates $S_\text{edge}$.
Note that such a remarkably simple model with clear physical interpretation in terms of the the hybridization of the edge states can in fact capture how the entanglement depends on the momentum.
This simple picture further demonstrates that the entropy of the edge states, that give the dominant contribution to the total entropy, can be treated independent of the bulk states and fully capture the dominant entropy contributions.

\section{Anatomy of Entanglement in the Gapless Phases and Critical Points}
\label{gapless phases tomography}

We now turn to the entanglement at the gapless phases and the critical points of Kitaev's honeycomb lattice model. When the bulk energy gap closes and the correlation length diverges, bulk contributions are expected to dominate over the edge as the system size increases. To investigate this, we write the total entanglement entropy as the sum of the three expected dominant contributions
\be \label{anatomy_critical}
  S \approx S_\text{c} + S_\text{qc} + S_\text{edge},
\ee
where $S_\text{c}$ describes contribution from critical 1D chains, $S_\text{qc}$ the quasi-critical contributions with long correlation lengths and $S_\text{edge}$ again the edge state contributions. The form of the first two bulk contributions has been analyzed in 1D systems and it is shown to take the leading forms~\cite{Cardy,Area}
\bq
  S_\text{c} & = & \frac{c}{3}\log(\frac{L_x}{\pi}\sin\frac{x\pi }{L_x})+\text{const.}, \label{eq:conformal} \\
  S_\text{qc} & = & \frac{c}{3}\log{\xi}+\text{const.}, \label{eq:quasicriticalcontribution}
\eq
where $c$ is the \emph{central charge} of the conformal field theory (CFT) that describes the universality class of the criticality, $x$ is the length of the partition in a chain of total length $L_x$ and $\xi$ is the correlation length. In other words, the critical chains at momenta $p_c$ where the bulk gap closes give extensive contributions in the system size, while quasi-critical chains give constant bulk contributions that depend on the correlation length $\xi(p) \sim G(p)^{-1}$ of the chain $p$. Contributions from edge states, that do appear also at some of the critical points, are given by \rf{eq:Sedgelinq}.

We begin by analysing the entropy scaling with system size to determine the central charge $c$ of each critical chain at each gapless phase and critical point in the phase diagram. The central charge is then used to explicitly show how the total entropy splits into the three components \rf{anatomy_critical} and to identify the universality classes of the topological phase transitions.

\begin{figure*}[t]
$\begin{array}{cccccc}
\multicolumn{3}{c}{\underline{\textrm{Vortex-free sector} \ (\theta=0)}} & \multicolumn{3}{c}{\underline{\textrm{Full-vortex sector} \ (\theta=1)}} \\
\includegraphics[scale=0.14]{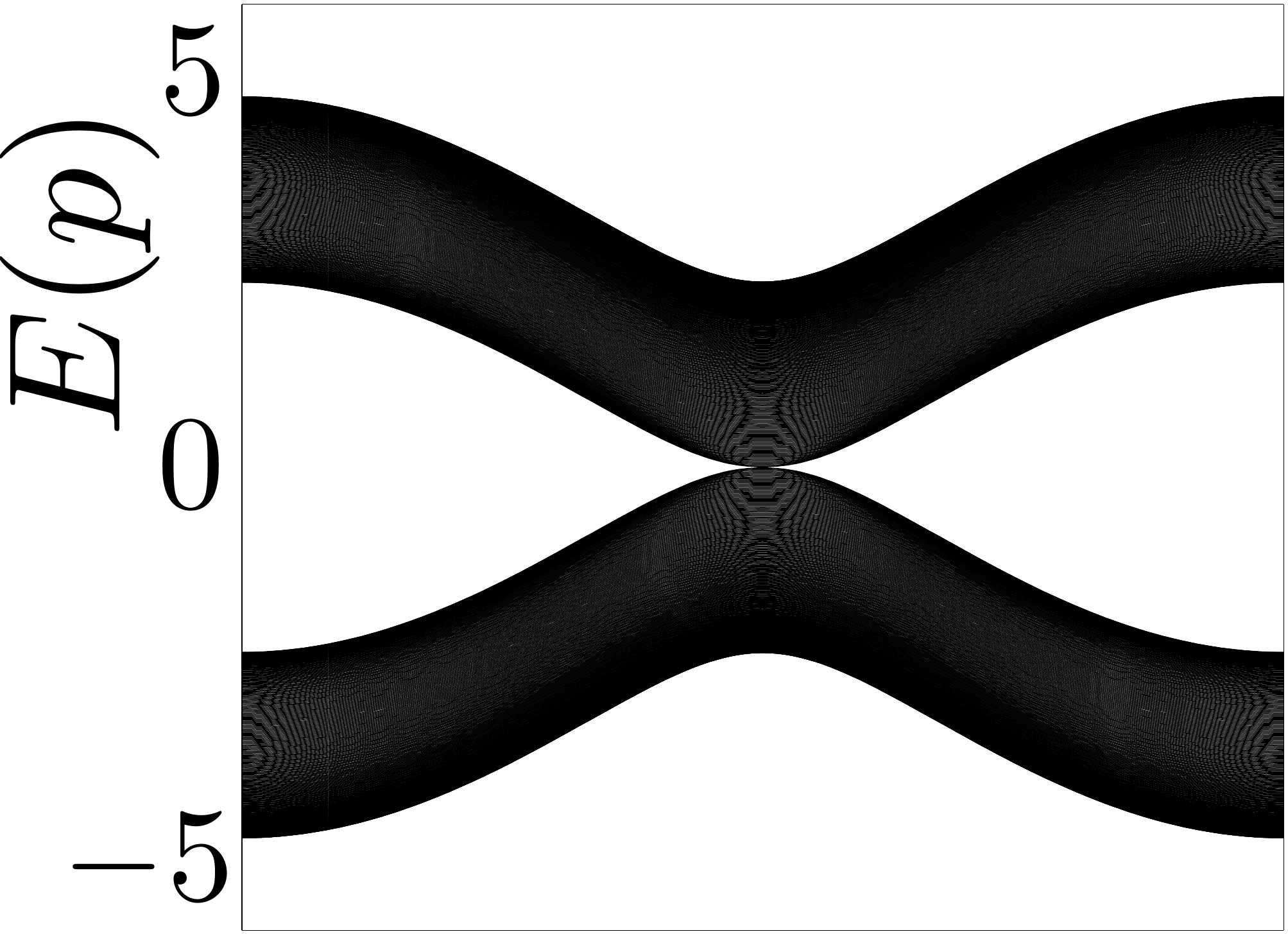}&\includegraphics[scale=0.14]{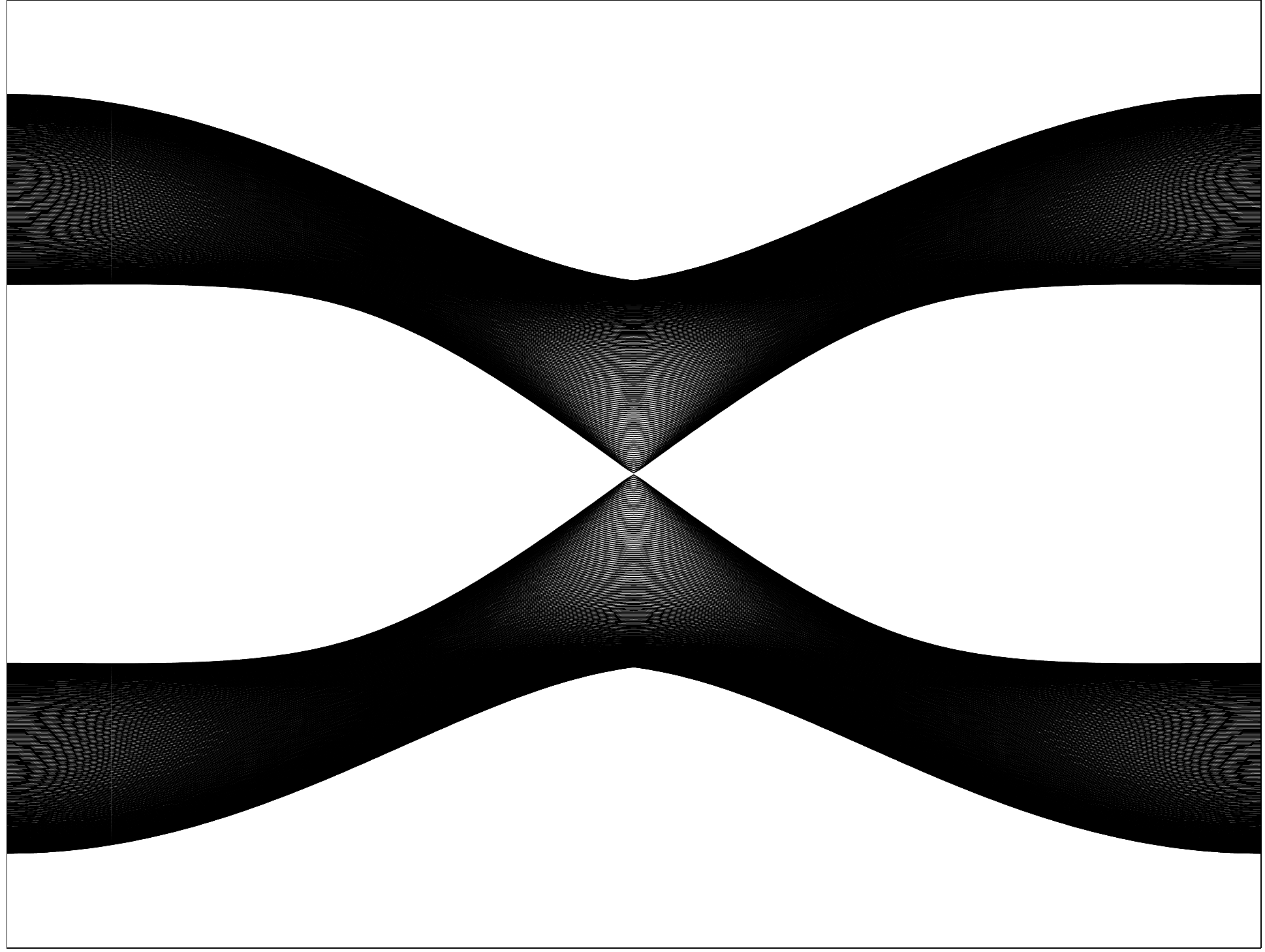}&\includegraphics[scale=0.14]{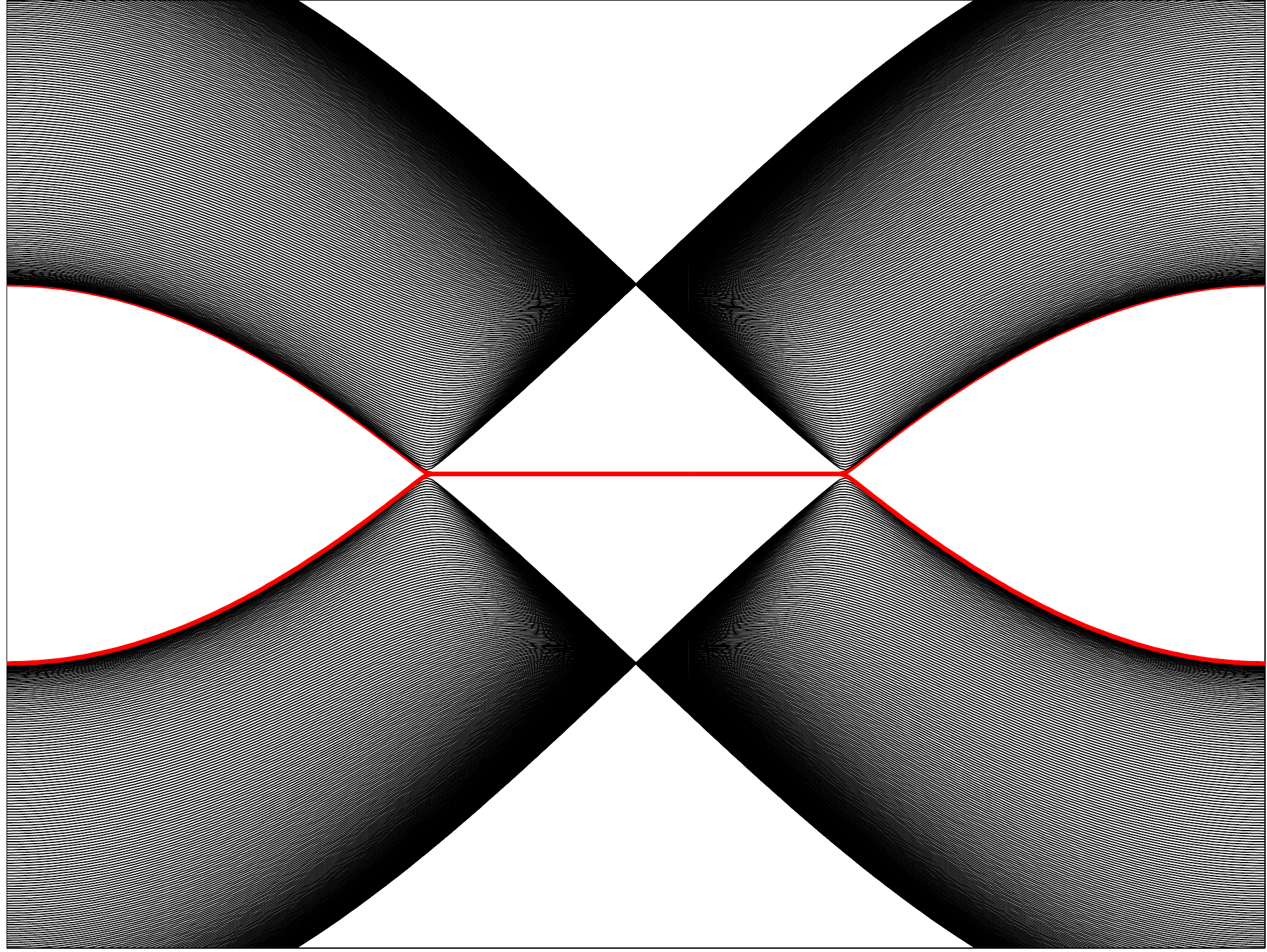}&\includegraphics[scale=0.14]{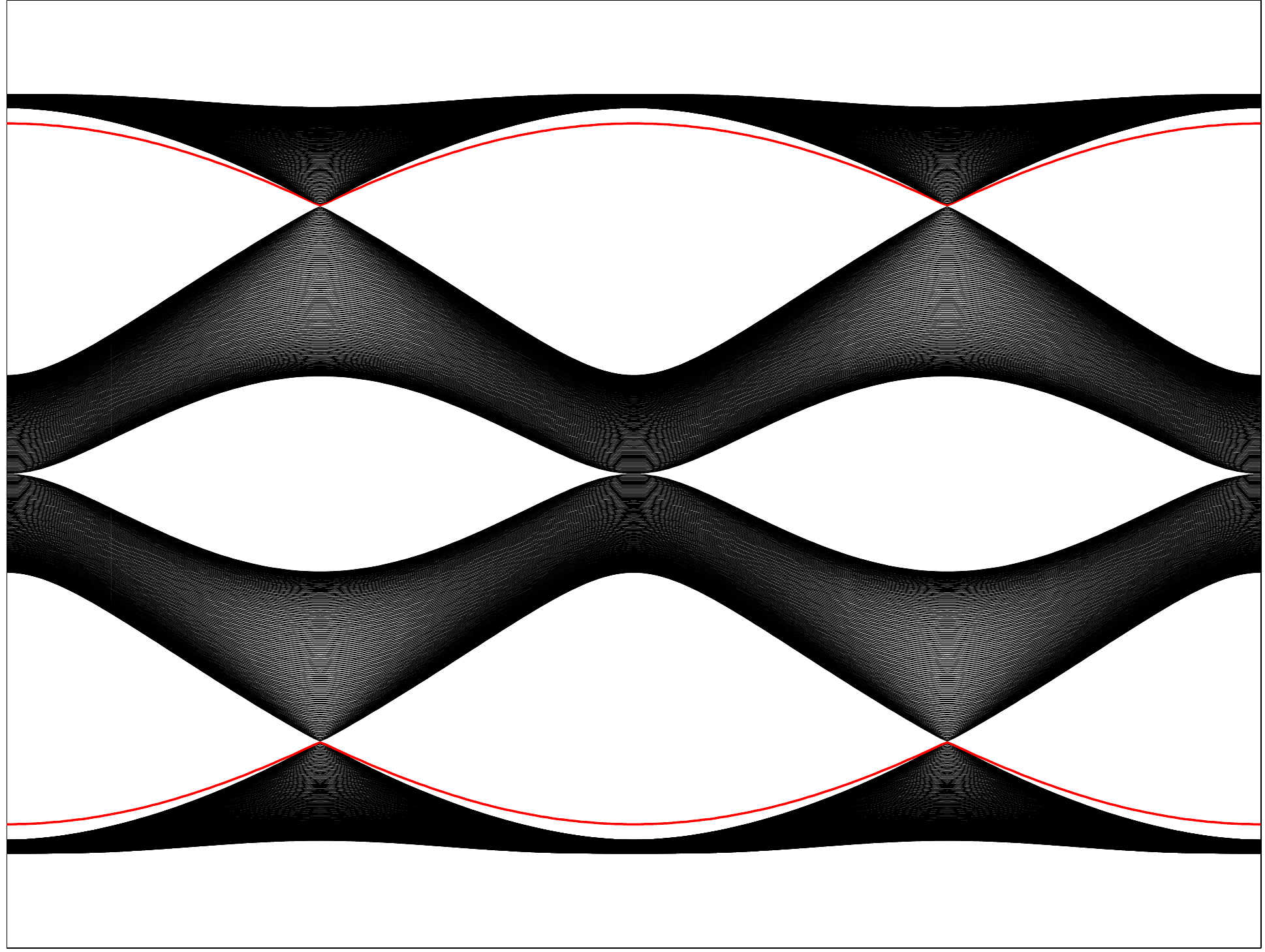}&\includegraphics[scale=0.14]{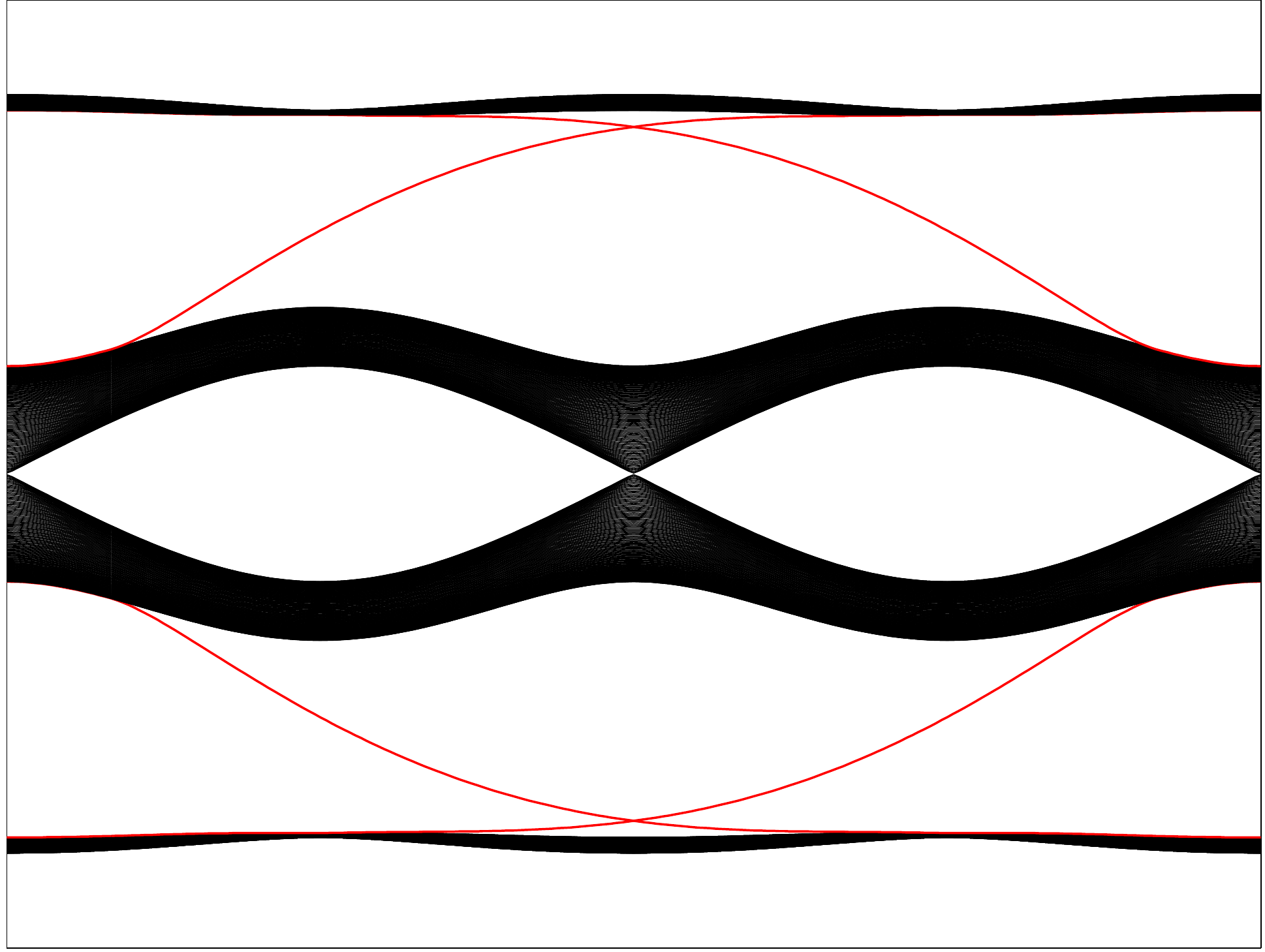}&\includegraphics[scale=0.14]{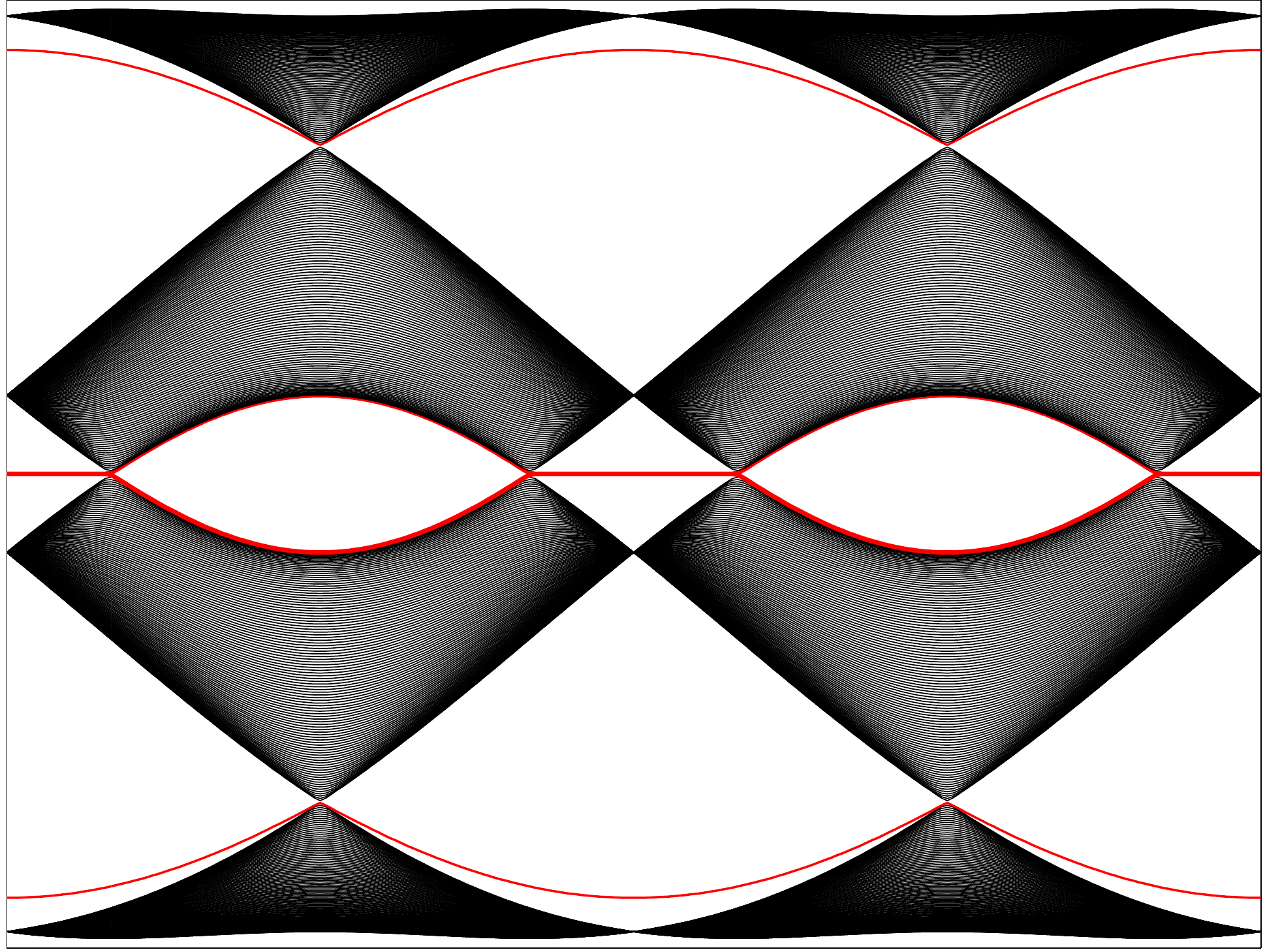}\\
\includegraphics[scale=0.14]{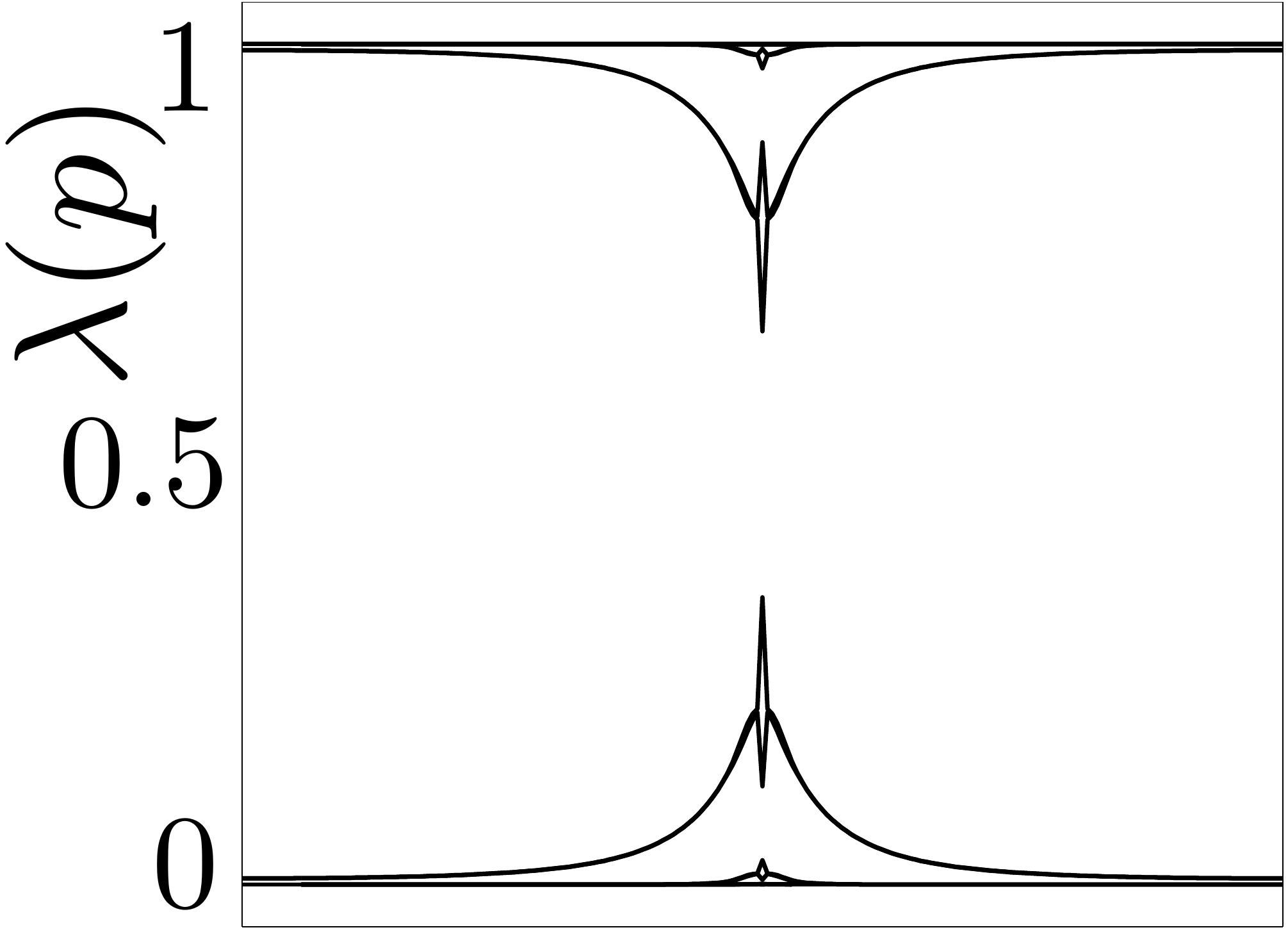}&\includegraphics[scale=0.14]{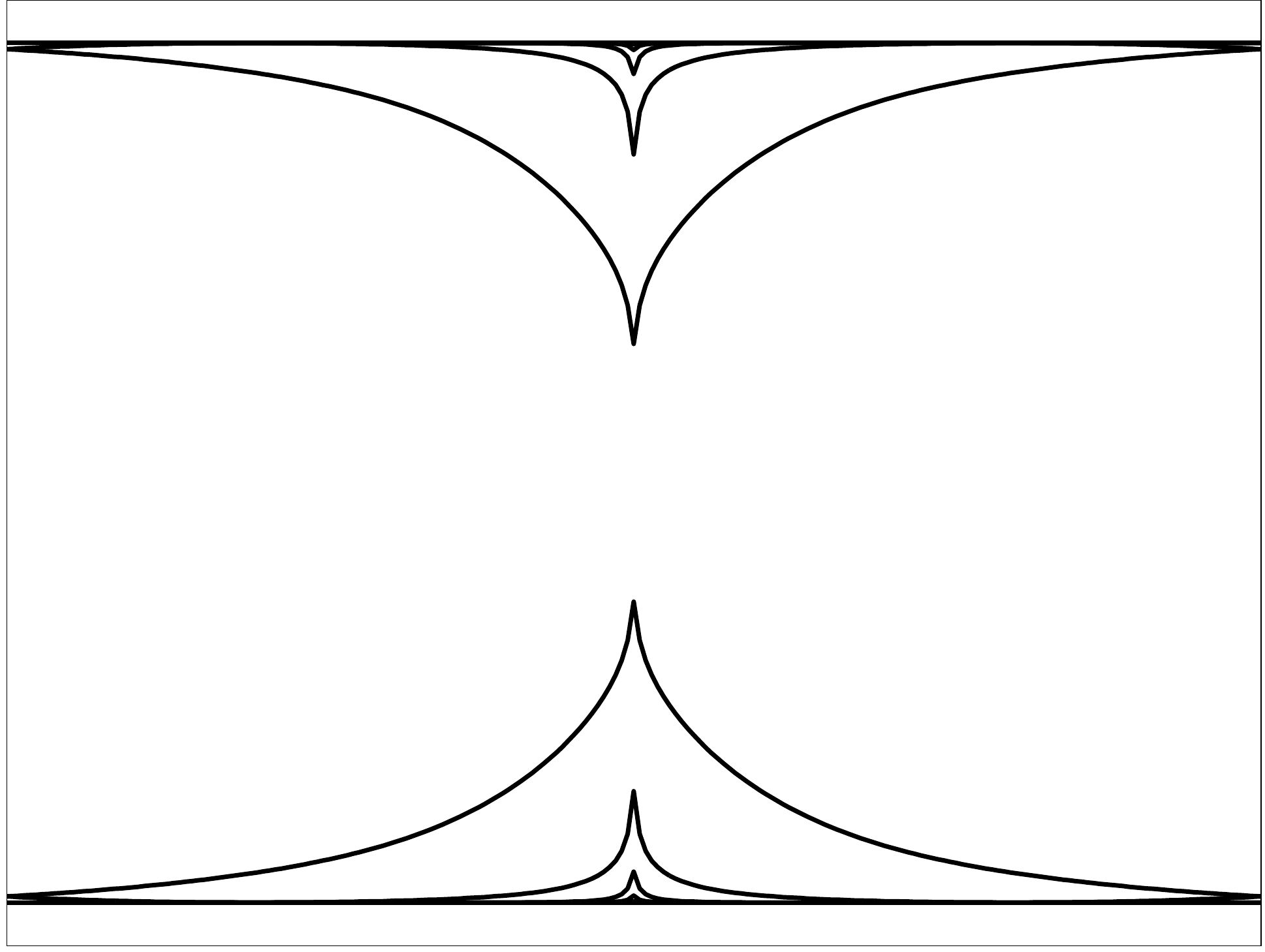}&\includegraphics[scale=0.14]{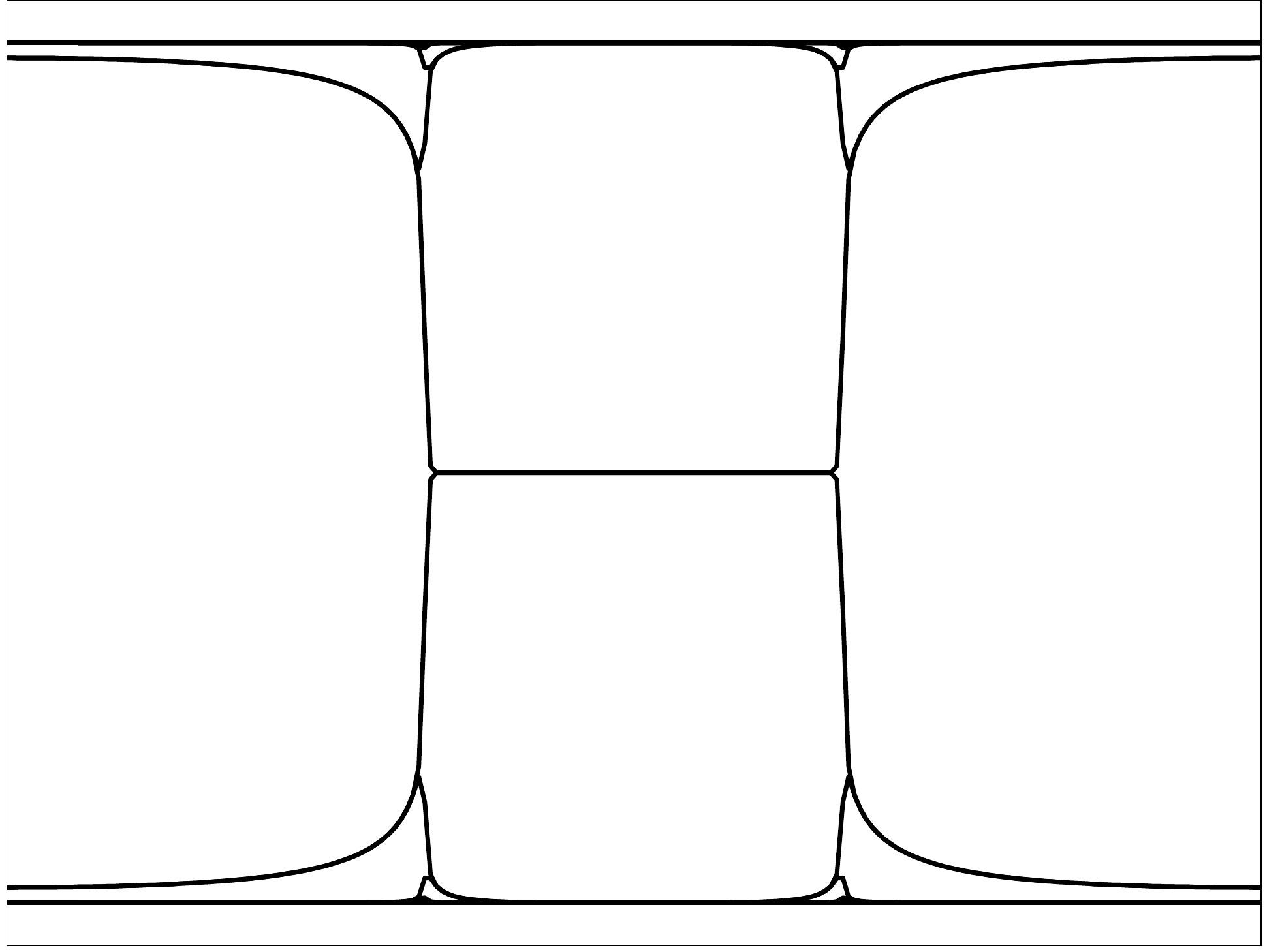}&\includegraphics[scale=0.14]{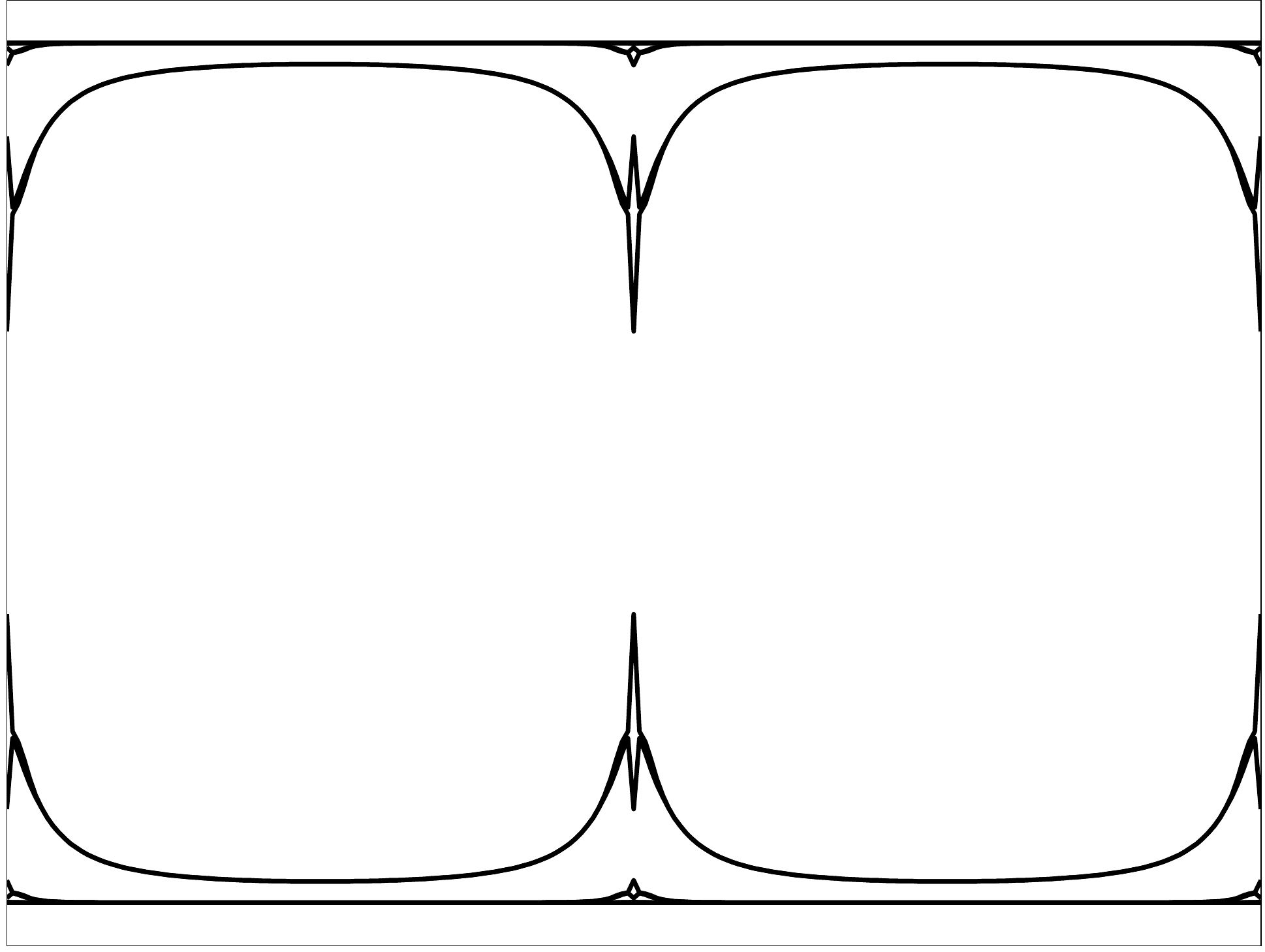}&\includegraphics[scale=0.14]{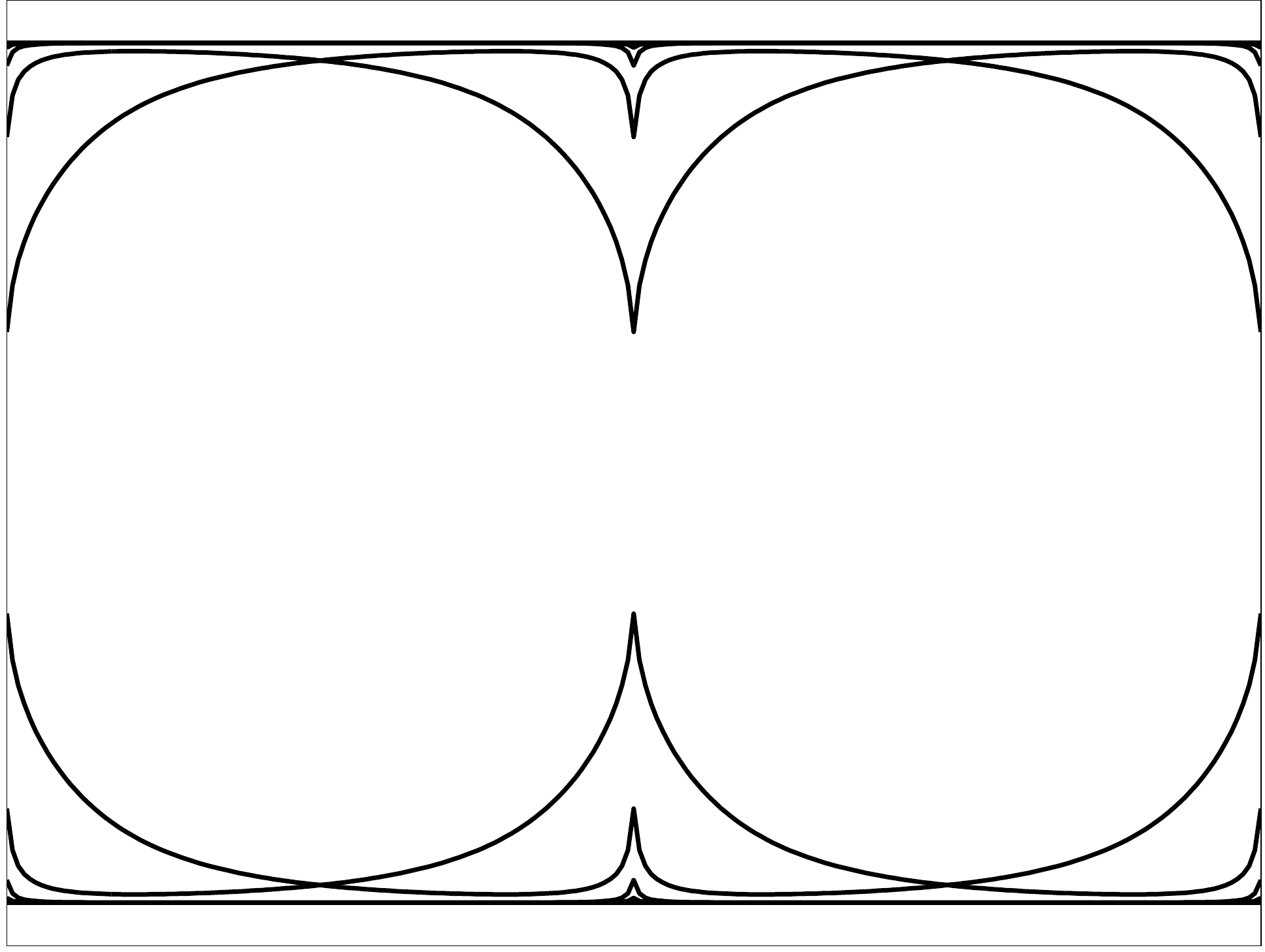}&\includegraphics[scale=0.14]{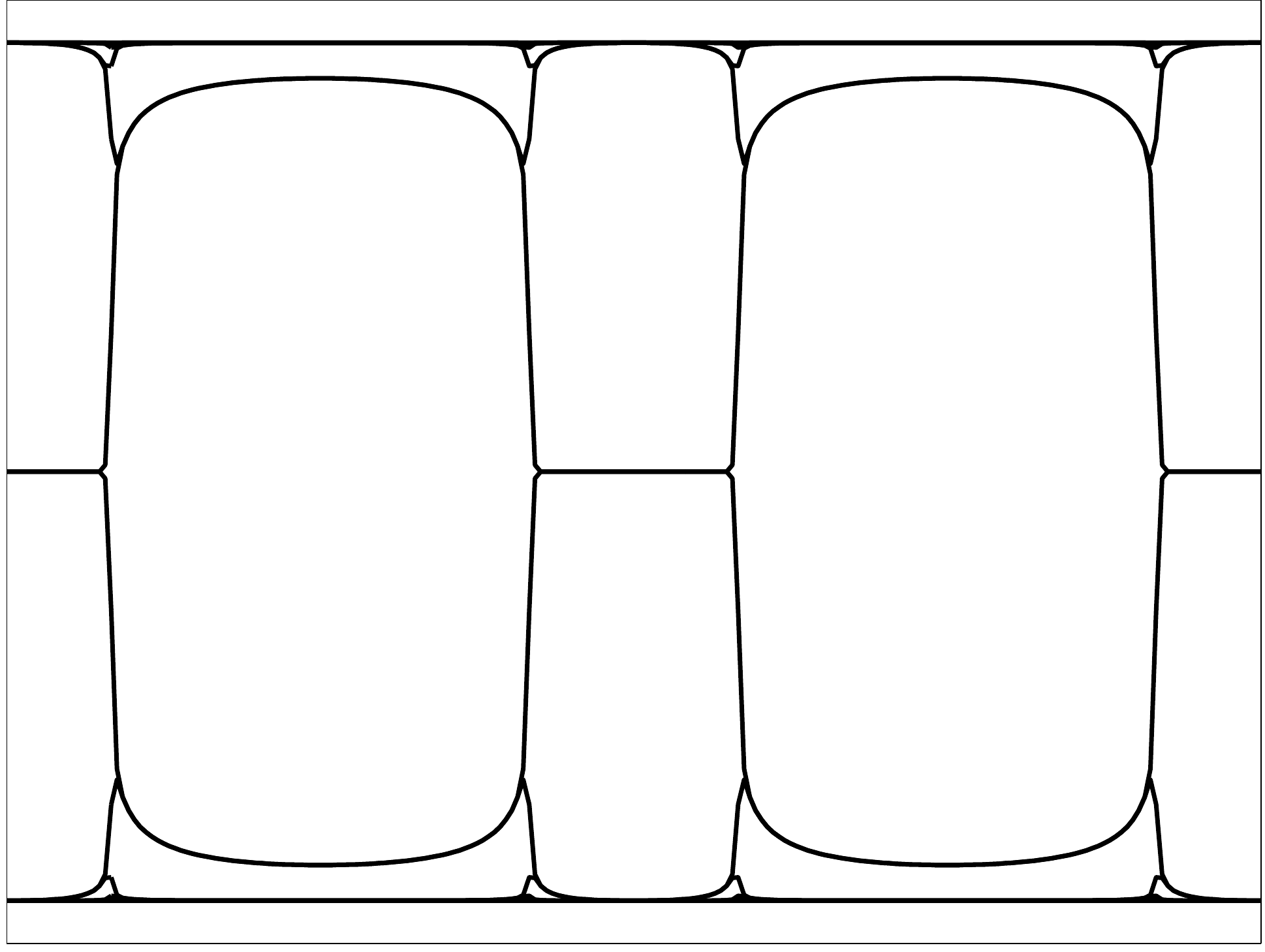}\\
\includegraphics[scale=0.14]{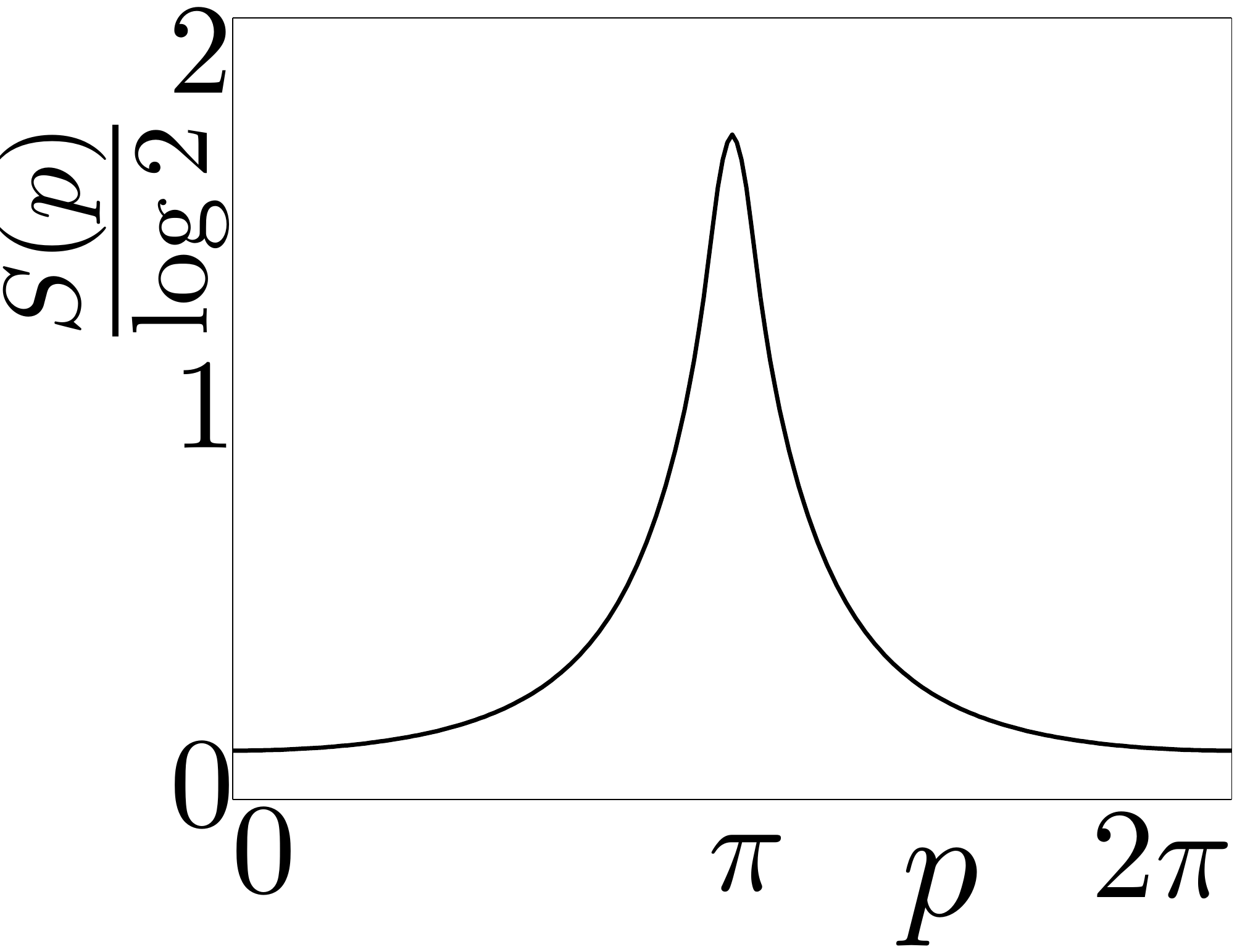}&\includegraphics[scale=0.14]{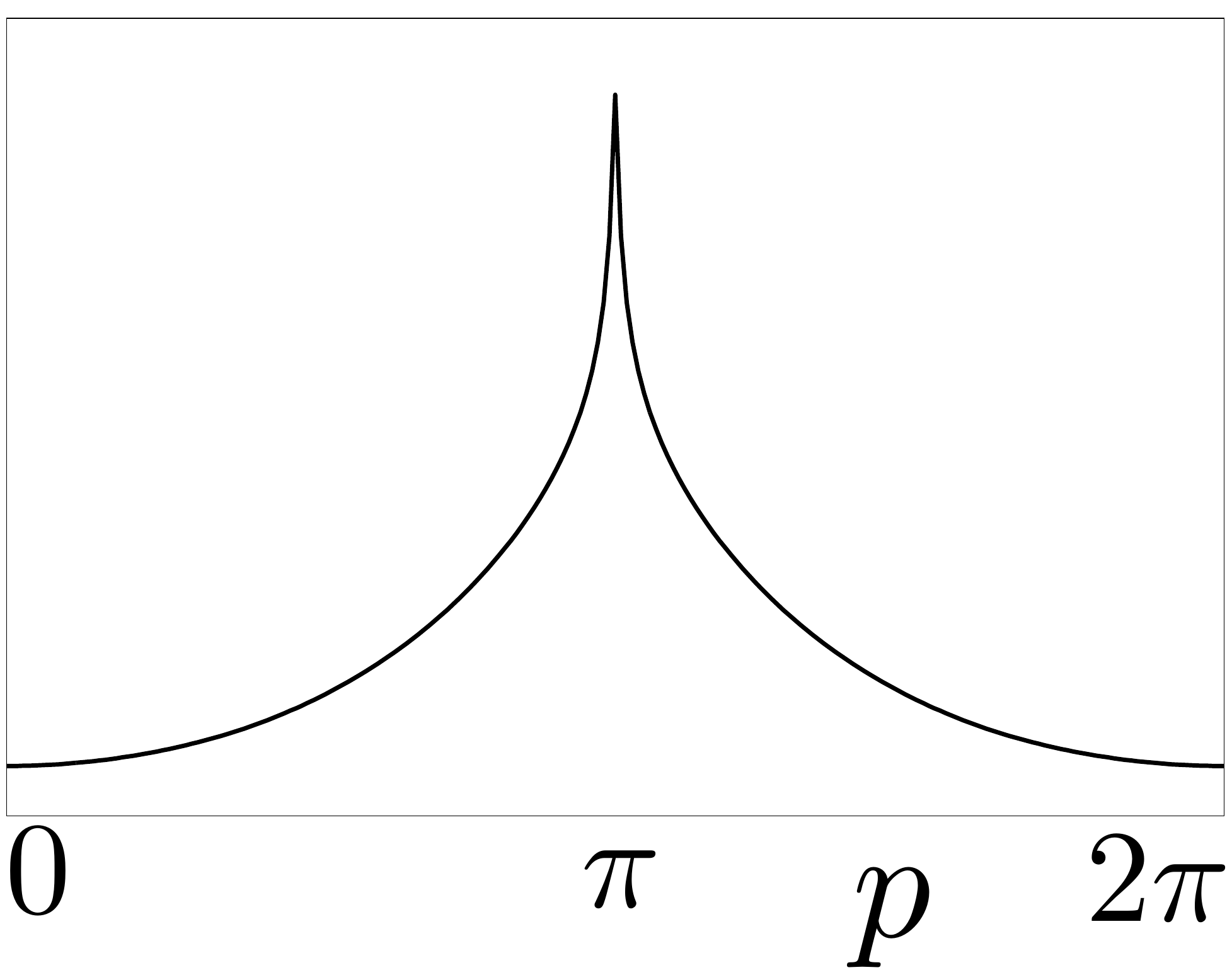}&\includegraphics[scale=0.14]{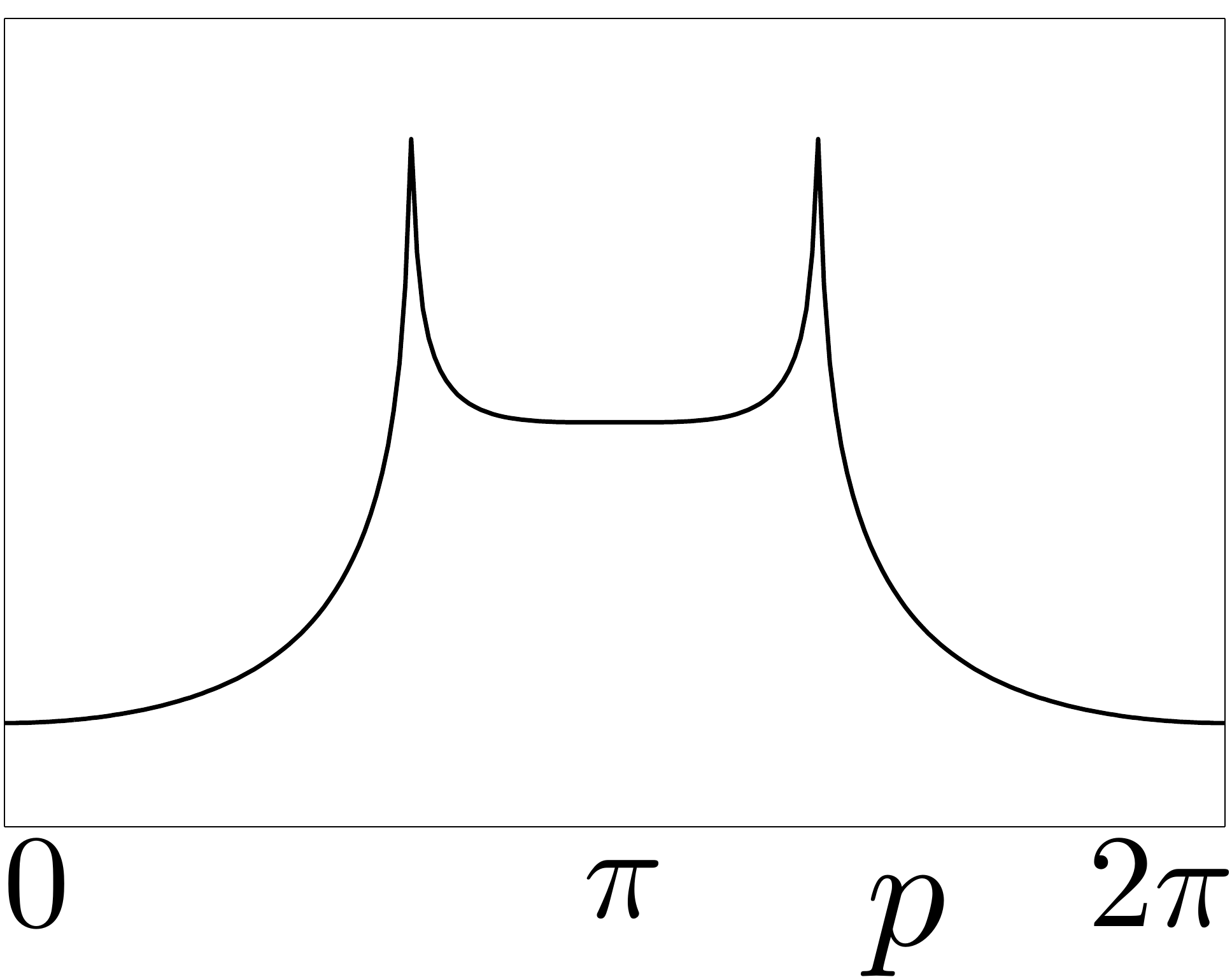}&\includegraphics[scale=0.14]{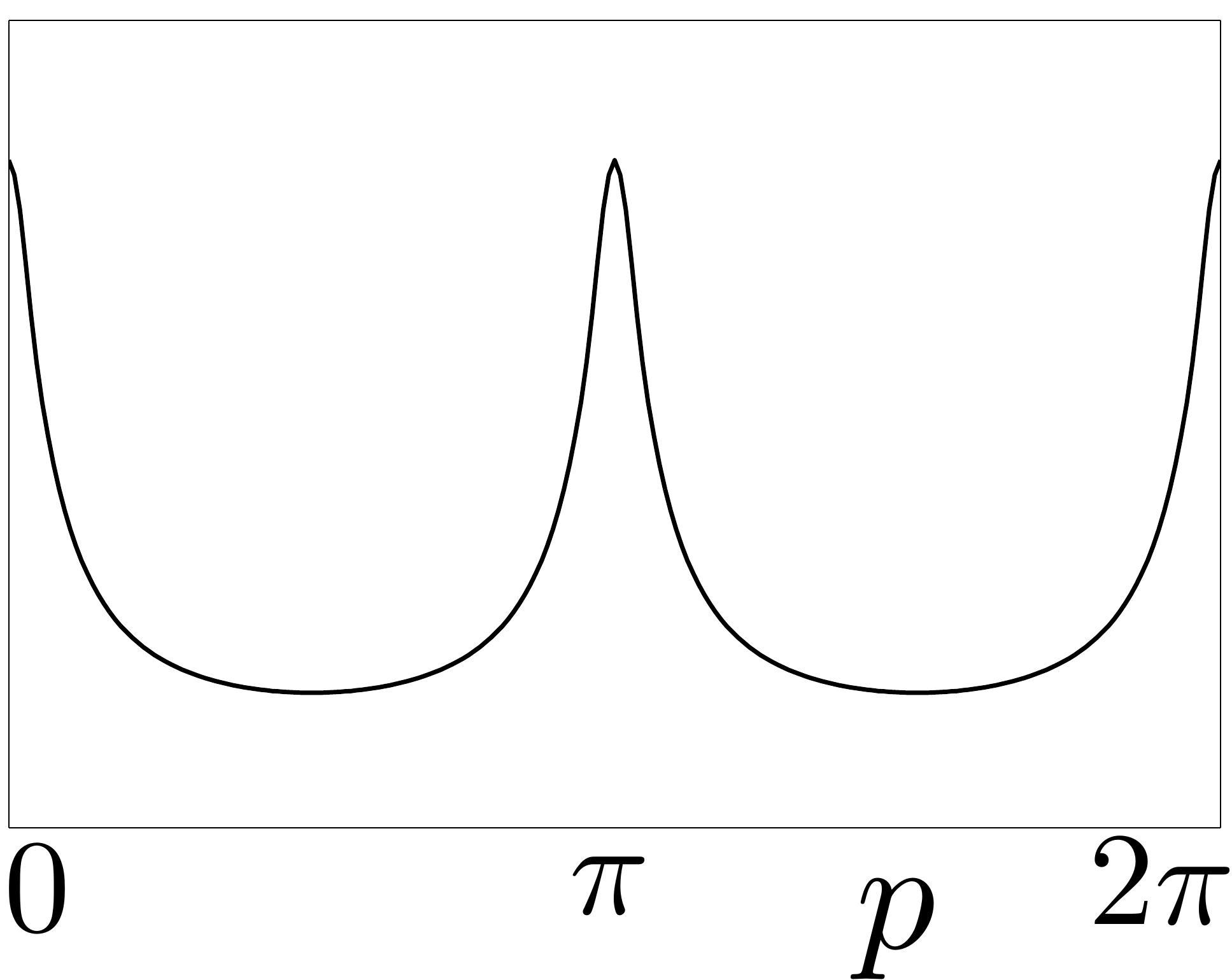}&\includegraphics[scale=0.14]{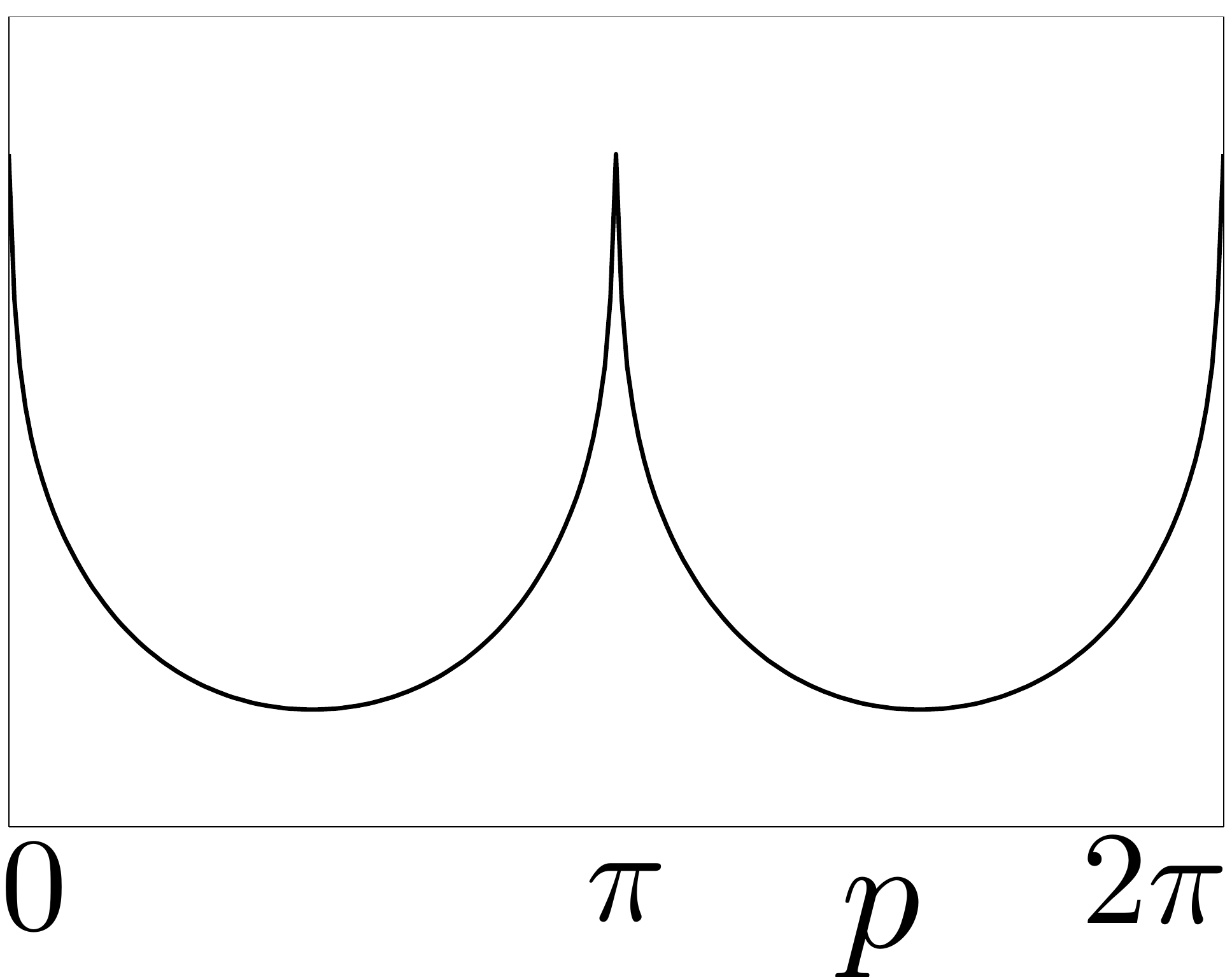}&\includegraphics[scale=0.14]{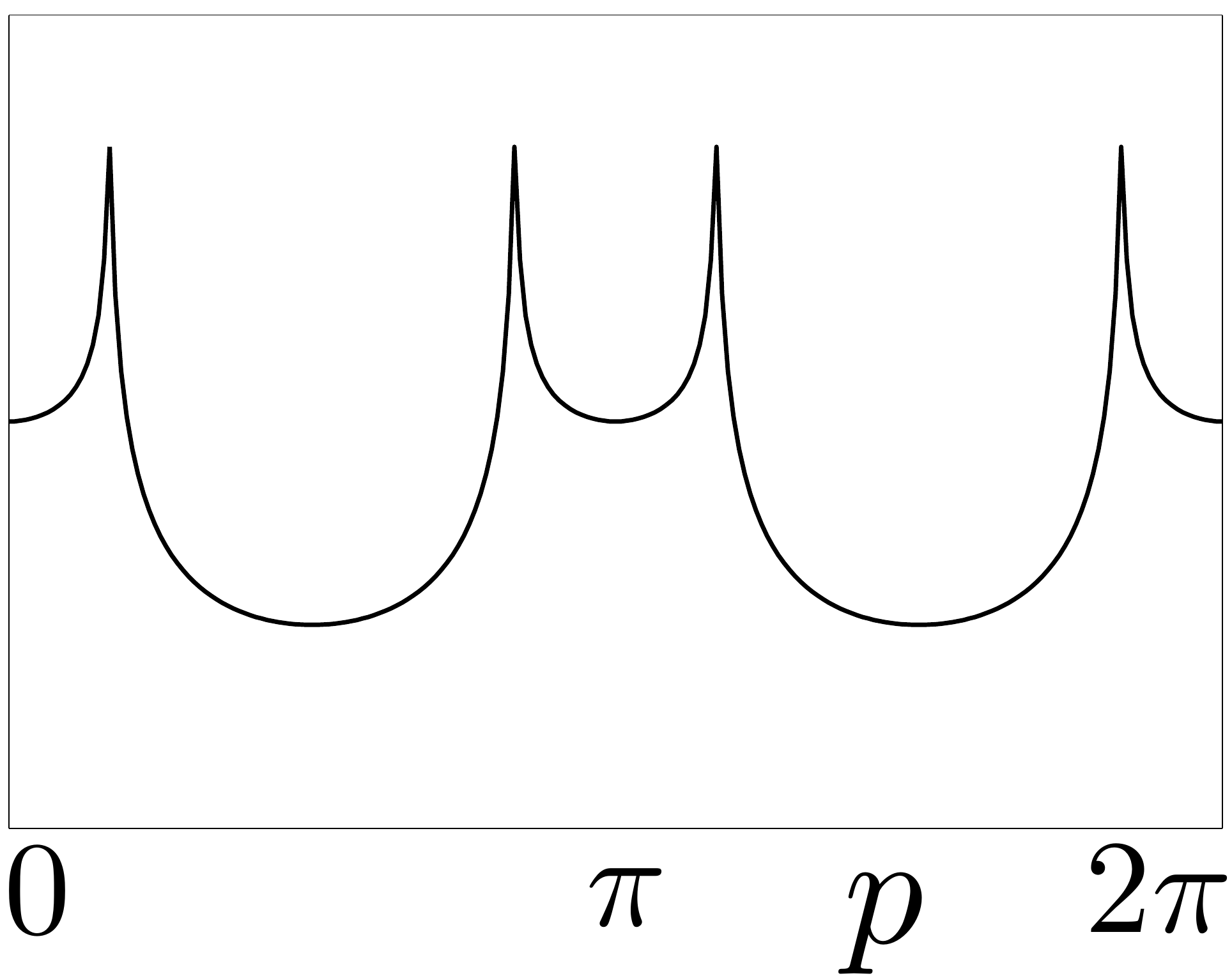} \\
(a) & (b) & (c) & (d) & (e) & (f)
\end{array}$
\caption{\label{fig:lambdaGapless}
The energy spectrum $E(p)$, the entanglement spectrum $\lambda(p)$ and the entanglement dispersion $S(p)$ at critical points in the (a)-(c) vortex-free sector ($\theta=0$) and (d)-(f) in the full-vortex sector ($\theta=1$). The transitions are: (a) $\nu=0$ to Majorana semi-metal with two Fermi points at $J=\frac{1}{2}, K=0$, (b) $\nu=0$ to $\nu=1$ at $J=\frac{1}{2}, K=0.15$, (c) $\nu=1$ to $\nu=-1$ at $J=1, K=0$, (d) $\nu=0$ to Majorana semi-metal with four Fermi points at $J=\frac{1}{\sqrt{2}}, K=0$, (e) $\nu=0$ to $\nu=2$ at $J=\sqrt{\frac{1-K^2}{2}}, K=0.15$ and (f) $\nu=2$ to $\nu=-2$ at $J=1, K=0$.}
\end{figure*}

\begin{figure}[t]
\includegraphics[scale=.39]{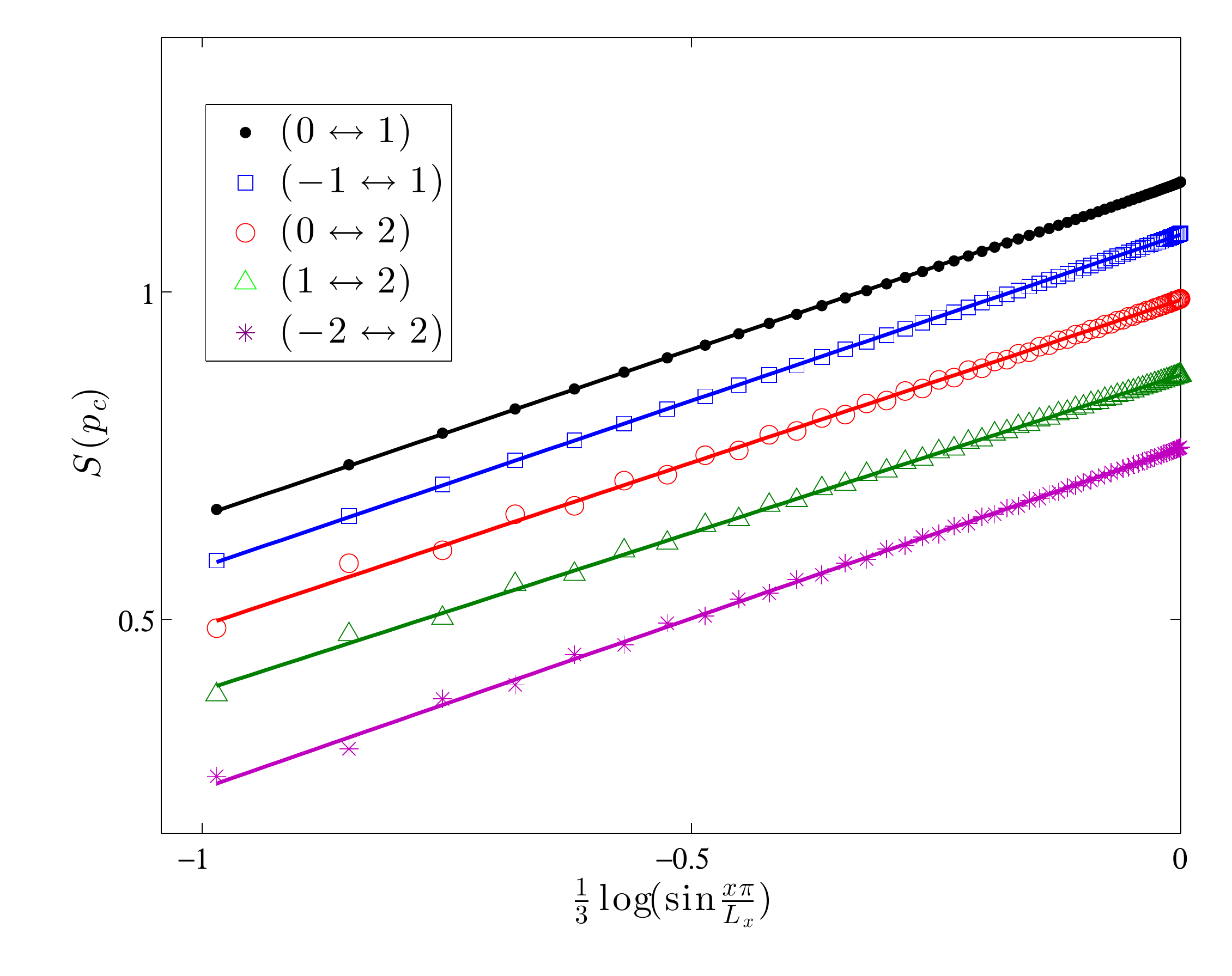}
\caption{\label{fig:c_pc} Scaling of the entanglement entropy $S(p_c)$ of the gapless chains at the transitions $(\nu \leftrightarrow \nu')$ between gapped topological phases characterized by Chern numbers $\nu$ and $\nu'$ as the function of the chord length \rf{eq:conformal}. For ${(0\leftrightarrow 1)}$ transition $p_\text{c}=\pi$, for ${(-1\leftrightarrow 1)}$ transition $p_\text{c}=\pm \frac{2\pi}{3}$, for ${(0\leftrightarrow 2)}$ transition $p_\text{c}=0, \pi$, for ${(1\leftrightarrow 2)}$ transition $p_\text{c}=0$, and for ${(-2\leftrightarrow 2)}$ transition $p_\text{c}=\pm \frac{\pi}{6}, \pm \frac{5\pi}{6}$. For each critical chain the scaling implies a central charge of $c=1/2$.}
\end{figure}

\subsection{Critical Entropy Scaling}

The critical momenta $p_c$ at which the bulk gap closes at each gapless point in the phase diagram can be obtained from the analytic solution presented in Appendix \ref{KitaevModelApp}. Fig. \ref{fig:lambdaGapless} shows representative energy band structures at the critical points and the corresponding entanglement spectra. We observe that in contrast to gapped topological phases with edge states giving rise to gapless entanglement spectrua, at criticality the entanglement spectra are gapped, but exhibit nesting of entanglement levels in the vicinity of the critical momenta $p_c$.  
%\kon{Correlations in critical systems obey a power law decay and this is responsible for the critical entanglement scaling. Since power law correlations imply the lack of a characteristic length scale, the correlations inherit a self similar structure. Thus, the observed nesting in the entanglement spectrum is a manifestation of the structure of the correlations in criticality.}

\begin{figure}[t]
\includegraphics[scale=0.37]{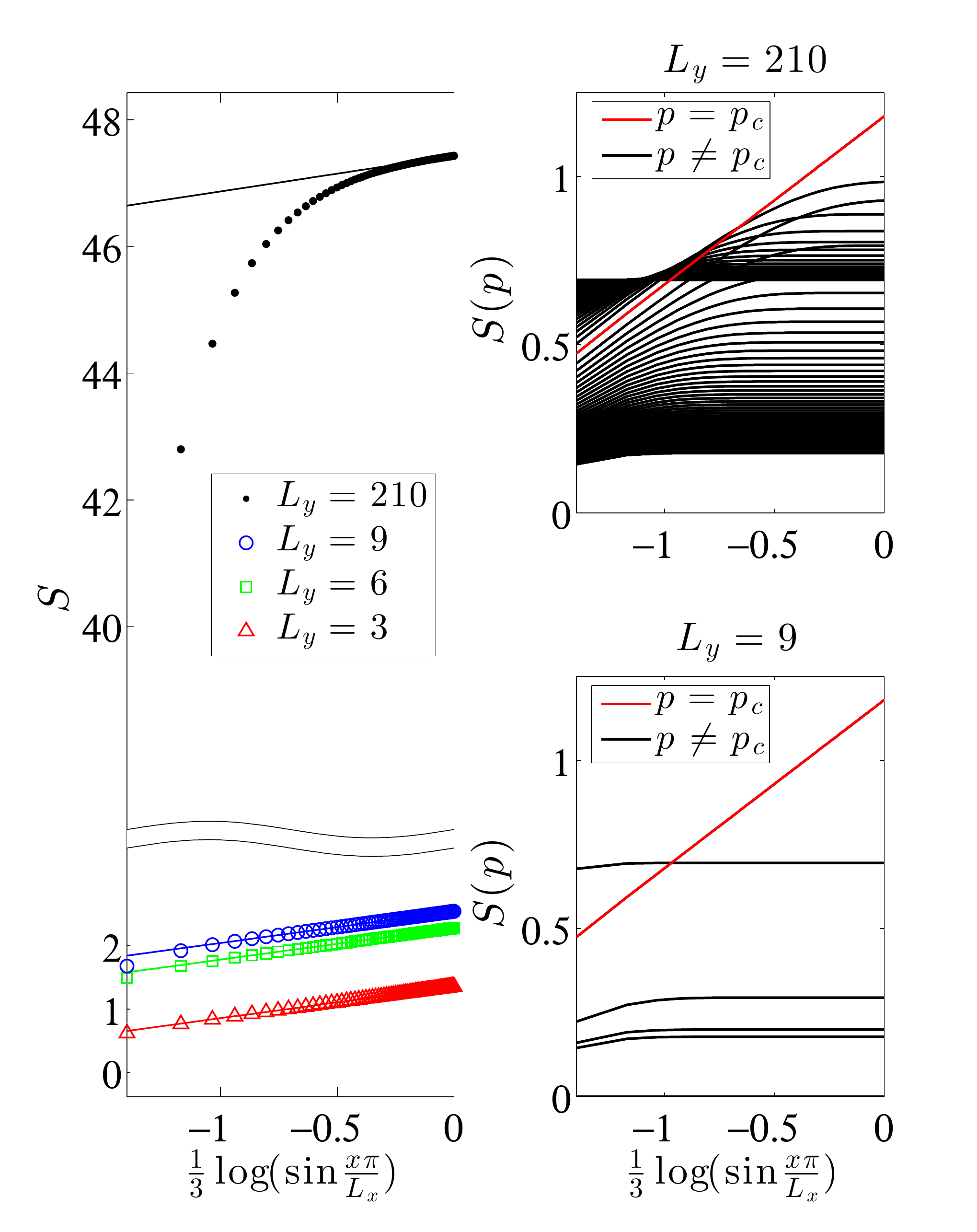}
\caption{\label{fig:EntropyScalingChordLength}  
Scaling of the entropy $S$ in the gapless phase in the vortex-free sector ($\theta=0$, $J=1$) as the function of the partition length $x$ for cut lengths $L_y$ commensurate with the critical momenta $p_c=\frac{2\pi}{3}$. (Left) In the thin torus limit $L_y \ll L_x$ the quasi-2D models scale precisely with $\tilde{c}=1/2$, while in the regime $L_y \approx L_x$ one only finds scaling close to this lower bound in the large chord length limit. (Right) This system size dependence of the scaling can be traced back to quasi-critical chains. In the thin-torus limit there are only few chains with large gaps giving constant contributions, while in a large system some quasi-critical chains show scaling all the way to the large $x \to L_x$ limit. Here $L_x=210$.}
\end{figure}

To extract the central charge characterizing the scaling according to \rf{eq:conformal}, in Fig. \ref{fig:c_pc} we plot the entropies $S(p_c)$ of critical chains as the function of the partition length. We find that all scale with $c=1/2$ implying that they all belong to the universality class of 1D quantum Ising model. However, one should keep in mind that due to the Majorana condition $\gamma_p^\dagger=\gamma_{-p}$, not all chains are independent. Using \rf{eq_HforChains_main} it is straighforward to verify that $H(p_c)=H(-p_c)$, meaning that critical chains at $\pm p_c$ describe the same critical contribution, while the chains $p_c=0,\pi$ are always independent. This implies that the entropy of the 2D honeycomb model at a critical point with $N_c$ critical chains in the momentum range $p \in [0,\pi]$ scales with an effective central charge 
\be
  \tilde{c} \geq c N_c.
\ee  
As shown in Fig.~\ref{fig:EntropyScalingChordLength}, the lower bound is saturated in the thin-torus limit $L_y \ll L_x$ when $L_y$ is commensurate with the critical momenta $p_c$. As $L_y \to L_x$, the obtained central charge is always $\tilde{c} > c N_c$, as the system contains more and more quasi-critical chains that add to the scaling at small partition lengths $x$. Thus in the thermodynamic limit, the quantized effective central charge can only be reliably obtained in the large partition limit $x \to L_x/2$, i.e. as the partition size approaches half the system size.\cite{Fendley}

%\begin{figure}[t]
%\includegraphics[scale=.26]{Chains_ChordLength_K0_version1}\includegraphics[scale=.26]{Chains_ChordLength_J12_version1}
%\includegraphics[scale=.26]{S_chordlength_K0}\includegraphics[scale=.26]{S_chordlength_J05}
%\caption{\label{fig:CriticalChains}
%Entropy $S(p)$ of each chain $H(p)$ as a function of the chord length $\log(\sin\frac{x\pi }{L_x})$ for the criticalities $(-1\leftrightarrow 1)$ (Left) and $(0\leftrightarrow 1)$ (Right). The entropy of the critical chains (black) shows linear scaling with $c=1/2$ and the entropy of the quasi-critical chains (red) shows linear scaling at the $x\rightarrow \frac{L_x}{2}$ limit. The rest of the chains (blue) obey the area law. In the thin slice limit $x \rightarrow 0$ we have all the chains scaling linearly with the chord length, as all of them appear effectively to be critical. (Bottom) (${-1 \leftrightarrow 1}$) criticality with $J=1$. Scaling of the total entropy $S$ with $L_y$ obeys the area law (here $L_x=120$). The finite size effect can be seen for small $L_y$ where $p_\text{c}$ being included in the BZ makes a significant contribution to the entropy, thus creating these oscillations which wash out for large $L_y$.}
%\end{figure}

\subsection{Anatomy of Critical Entanglement}

Having determined the central charges of the critical chains, we now turn to analyze the anatomy of critical entanglement \rf{anatomy_critical} in the vortex-free sector ($\theta=0$) of the honeycomb model. When time-reversal symmery is not broken ($K=0$), the regime $J>1/2$ supports a gapless Majorana semi-metal with two Fermi points, as shown in Fig.~\ref{fig:lambdaGapless}. Even if the spectrum is gapless, the phase also exhibits edge states between the Fermi points that appear as a two-fold degenerate flat band. This follows from the topological nature of the chains, such as \rf{eq:Hp=pi} that depends only on $J$. Thus the entanglement per momentum $p$ can be approximated by the three components
\bq\label{eq:Sapp}
    S_\text{comp}(p)& = & \left\{ \begin{array}{lr} S_\text{edge} = \log{2} &  p\in\Delta p, \\ S_\text{c} & p = p_c, \\ S_\text{qc} &  p \notin\{\Delta p \cup p_c \}. \end{array} \right. 
\eq
The range $\Delta p$ for the edge state contributions, that now coincides with the momenta between the critical momenta $\pm p_c$, can be obtained analytically for $K=0$ and it is given by (see Appendix C)
\be \label{Dp}
\Delta p= 2 \arccos\left(\frac{1}{2 J} \right).
\ee

We approximate the correlation length of each quasi-critical chain $p$ as $\xi(p)\approx 1/G(p)$ with the effective gap given by ${G(p)=\min_{q \in [0,2\pi]}{|2(1+J e^{i q} + J e^{i p})|}}$ (see Appendix \ref{KitaevModelApp}).
In Fig. \ref{fig:S_approx} we show that under these assumptions we recover to high accuracy the total entropy.
The presence of edge states implies that also the gapless phase exhibits a lower bound of entanglement, that is now given by $S \geq \frac{L_y}{2\pi} \Delta p \log 2$.
We also find that even if the system is critical, for a fixed partition length $x$ the entanglement entropy still obeys the area law.

We have verified that similar anatomy applies also to the full-vortex phase that also exhibits a lower bound that depends on the two momentum ranges where the edge states exist. The same applies also to the critical points between gapped topological phases for $K \neq 0$, that in the absence of edge states consist only of critical and quasi-critical contributions. 

% Similar analysis can be carried out in the full-vortex sector ($\theta=1$), where a gapless phase with with four Fermi points appears for  $J>1/\sqrt{2}$, as shown in Fig.~\ref{fig:lambdaGapless}. Again, the spectrum exhibits edge states that now come in pairs for each edge. The total entropy can again be broken into parts \rf{eq:Sapp} by choosing the appropriate momentun ranges $\Delta p$ for the edge states and evaluating the correlation lengths using the analytic expression for the full-vortex gap given in Appendix \ref{KitaevModelApp}. % Figure~\ref{fig:S_approx} shows that the total entropy is again accurately reproduced.

%
%\begin{figure}[t]
%\includegraphics[scale=.23]{arealaw}
%\caption{\label{fig:arealaw}
% (${-1 \leftrightarrow 1}$) criticality with $J=1$. Scaling of the total entropy $S$ with $L_y$ obeys the area law (here $L_x=120$).
%The finite size effect can be seen for small $L_y$ where $p_\text{c}$ being included in the BZ makes a significant contribution to the entropy, thus creating these oscillations which wash out for large $L_y$.
%}
%\end{figure}

\begin{figure}[t]
\includegraphics[scale=.3785]{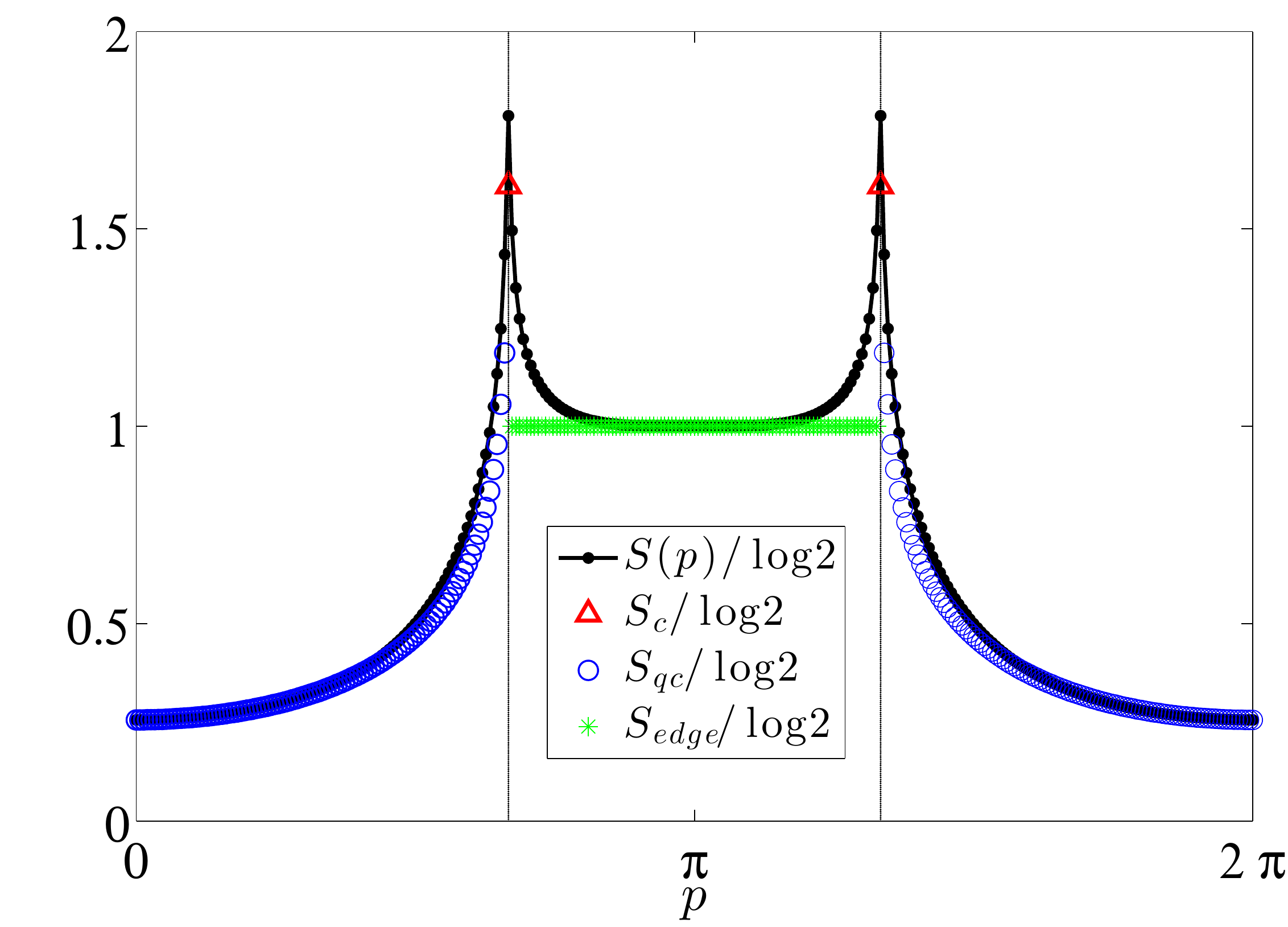}
\begin{picture}(0,0)
\put(46.2,110){\includegraphics[scale=0.1775]{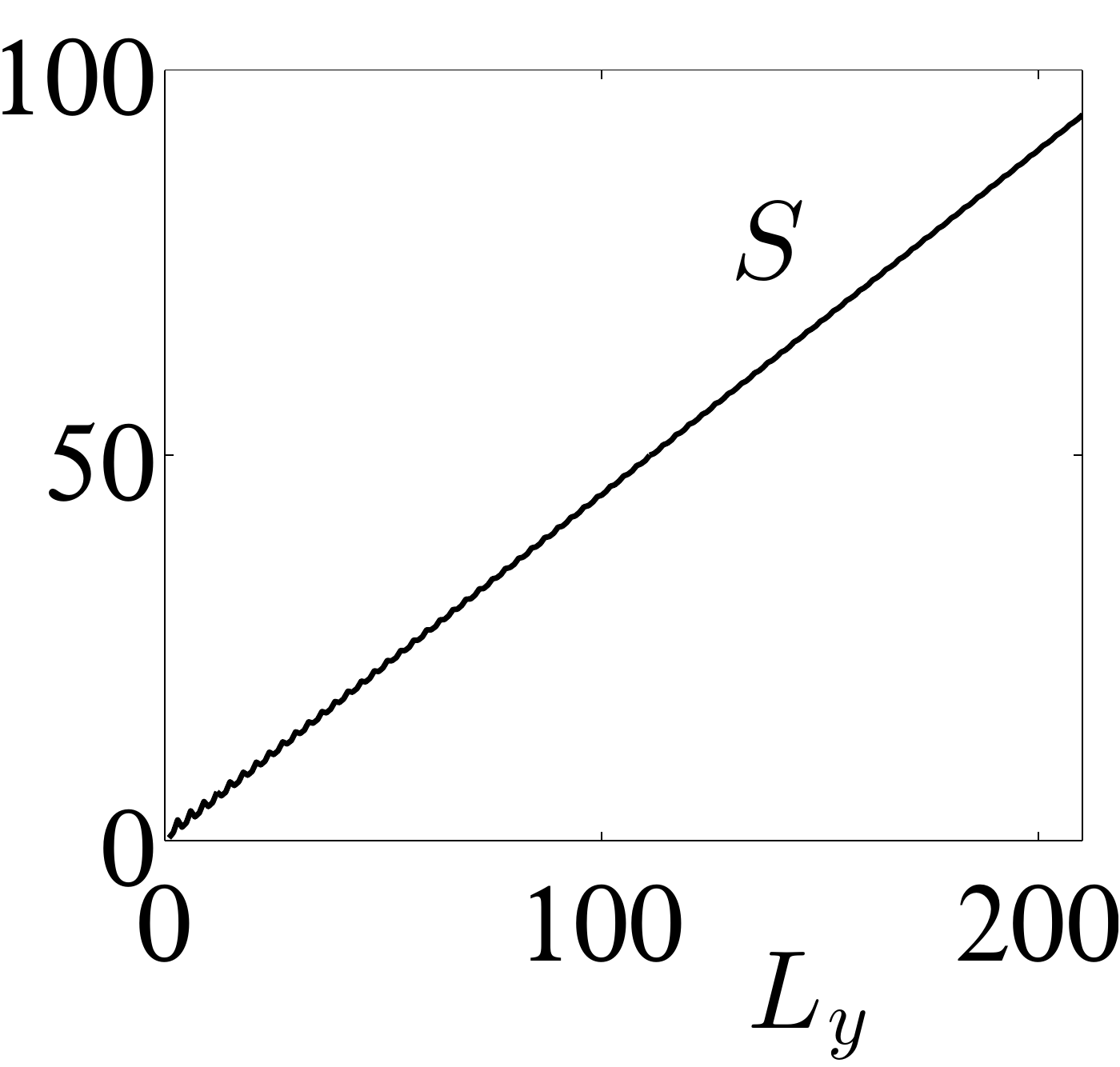}}
\end{picture}
\caption{\label{fig:S_approx} Composite entropy, $S_\text{comp}(p)$, as given by \rf{eq:Sapp} in the gapless phase of the vortex-free sector. The flat band of edge states contributes $S_\text{edge}$ between the Fermi cones at $p_c =\pm \frac{2\pi}{3}$, the critical contributions $S_\text{c}$ only arise at $p_c$ and quasi-critical chains contribute $S_\text{qc}$ outside them. The non-universal constant in \rf{eq:conformal} and \rf{eq:quasicriticalcontribution} is chosen such that $S_\text{qc}(0)=S(0)$. Here $J=1$ and $L_x=L_y=300$. %Neglecting the quasi-critical contributions between the Fermi points, we find the composite entropy approximating the total entropy within the accuracy of $\frac{|S-S_\text{comp}|}{S}\approx 6\%$.
(Inset) Even if the system is critical, for a fixed partition size $x$ the system still obeys the area law. Here $x=105$.}
\end{figure}

\subsection{Universality classes of phase transitions}

Finally, we turn to the universality classes that describe the model independent behavior at phase transitions. In 1D quantum systems second-order phase transitions with linearly dispersing gapless excitations are in general described by CFTs that capture the universal behavior of all correlations functions.\cite{diFrancesco} In 2D such special exactly solvable points do not in general exist unless the critical points exhibit 2D conformal symmetry\cite{Ardonne04} or exhibit effective 1D behavior that can be identified with a known 1D universality class. The latter property enables us to identify the universality classes of the phase transitions in Kitaev's honeycomb model and show that they are consistent with the topologically ordered phases of the model.

In the vortex free sector the single independent critical chain corresponding to the critical momenta $p_c$ is given by
\be
	H_{p_c} = \frac{i}{2}\sum_{j}\Big[(J e^{-i p_c} + 1) a^\dagger_{j,p_c} b_{j,p_c}+ J a^\dagger_{j,p_c} b_{j-1,p_c} + \textrm{h.c.} \Big].
\ee 
At the transition to the gapped $\nu=0$ phase at $J=1/2$ the gap closes at $p_c=\pi$ only, as shown in Fig. \ref{fig:lambdaGapless}, and fermionic operators regain their Majorana nature. The corresponding critical chain is then nothing but Kitaev's Majorana chain at the critical point between the strong- and weak-pairing phases.\cite{KitaevWire} The low-energy theory is well known to be of a single linearly dispersing fermion and hence belongs to the Ising universality class (described by Ising CFT), consistent with the central charge $c=1/2$.\cite{Lahtinen14} For $J>1/2$ the gap closes at some momentum $p_c$ and the critical chain describes complex fermions with staggered complex hopping. The low-energy dispersion is again given by single linearly dispersing fermion though (e.g. at $J=1$ and $p_c= 2\pi /3$ the dispersion is $E=\pm |\sin(\frac{q-2\pi/3}{2})|$ as the function of the momentum $q \in [0,2\pi]$ along the chain) and hence the whole gapless phase belongs to the Ising universality class.

The universality class described by Ising CFT is consistent with the gapless phase being the critical point between the chiral $\nu=\pm 1$ topological phases. As these chiral phases support Ising non-Abelian anyons, their edge theories are given by the opposite chiral sectors of Ising CFT.\cite{KitaevHoneycomb} For $K=0$ both chiral sectors coexist giving rise to the full Ising CFT with both left- and right-moving modes. For $K \neq 0$ the Majorana semi-metal gaps into the $\nu=\pm 1$ phases, but the critical point between them and the $\nu=0$ phase remains at $J=1/2$, as shown in Fig. \ref{fig:lambdaGapless}. The analytic solution presented in Appendix \ref{KitaevModelApp} shows that all $K$ dependent contributions vanish at $p_c=\pi$, which implies that the transition is still in the Ising universality class. This can be again understood in terms of the edge theory of the topological phase. The fermionic degrees of freedom in the non-chiral $\nu=0$ phase are topologically trivial and hence the $J=1/2$ is equivalent to a physical boundary of the topological phase. However, as the bulk gap closes the chiral edge states are no longer localized at the edges, but span the whole system and manifest as coexisting chiral parts of the Ising CFT at the critical momentum.

%\begin{figure}[t]
%$\begin{array}{cc}
%\includegraphics[scale=0.25]{./E_Kitaev_vf_cylinder_K0_J05}&\includegraphics[scale=0.25]{./E_Kitaev_vf_cylinder_K0}\\
%\includegraphics[scale=0.25]{./ES_Kitaev_vf_torus_K0_J05}&\includegraphics[scale=0.25]{./ES_Kitaev_vf_torus_K0}\\
%\includegraphics[scale=0.25]{./S_Kitaev_vf_torus_K0_J05}&\includegraphics[scale=0.25]{./S_Kitaev_vf_torus_K0}\\
%\end{array}$
%\caption{\label{fig:lambdaGapless}
%The energy spectrum $E(p)$, the entanglement spectrum $\lambda(p)$ and the entanglement dispersion $S(p)$ in the gapless Majorana semi-metals in the vortex-free ($\theta=0$, Left) and full-vortex sectors ($\theta=1$, Right). Here $J=1$ and $K=0$.}
%\end{figure}

%
%\begin{figure}[t]
%$\begin{array}{cc}
%\includegraphics[scale=0.25]{./E_Kitaev_vf_cylinder_K015_J05}&\includegraphics[scale=0.25]{./E_Kitaev_torictofv_cylinder_K015}\\
%\includegraphics[scale=0.25]{./ES_Kitaev_vf_torus_K015_J05}&\includegraphics[scale=0.25]{./ES_Kitaev_torictofv_torus_K015}\\
%\includegraphics[scale=0.25]{./S_Kitaev_vf_torus_K015_J05}&\includegraphics[scale=0.25]{./S_Kitaev_torictofv_torus_K015}
%\end{array}$
%\caption{\label{fig:lambdaGapless2}
%The energy spectrum $E(p)$, the entanglement spectrum $\lambda(p)$ and the entanglement dispersion $S(p)$ at the transitions from the gapped $\nu=0$ phase into the gapped topological $\nu=1$ phase in the vortex-free sector ($\theta=0$ and $J=1/2$, Left) and the gapped $\nu=2$ phase inthe full-vortex sector ($\theta=1$ and $J=\sqrt{(1-K^2)/2}$, Right). Here $K=0.15$ in both. }
%\end{figure}

In the full-vortex sector the gapless phase has four Fermi points. At $J=1$, only two of them, $p_c= \frac{\pi}{6}, \frac{5\pi}{6}$, are independent with the corresponding critical wires being given by
\be
	H_{p_c} = \frac{i}{2}\sum_{j}\Big[(e^{-i p_c} + (-1)^j) a^\dagger_{j,p_c} b_{j,p_c}+ J a^\dagger_{j,p_c} b_{j-1,p_c} + \textrm{h.c.} \Big].
\ee
At $J=1$ the dispersion is given by 
\be
	E=\pm\sqrt{3\pm\sqrt{6+2\cos (2p_c)+\cos(2p_c-2q)-\cos 2q}}, \nonumber
\ee
($q \in [0,\pi]$ is again the momentum along the wire), which for $p_c= \frac{\pi}{6}, \frac{5\pi}{6}$ vanishes linearly for $q=\frac{\pi}{3},q=\frac{2\pi}{3}$, respectively. This means that each wire describes an independent linearly dispersing fermion, consistent with the total central charge of $c=1$. Thus the gapless phase in the full-vortex sector is in the universality class of the 1D quantum XY model described by $so(2)_1$ CFT.\cite{Lahtinen14} Indeed, the edge spectra of the gapped $\nu=\pm2$ phases for $K \neq 0$ are again described by the chiral parts of this CFT that must coexist at the critical gapless phase.\cite{KitaevHoneycomb}
 
The transition between the gapped $\nu=0$ phase and the gapless phase occurs at $J=1/\sqrt{2}$, where the gap closes at the independent momenta $p_c=0,\pi$. At these momenta the critical chains reduce to staggered Majorana chains that via a Jordan-Wigner transformation map into critical Ising chains with staggered transverse fields. The dispersion is given by $E=\pm\sqrt{2\pm\sqrt{(7-\cos q)/2}}$, implying that both describe again independent linearly dispersing fermions. This further confirms that the whole gapless phase is in the XY universality class. When $K \neq 0$, the transition shifts smoothly to $J=\sqrt{\frac{1-K^2}{2}}$, but as there are still two Fermi points, also the transition between $\nu=0$ and $\nu=\pm 2$ phases belongs to the XY universality class.

Finally, the spectrum at the transition from the $\nu=1$ to the $\nu=2$ topological phase, that for $J=1$ occurs at $\theta=\frac{3+K^2}{4}$, exhibits a band gap closure only at $p_c=0$. Thus this critical point belongs also to the Ising universality class, as confirmed by the observation of $c=1/2$. This is consistent with the coset $\frac{so(2)_1}{\textrm{Ising}} \simeq \textrm{Ising}$ CFT that is predicted to describe the edge theories of nucleated topological phases.\cite{Ludwig11} As argued in Ref. \onlinecite{Lahtinen}, the $\nu=2$ phase nucleates from the $\nu=1$ phase due to anyon-anyon interactions and thus the transition between them should indeed be described by the Ising CFT.

%\begin{figure}[t]
%$\begin{array}{cc}
%\includegraphics[scale=0.29]{./E_Kitaev_vftofv_cylinder_K015}&\includegraphics[scale=0.28]{./E_Kitaev_vftofv_cylinder_K0_version1}\\
%\includegraphics[scale=0.29]{./ES_Kitaev_vftofv_torus_K015}&\includegraphics[scale=0.29]{./ES_Kitaev_vftofv_torus_K0}\\
%\includegraphics[scale=0.29]{./S_Kitaev_vftofv_torus_K015}&\includegraphics[scale=0.29]{./S_Kitaev_vftofv_torus_K0}
%\end{array}$
%\caption{\label{fig:lambdaGapless3} Criticality ${(1\leftrightarrow 2)}$ between topological phases of vortex-free, $\theta=0$, and full-vortex, $\theta=1$, sectors due to sign staggering in $J_z$. Here $ J=1$ and $K>0$ (Left) or $K=0$ (Right). }
%\end{figure}

\section{Conclusions}

We have shown that in Kitaev's honeycomb model the topological character of the gapped phases as well as the criticality of the gapless ones is encoded in few 1D subsystems. By analyzing them, we showed that in the gapped phases the topological edge states dominate the area law behavior, with the precise non-universal contribution depending on the edge state velocity. The edge states also give rise to a universal lower bound of entanglement entropy that depends on their number and character. At the critical points between topological phases, we showed that only a few 1D subsystems are responsible for the critical entropy scaling of the 2D model. By determining the central charges of these chains, we argued that the criticality of the 2D model over the two considered vortex sectors is in the universality class of either the 1D quantum Ising or the XY model.
Other vortex sectors can support topological phases\cite{Lahtinen} with $|\nu|>2$ and phase transitions between them are expected to be more generally in the $so(N)_1$ hierarchy of universality classes, that the Ising and XY classes are examples of.\cite{Lahtinen14} 

First and foremost, our results provide a detailed study how band topology in free fermion systems affects entanglement due to the presence of topological edge states. The universal lower bound is a direct entropic consequence of the band topology\cite{Klich, Fidkowski10} that can in principle be used to numerically distinguish different states of matter also in interacting or disordered systems.\cite{Mei15}
Topological entanglement entropy has been demonstrated to be a useful tool to identify the presence of intrinsic topological order,\cite{Isakov11,Jiang12} but it does not necessarily distinguish between different types of topological orders if it takes the same value in distinct topological phases. This is precisely what happens in the honeycomb model.\cite{Yao10} We showed that fermionic entanglement distinguished the different phases and thus provides valuable complementary information about the topological nature of the ground state.
It would thus be interesting to study the entropy due to edge states in fractional topological insulators that may harbour different topologically ordered states.\cite{Bergholtz13}

Kitaev's honeycomb model itself is again topical due to recent progress in experimentally identifying materials that may be described by it.\cite{Chun15,Banerjee16} The model can be generalized to 3D lattices, where it realizes a Majorana analogue of Weyl semi-metal\cite{Hermanns15}. It remains an open question what is the nature of the criticality at topological phase transitions in such 3D systems or how the Fermi arc-like surface states affect their entropic properties. In principle the route taken here, a decomposition into lower dimensional subsystems, could also be applied to these models. Ideally, one would like to go beyond the exactly solvable variants of the honeycomb model and analyze the entropy signatures of topological order and phase transitions in the presence of exact solvability breaking perturbations present in the candidate materials.\cite{Jackeli09} The study of such frustrated systems could be significantly simplified if it were sufficient to consider the effect of perturbations only in a subset of lower dimensional subsystems.

\section*{Acknowledgements}
K.M. would like to thank Chris Self for inspiring discussions and Kris Patrick for useful comments on the manuscript. This work was partially supported by the EPSRC grant EP/I038683/1 (K.M.) and the DRS POINT fellowship program (V.L.).

\newpage

\bibstyle{plain}

\newpage

\appendix
\section*{Appendix}
\section{Entropy from correlation matrix}\label{EntropyCorrMat}

Following Refs \onlinecite{Peschel, Fidkowski10}, we review the entanglement spectrum in free fermion models and how to obtain it from covariance matrices.

Consider a gaussian state $\rho\propto e^{-H}$, where $H=H_{ij}f_i^\dagger f_j$ is a free fermion Hamiltonian,  that can be interpreted as a thermal state. The Hamiltonian can be brought to the diagonal form $H=\sum_k \epsilon_k \tilde{f}^\dagger_k \tilde{f}_k$, where $\epsilon_k$ is the energy associated with the fermionic mode $\tilde{f}_k$. The eigenvalues $\lambda_k$ of the correlation matrix $C_{ij}=\tr{\rho f^\dagger_i f_j}$, that contains all the information about the state, are then related to the eigenvalues of $H$ via 
\be \label{corr}
	e^{-\epsilon_k}=\frac{\lambda_k}{1-\lambda_k}.
\ee	
The entropy of the state $\rho$ can then be expressed in terms of the correlation matrix eigenvalues as
\be
S=-\sum_k(1-\lambda_k)\log(1-\lambda_k)+\lambda_k\log\lambda_k.
\ee
Likewise, when restricting the state to a subsystem $A$,  the eigenvalues $\lambda^A_k$ of the restricted correlation matrix $C^A_{ij}=\text{tr}_{A}(\rho_A f_i^\dagger f_j)$ are in the correspondence \rf{corr} with the eigenvalues of the reduced state $\rho_A=\text{tr}_{\bar{A}}\rho$. Thus the entanglement entropy between $A$ and its complement is given by
\be
\label{SS}
S=-\sum_k(1-\lambda^A_k)\log(1-\lambda^A_k)+\lambda^A_k\log(\lambda^A_k).
\ee

%Alternatively, consider $\rho_A=\frac{e^{- H_A}}{Z}$ where $Z=\text{tr}(e^{- H_A})$. We can write $\rho_A=\otimes_k \rho^A_k$ with $\rho^A_k=\frac{e^{-\epsilon^A_k \hat{n}_k}}{Z_k}$where $Z_k=\text{tr}(e^{-\epsilon^A_k \hat{n}_k})$ and $\hat{n}_k=\tilde{f}^\dagger_k \tilde{f}_k$. This structure is a consequence of the fermionic Fock space associated with the number operator $\hat{n}_k=\tilde{f}^\dagger_k\tilde{f}_k$. Consequently, the Hilbert space is a tensor product of two-level systems so that its entropy is given by $S=-\sum_{k,n_k} s^{n_k}_k\log{s^{n_k}_k}$ where $s_k^{n_k}=\bra{n_k}\rho^A_k\ket{n_k}$, or explicitely $s^0_k=\frac{1}{1+e^{-\epsilon^A_k}}$ and $s^1_k=\frac{e^{-\epsilon^A_k}}{1+e^{-\epsilon^A_k}}$. From the identity $e^{-\epsilon^A_k}=\frac{\lambda^A_k}{1-\lambda^A_k}$ we find $s_k^0=\lambda_k$ and $s_k^1=(1-\lambda_k)$.In any case, we get
%\be
%\label{SS}
%S=-\sum_k(1-\lambda^A_k)\log(1-\lambda^A_k)+\lambda^A_k\log(\lambda^A_k).
%\ee
%Regarding the correlation matrix, it is diagonal in the basis that diagonalizes the Hamiltonian $H=-\log(\rho)$, i.e. $C_{nm}=\tr{\rho \tilde{f}^\dagger_n \tilde{f}_m}=\lambda_n \delta_{nm}$. 

To treat all free fermion models including superconducting ones on equal footing, it is advantageous to work in terms of covariance matrices (Majorana correlators) instead of correlation matrices (complex fermion correlators). For a system of $N$ fermions, one can always define $2N$ Majorana fermions through $\tilde{a}_n=\tilde{f}_n^\dagger+\tilde{f}_n$ and $\tilde{b}_n=\frac{1}{i}(\tilde{f}_n-\tilde{f}^\dagger_n)$ and consider the matrix of their correlators $\tilde{C}_{mn}=\frac{1}{2}\tr{\rho \gamma_m \gamma_n}$. Using the correlators for the fermion eigenmodes, one observes that $\tr{\rho \tilde{a}_n \tilde{a}_m}=\tr{\rho \tilde{b}_n \tilde{b}_m}=\delta_{nm}$ and $\tr{\rho \tilde{a}_n \tilde{b}_m}=-i(2 \lambda_n-1)\delta_{nm}$. Thus the covariance matrix $\tilde{C}$ decomposes into diagonal blocks given by 
\be
  \frac{1}{2}\left(\begin{array}{cc}1&-i(2 \lambda_n-1)\\i(2 \lambda_n-1)&1\end{array}\right).
\ee  
These have eigenvalues $t_n$ given by $(\frac{1}{2}-t_n)=\pm(\lambda_n-\frac{1}{2})$, which implies that the eigenvalues of the correlation matrix $C$ and the covariance matrix $\tilde{C}$ are related by $t_n=\lambda_n,(1-\lambda_n)$. Thus if one calculates the eigenvalues of $\tilde{C}^A$ and use (\ref{SS}) to calculate the entropy, one obtains twice the value that one would have obtained from $C^A$ . The correct entropy from the covariance matrix is thus given by
\be
S=-\frac{1}{2}\sum_k(1-t_k)\log(1-t_k)+t_k\log(t_k).
\ee

This expression shows that entanglement modes $t_k=1/2$ contribute a maximal entropy of $\log \sqrt{2}$, while all other modes contribute less to the total entropy. If $M$ such modes exist in the spectrum, and if they are topological in the sense that their value can not be changed by adiabatically varying the microscopic parameters of the Hamiltonian $H$, then the total entropy in the topological phase exhibits a lower bound $S \geq M \log \sqrt{2}$.

\section{Kitaev's Honeycomb Model} \label{KitaevModelApp}

\begin{figure}[t]
\begin{center}
\includegraphics[scale=0.3]{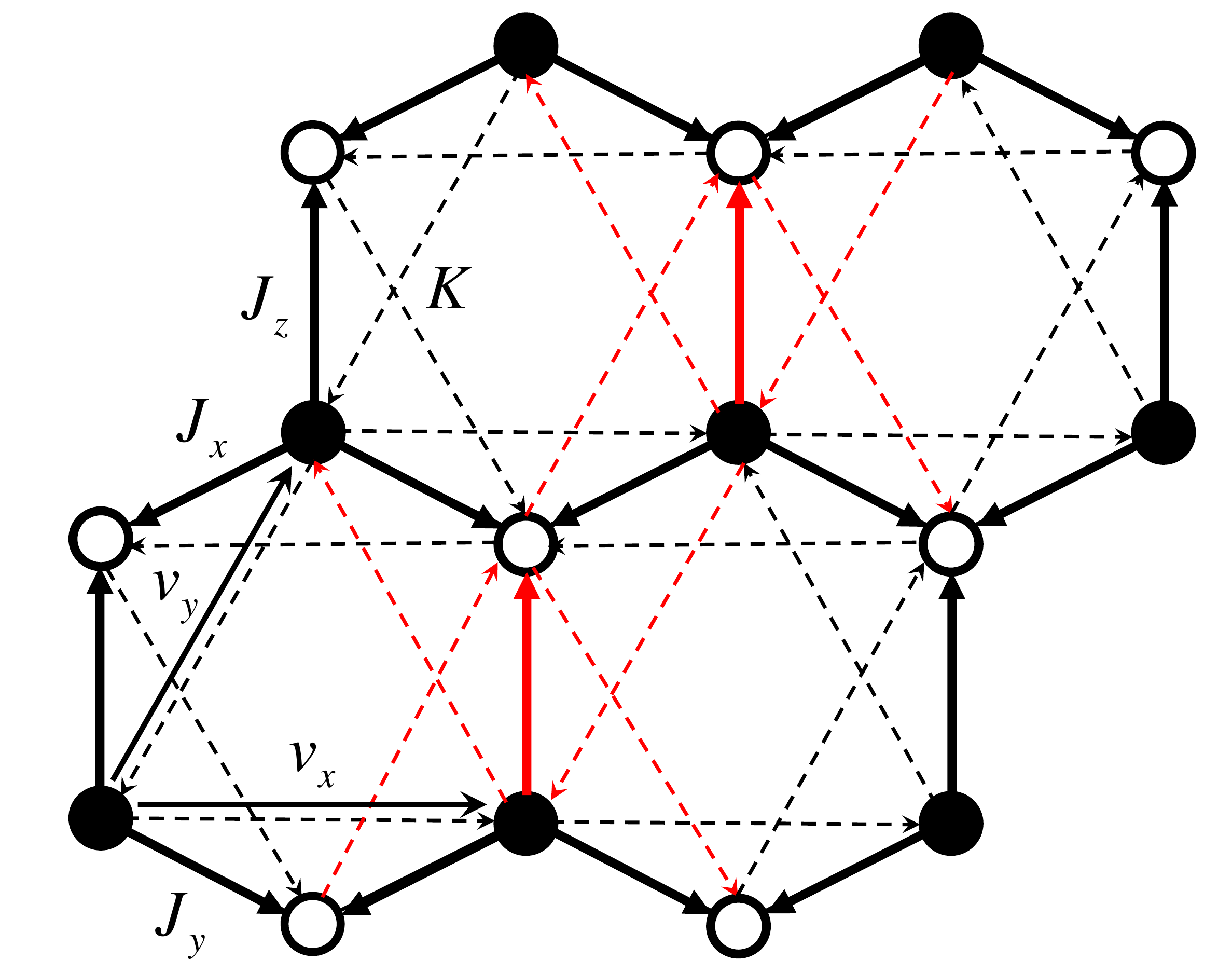}
\end{center}
\caption{\label{fig:honeyhoppings} 
Anisotropic nearest neighbour couplings $J_x, J_y, J_z$ (solid arrows) and chiral next nearest neighbour couplings $K$ (dashed arrows) of the honeycomb lattice model with lattice vectors $v_x$ and $v_y$ (dash-dotted arrows). The black and white sites denote the two triangular sublattices of the honeycomb lattice. In the Majorana picture the tunneling is chosen to be positive along the arrows. In red we denote every other $z$-link along rows of the honeycomb whose value is dictated by $\theta$, enabling the interpolation between the vortex-free and full-vortex sectors.}
\end{figure}

Kitaev's honeycomb model describes spin $1/2$-particles residing on the vertices of a honeycomb lattice~\cite{KitaevHoneycomb}. The spins interact according to the Hamiltonian
\be \label{H_honey}
	H = \sum_{r =x,y,z} \sum_{\langle i,j \rangle} J_r \sigma_i^r \sigma_j^r + K \sum_{\langle\langle i,j,k \rangle\rangle} \sigma_i^x \sigma_j^y \sigma_k^z,
\ee
where $J_r>0$ are nearest neighbour spin exchange couplings along links of orientation $r$, as illustrated in Fig.~\ref{fig:honeyhoppings}, and $K$ is the magnitude of a three spin terms that explicitly breaks time reversal symmetry. The latter is required for the model to support gapped topological phases characterised by non-zero Chern numbers. They are chosen such that for every hexagonal plaquette $\hexagon$ one can associate a $\Z_2$ valued six spin operator $\hat{W}_{\hexagon}=\sigma_1^x \sigma_2^y \sigma_3^z \sigma_4^x \sigma_5^y \sigma_6^z$ that describes a local symmetry $[H,\hat{W}_{\hexagon}]=0$. The Hilbert space of the spin model thus breaks into sectors labeled by the patterns $\{ W_{\hexagon}=\pm 1 \}$ of the eigenvalues of $\hat{W}_{\hexagon}$. We refer to these sectors as {\it vortex sectors}, since $W_{\hexagon}=-1$ corresponds to having a $\pi$-flux vortex on plaquette ${\hexagon}$.

The interacting spin system \rf{H_honey} can be mapped to free Majorana fermions, by introducing four Majorana fermions per vertex through $\sigma^r_i=i b^r_i c_i$. The corresponding Hamiltonian is given by (positive tunneling assumed along the oriented links shown in Fig.~\ref{fig:honeyhoppings})
\be \label{H_honey_Maj}
	H = \frac{i}{2}\sum_{\langle ij \rangle} J_{ij} \hat{u}_{ij} \gamma_i \gamma_j + \frac{i}{2} K \sum_{\langle \langle ij\rangle\rangle} \hat{u}_{ik} \hat{u}_{k j} \gamma_i \gamma_j,
\ee
where the first sum is over nearest neighbour sites $\langle ij \rangle$, the second over next nearest neighbors $\langle\langle ij \rangle\rangle$ with $k$ denoting the connecting site and $J_{ij}=J_x, J_y, J_z$ depending on the orientation of $(i,j)$-link. The operators $\hat{u}_{ij}=i b_i^r b^r_j$ describe $Z_2$ valued gauge fields living on the links. They are static, i.e., they satisfy $[H,\hat{u}_{ij}]=0$, but the mapping requires all physical states to obey $D_i=ib_i^x b^y_ib^z_i c_i=1$. As the operator $D_i$ commutes with the Hamiltonian, but anti-commutes with the three gauge potentials at the vertex, it acts as a local gauge transformation on the $\hat{u}_{ij}$ fields. The physical gauge-invariant operators are the the plaquette operators that take the form of $\Z_2$ valued Wilson loop operators
\be
\hat{W}_{\hexagon}=\prod_{(ij)\in {\hexagon}} \hat{u}_{ij}. 
\ee
	
By choosing a gauge, i.e., replacing the operators $\hat{u}_{ij}$ with their eigenvalues $u_{ij}=\pm1$, one restricts to a particular vortex sector $\{W(\{u\})\}$. In each sector the Hamiltonian $H_{\{W(\{u\})\}}$ is quadratic in the Majorana fermions and hence readily diagonalisable, with the resulting spectrum of free fermions depending only on the vortex sector $W$. Thus even if physical states need to be in principle symmetrized over all compatible gauge configurations, every gauge sector is orthogonal to each other and the physical spectrum of the fermions can be obtained by working in a single gauge. The only aspect lost when working in a single gauge is the long-range entanglement responsible for the topological entanglement entropy.\cite{Yao10} We do not concern us with it here as it takes the same value in all topological phases.

\subsection{Mapping into 1D chains}\label{chaindecomp}

%The structure of the honeycomb lattice allows us to define a unit-cell of two Majorana fermions $c^a$ and $c^b$. We can then explicitly write
%\bq
%\label{eq:H}
%H&&=\frac{i}{2}\sum_r \Big[\big( J_x u^x_{r,r-s_1}c^a_r c^b_{r-s_1} + J_y u^y_{r,r-s_2} c^a_r c^b_{r-s_2} \nonumber\\ 
%&&~~~~~~~~~~~~~~~+ J_z u^z_{r} c^a_r c^b_{r} \big)\nonumber\\
%&&-K \big(u^y_{r,r-s_2}u^x_{r+s_1-s_2,r-s_2}c^a_r c^a_{r+s_1-s_2}\nonumber\\
%&&~~~~~~+u^z_{r}u^y_{r+s_2,r}c^a_r c^a_{r+s_2}+u^x_{r,r-s_1}u^z_{r-s_1}c^a_r c^a_{r-s_1}\nonumber\\
%&&~~~~~~+u^z_{r}u^y_{r,r-s_2}c^b_r c^b_{r-s_2}+u^x_{r+s_1,r}u^z_{r+s_1}c^b_r c^b_{r+s_1}\nonumber\\
%&&~~~~~~+u^y_{r+s_2,r}u^x_{r+s_2,r+s_2-s_1}c^b_r c^b_{r+s_2-s_1}\big)\Big]+\text{h.c.},
%\eq
%where $r$ labels unit-cells where the Majorana fermions $c^a_r$ and $c^a_r$ live, and $s_1=(\frac{\sqrt{3}}{2},\frac{3}{2})$ and $s_1=(-\frac{\sqrt{3}}{2},\frac{3}{2})$ are the primitive lattice vectors. The notation $u^{x,y,z}_{r_1,r_2}$ represents the link between the Majorana fermion  at position $r_1$ and the Majorana fermion  at position $r_2$. To avoid confusion the $u$s also have the information about the direction of the link ($x,y,z$). In fact it is always the case that $u^{x}_{r_1,r_2}=u^{x}_{r_1,r_1-s_1}$, $u^{y}_{r_1,r_2}=u^{y}_{r_1,r_1-s_2}$ $u^{z}_{r_1,r_2}=u^{z}_{r_1}$.

The basis vectors of the honeycomb lattice are given by $\mathbf{v_x}=(\frac{1}{2},\frac{\sqrt{3}}{2})$ and $\mathbf{v_y}=(-\frac{1}{2},\frac{\sqrt{3}}{2})$. Numerically, it is more convenient to work with a brickwall variant by taking them as $\mathbf{v_x}=(1,0)$ and $\mathbf{v_y}=(0,1)$, which only deforms the spectrum without affecting the physics.

In the main text we focus on the vortex-free and full-vortex sectors of the model. When working on a fixed gauge, the local gauge variable $u_{ij}$ is equivalent to the signs of the couplings $J_{ij}$ and $K_{ijk}$. Thus one can effectively interpolate between different vortex sectors by tuning the couplings.\cite{Lahtinen08} To interpolate smoothly between them, we vary only couplings along the $z$-links, as shown in Figure \ref{fig:honeyhoppings}, by multiplying each $J_z$ and the corresponding $K$ couplings with $\Theta_j=(1-\theta)+\theta(-1)^j$, where the index $j$ runs enumerates unit cells containing a single black and a white site along $\mathbf{v_x}$. For $\theta=0$ all the couplings have same signs and hence the system is in the vortex-free sector, while for $\theta=1$ the signs alternate and the system describes the full-vortex sector.

Under these conventions and setting $J_x=J_y=J$ and $J_z=1$, the Hamiltonian \rf{H_honey_Maj} becomes
\bq
\label{eq_realspaceMajorana}
H&&=\frac{i}{2}\sum_{\bf r}\Big[J a_{\bf r} b_{{\bf r}-{\bf v}_y}+ J a_{\bf r} b_{{\bf r}-{\bf v}_x} +\Theta_j a_{\bf r} b_{\bf r}\nonumber\\
&&+K\Big( a_{\bf r} a_{{\bf r}+{\bf v}_x-{\bf v}_y} + \Theta_j a_{\bf r} a_{{\bf r}+{\bf v}_y} + \Theta_{j+1}  a_{\bf r} a_{{\bf r}-{\bf v}_x}   \nonumber\\
&&+ b_{\bf r} b_{{\bf r}-{\bf v}_x+{\bf v}_y}+\Theta_j b_{\bf r} b_{{\bf r}-{\bf v}_y}+\Theta_{j+1} b_{\bf r} b_{{\bf r}+{\bf v}_x} \Big)\Big]+\text{h.c.},\nonumber\\
\eq
where $\mathbf{r}={j,k}$ is a vector in the basis $\mathbf{v}_{x,y}$. The Majorana operators $\gamma$ are labelled $a$ or $b$ depending whether they live on the black or white sublattice, respectively, as illustrated in Figure \ref{fig:honeyhoppings}. The system always has full translational invariance along $\mathbf{v_y}$ along which the Fourier transform of the Majorana operators is given by
\be
\gamma_\mathbf{r}= \sum_{p} e^{ -i p k} \gamma_{j,p}~~,~~~~~p\in[0,2\pi).
\ee
Inserting this into \rf{eq_realspaceMajorana} gives the Hamiltonian \rf{eq_HforChains_main}, which describes a set of 1D systems indexed by the momentum $p$.

\subsection{Analytic forms of the energy gaps}

The energy gaps for vortex-free and full-vortex sectors have been solved in Ref. \onlinecite{Lahtinen08}. Fourier transforming also along $\mathbf{v}_x$, in the vortex-free sector it is given by
\bq
	G(p) & = & \min_q{\sqrt{|f(q,p)|^2+g(q,p)^2}}, \\
	f(p,q) & = & 2 i (1+\frac{1}{2} e^{i q} + \frac{1}{2} e^{i p}), \nonumber \\
	g(p,q) & = & 4 K (\sin{p}-\sin{q}+\sin{(q-p)}), \nonumber
\eq 
where $q \in [0,2\pi]$. For $K=0$, in the full-vortex sector the gap is given by 
\bq
	G(p) & =& \min_q {2 \sqrt{1+2 J^2-J\sqrt{2 g(q,p)}}}, \\
	g(p,q) & =& 2 + \cos{2 p} -\cos{2 q} + J^2(1+ \cos{2 (p-q)}), \nonumber
\eq
with $q \in [0,\pi]$ since the unit cell is doubled due to sign staggering of the couplings.

\subsection{Vortex-free sector and Kitaev's Majorana chain}
\label{Kitaev chain in the topological phase of the vortex-free sector}

Kitaev's Majorana chain is describes spinless 1D superconducting fermions $c$ with the Hamiltonian
\begin{equation}
H_{\text{KC}}=-\mu\sum_j c^\dagger_j c_j-\frac{1}{2}\sum_j (t c^\dagger_{j}c_{j+1}+\Delta e^{i\phi}c_{j}c_{j+1}+\text{h.c.}).
\end{equation}
Writing $c_j=\frac{1}{2}(b_j+i a_j)$, in terms of Majorana fermions $a_j$ and $b_j$ it takes the form
\bq
H_{\text{KC}}&=&\frac{\mu}{2}\sum_j ( i a_j b_j +1)+\frac{t+\Delta_R}{4} (i a_j b_{j-1}) \nonumber\\
&&-\frac{t-\Delta_R}{4}(i a_j b_{j+1})
+\frac{\Delta_I}{4} (i b_j b_{j-1}-i a_j a_{j-1}),\nonumber\\
\eq
where $\Delta_R=\Delta \cos{\phi}$ and $\Delta_I=\Delta \sin{\phi}$. By direct comparison to \rf{eq:Hp=pi}, one finds that the chain in the vortex-free sector that corresponds to the crossing of the edge states is equivalent to the Majorana chains through ${\mu=2(1-J)}$, ${t=\Delta_R=2 J}$ and ${\Delta_I=-4 K}$. Since the pairing function can be chosen real via the gauge transformation $c_j \to e^{-i\phi/2} c_j$, one can set $\phi=0$, which implies that the model does not depend on $\Delta_I$. Thus the parameter $K$ of the honeycomb model is irrelevant to the behavior of this chain.

\section{Edge-mode dispersion in the vortex-free sector}
\label{GenFunMethod}

On the cylinder after a partial Fourier transformation the Hamiltonian reads $H=\sum_p H(p)$where $p$ is the momentum in the periodic direction. We impose that the cylinder is semi-infinite with the boundary lying at $j=1$. The unit cell contains two sites indicated by the two species of fermionic operators $a,b$. The Hamiltonian reads
\bq
H(p)&=&\sum_{j=2}\chi_j^\dagger \Gamma_1\chi_j+\chi_j^\dagger \Gamma_2\chi_{j-1}+\chi_j^\dagger \Gamma_2^\dagger\chi_{j+1}\nonumber\\
&&+\chi_1^\dagger \Gamma_1\chi_1+\chi_1^\dagger \Gamma_2^\dagger\chi_2,
\eq
where $\chi_j={\left(\begin{array}{cc}a_j&b_j\end{array}\right)}^T$ and $\Gamma_1= \left(\begin{array}{cc}\zeta&\omega\\{\omega}^*&-\zeta\end{array}\right)$ and $\Gamma_2=\left(\begin{array}{cc}\gamma&J\\0&-\gamma\end{array}\right)$
with
$\zeta=- K \sin{p}$, $\omega=i(J e^{i p}+1)$, and $\gamma=-\frac{i}{2}K(e^{-i p}-1)$.
For any chain, the Schr\"odinger equation $H\Psi=E\Psi$, where $\Psi=\sum_p \chi_j^\dagger\Psi_j(p)\ket{0}$ and $\Psi_j(p)=(\Psi_j^1~\Psi_j^2)^T$,
leads to the following recursive equation and boundary condition
\bq\Gamma_2\Psi_{j-1}+(\Gamma_1-E)\Psi_j+\Gamma_2^\dagger\Psi_{j+1}&=&0\nonumber\\
(\Gamma_1-E)\Psi_1+\Gamma_2^\dagger\Psi_{2}&=&0.
\eq
We now multiply by $z^j$ where $z\in\mathbb{C}$, sum over $j$, and use the boundary condition to obtain
\be\label{eq:GenFunction}
(z^2 \Gamma_2+z(\Gamma_1-E)+\Gamma_2^\dagger)G(z)=\Gamma_2^\dagger\Psi_1,
\ee
where we introduced the generating function 
\be
\label{eq:GGz}
G(z)=\sum z^{j-1}\Psi_j.
\ee
By inverting (\ref{eq:GenFunction}) we solve for the generating function
\be
G(z)=(z^2 \Gamma_2+z(\Gamma_1-E)+\Gamma_2^\dagger)^{-1}\Gamma_2^\dagger \Psi_1,
\ee
or, explicitly
\bq
\label{eq:G}
&G(z)=\frac{1}{\Delta}\times\\
&\times\left(\begin{array}{c}
z^2(-\gamma \phi_1-J \phi_2)+z[(-\zeta-E)\phi_1-\omega\phi_2]-{\gamma}^*\phi_1\\
z^2(\gamma \phi_2)+z[(\zeta-E)\phi_2-{\omega}^*\phi_1]-J \phi_1+{\gamma}^*\phi_2
\end{array}\right),\nonumber
\eq
where  $\Gamma_2^\dagger \Psi_1=(\phi_1~\phi_2)^T$ and $\Delta(z)=\text{det}(z^2 \Gamma_2+z(\Gamma_1-E)+\Gamma_2^\dagger)$.

The case $\det{\Delta}(z)=0$ with $z=0$ only happens if $\gamma=0$. Such a case implies either $K=0$ or $p=0$. For $p=0$ we do not expect end modes since the chain is topologically trivial and we are keeping the model gapped $K>0$. If $\Delta(z_0)=0$ with $z_0\neq 0$ we notice that $\Delta(\frac{1}{z^*_0})=\frac{1}{{z^*}^2}{\Delta}^*(z_0)=0$, so the poles of the generating function come in pairs, $(z_{1,2},\frac{1}{z_{1,2}})$. An end mode exists provided that~\cite{Pershoguba} the poles have norm bigger than $1$. Since the poles of the generating function come in pairs, $z_{1,2}$, at least two of them are such that $|z_{1,2}|<1$.

For an end mode to exists, the poles $z_{1,2}$ must be simplified by zeros in the numerator in the expression for $G$. This means that both rows of the vector defining $G$ have to be proportional to $(z-z_1)(z-z_2)$. As a consequence, we can write the following proportionality relations (for the terms proportional to $z^2$ and identity)
\bq
\label{eq:necessaryCond}
-{\gamma}^*\phi_1&=&t(-J\phi_1+{\gamma}^*\phi_2),\nonumber\\
-\gamma\phi_1-J\phi_2&=&t \gamma \phi_2,
\eq
where $t\in\mathbb{C}$ is a proportionality constant. By multiplying both sides together this leads to
\be
|J|^2\phi_1\phi_2+\gamma J \phi_1^2-{\gamma}^*J\phi_2^2=0,
\ee
which imposes the following constraint
\be
\phi_1=\xi_\pm \phi_2,
\ee
with 
\be
\label{eq:xi}
\xi_\pm=\frac{-|J|^2\pm\sqrt{|J|^4+4|\gamma|^2|J|^2}}{2\gamma J},
\ee
with the property $\frac{1}{\xi_\pm}=-{\xi}^*_\mp$.
We can now insert this constraint back in (\ref{eq:necessaryCond}) to get yet two other necessary condition
\bq
\label{eq:ForFootnote1}
t_\pm &=&-\frac{{\gamma}^*\xi_\pm}{{\gamma}^*-J\xi_\pm},\nonumber\\
t_\pm&=&-\frac{\gamma \xi_\pm+J}{\gamma}.
\eq
These two conditions have to be compatible and by equating the two equations we get the following compatibility condition
\be
\label{eq:quadraticXi}
J \gamma\xi_\pm^2+|J|^2\xi_\pm-J{\gamma}^*=0,
\ee
which, in fact, has solutions given by (\ref{eq:xi}).
We now go back to (\ref{eq:G}) and impose the proportionality for the terms proportional to $z$ to get
\be
\label{eq:EnergyBoundaryModes}
E_\pm=\frac{t_\pm(\zeta-{\omega}^*\xi_\pm)+(\zeta\xi_\pm+\omega)}{t_\pm-\xi_\pm}.
\ee
%\begin{figure}[t]
%\begin{center}
%\includegraphics[scale=0.5]{pplot}
%\end{center}
%\caption{\label{fig:plotAnalytical} Plot of energy of the edge-modes as a function of momentum for arbitrary value of $J$ and $K$. In red, the edge-modes following  (\ref{eq:EnergyBoundaryModes}) and in blue their linear approximation at $p=\pi$ as described in (\ref{eq:EnergyBoundaryModesLinear}). The plot was calculated for $J=0.6$ and $K=0.1$. These dispersions are only valid in the range between $p=\pi\pm \frac{\Delta p}{2}$, see (\ref{eq:DeltaP}) and Fig.~\ref{fig:11}.}
%\end{figure}

We now linearise around $p=\pi$ since we are looking for an approximate solution.
Noting that the derivative of $|\gamma|^2$ with respect to $p$ is null at $p=\pi$ we can write in first order in $p-\pi$
\be
\label{eq:ForFootnote2}
\xi_\pm=\frac{\xi_\pm^0}{\gamma}=\xi_\pm^0(-\frac{i}{K}+\frac{1}{2K}(p-\pi)),
\ee
where $\xi_\pm^0=\frac{i}{4}(-J\pm\sqrt{J^2+16  K^2})$ and similarly
\be
\label{eq:ForFootnote3}
t_\pm=t_\pm^0 {\gamma}^*=t_\pm^0(-i K +\frac{K}{2} (p-\pi)),
\ee
where $t_\pm^0=\frac{2i(J\pm\sqrt{J^2+16 K^2})}{(J^2+8  K^2\pm J\sqrt{J^2+16 K^2})}$
with the property $\xi^0_\pm t_\pm^0=1$.
The energy dispersion around $p=\pi$ then is
\be
\label{eq:EnergyBoundaryModesLinear}
E_\pm=\pm (2J+1) \cos\left(\arctan{\left(\frac{J}{4K}\right)}\right) (p-\pi)+\ldots 
\ee

We make the cylinder semi-infinite in the other direction by puttting a boundary at $i=N$. Similarly to what has been done above for the other boundary, we have
\be
G_N(z)=( \Gamma_2+z(\Gamma_1-E)+z^2\Gamma_2^\dagger)^{-1}\Gamma_2\Psi_N.
\ee
By defining $\Gamma_2^\dagger \Psi_N=(\phi_1~\phi_2)^T$ we can repeat the procedure given in the previously. Since the zeros of the determinant $\Delta_N=\det( \Gamma_2+z(\Gamma_1-E)+z^2\Gamma_2^\dagger)$ are invariant under $z\rightarrow1/z$ (for $z\neq 0$) then the two sets $S=\{z\in\mathbb{C}:\det(\Delta)=0\}$ and  $S_N=\{z\in\mathbb{C}:\det(\Delta_N)=0\}$ are equivalent $S=\tilde{S}$. Similarly, the zeros of the numerators of $G_N(z)$ are inverted with respect to the ones of $G(z)$: if $z_0$ is a common zero of the numerators of $G_N$ associated with an energy $E_\pm$ then $1/z_0$ is associated with a common zero of the numerators of $G$ with the same energy $E_\pm$.

\begin{figure}[t]
\includegraphics[scale=0.55]{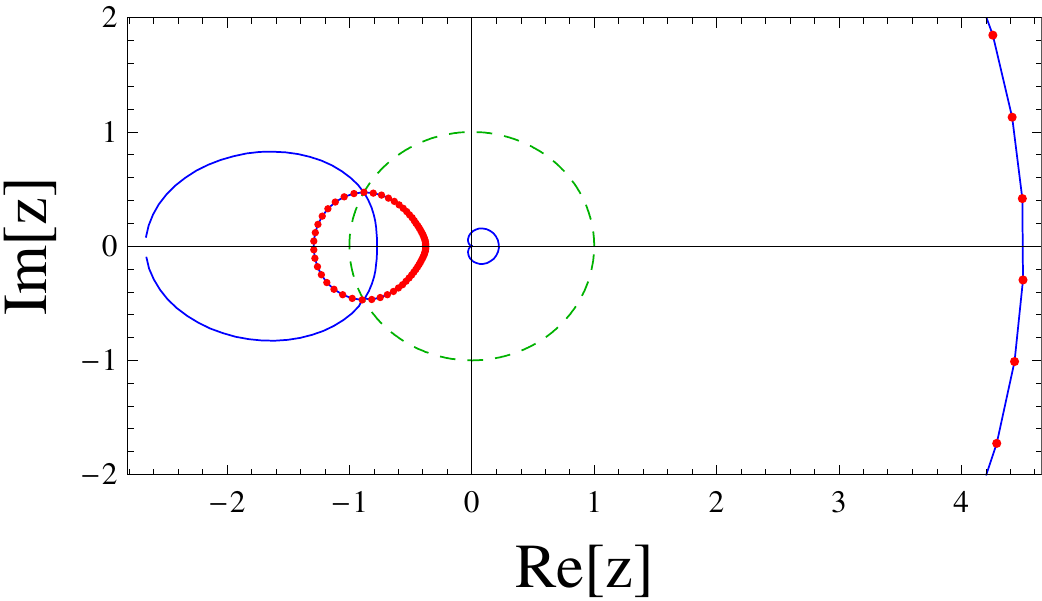}
\includegraphics[scale=0.55]{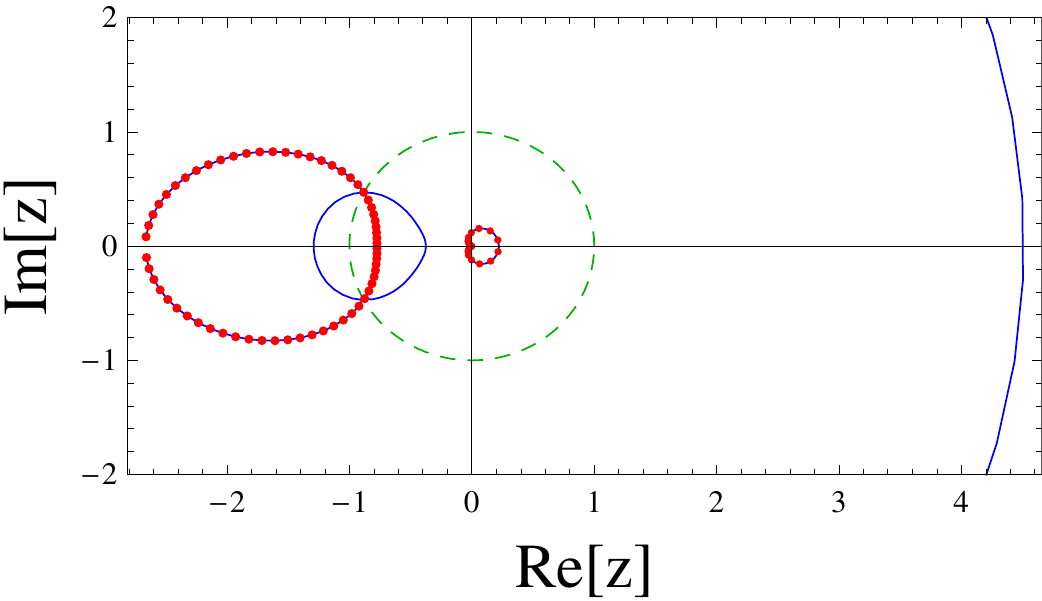}
\caption{\label{fig:9} The set of zeros of the of the denominator $\{z\in D(p)\}$ (blue) and the set of zeros of the numerator $\{z\in N_\pm(p)\}$ (red) of $\tilde{G}_\pm(z)$ on the complex plane (for $J=0.6$ and $K=0.15$). The unit circle $|z|=1$ is depicted in green. Simplification among zeros in the numerator and denominator of $\tilde{G}$ occurs when the red dots cover the blue lines. End modes are possible only if portions of \emph{both} of the two connected blue lines not covered with red dots lie completely outside or inside the green circle. Consistently with Fig.~\ref{fig:10}, the former (latter) situation is possible only for $G_+$ ($G_-$) on the bottom (top) plot. This  implies the existence of edge-modes with dispersion $E_+$ at the boundary $j=1$ and $E_-$ at the boundary $j=N$.}
\end{figure}

\begin{figure}[t]
\begin{center}
$\begin{array}{rr}
\includegraphics[scale=0.35]{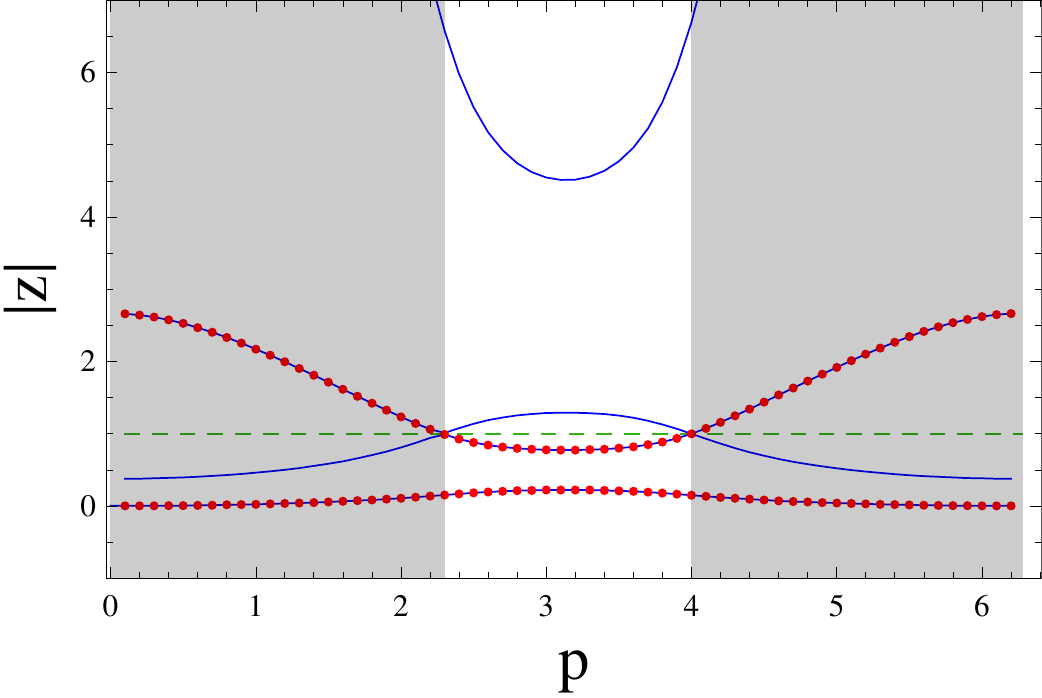}&\includegraphics[scale=0.35]{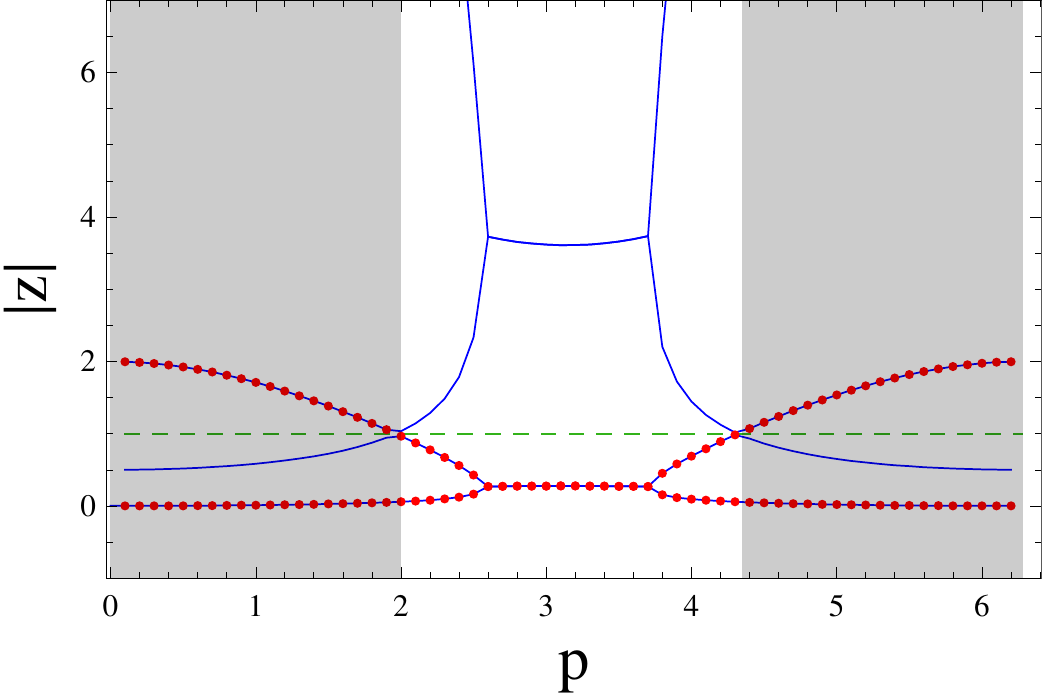}\\
\includegraphics[scale=0.35]{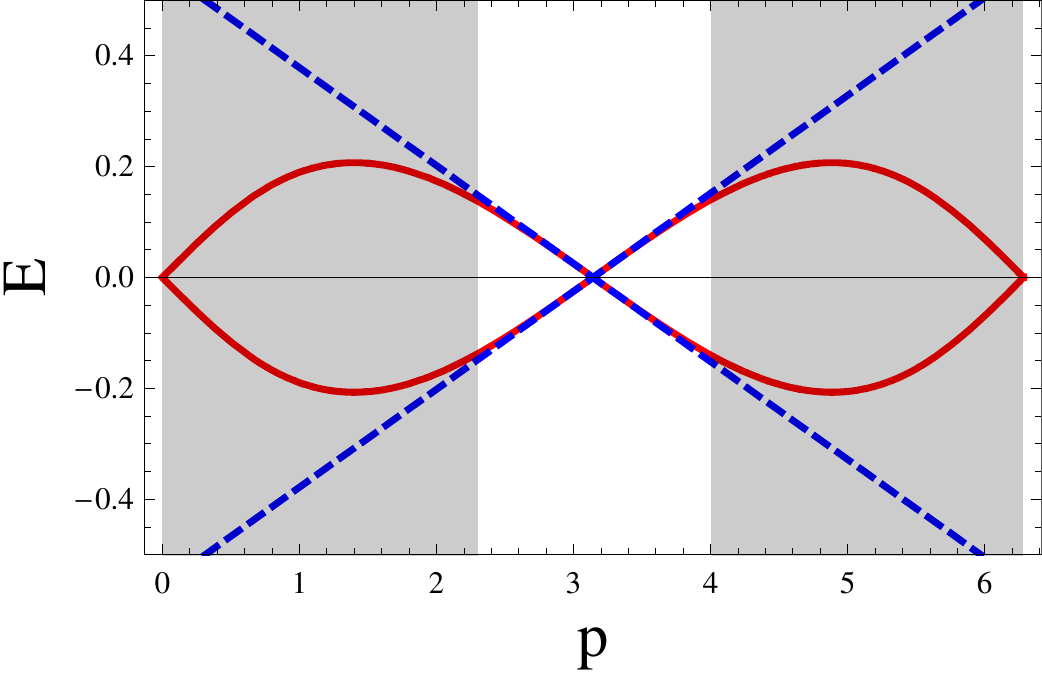}&\includegraphics[scale=0.35]{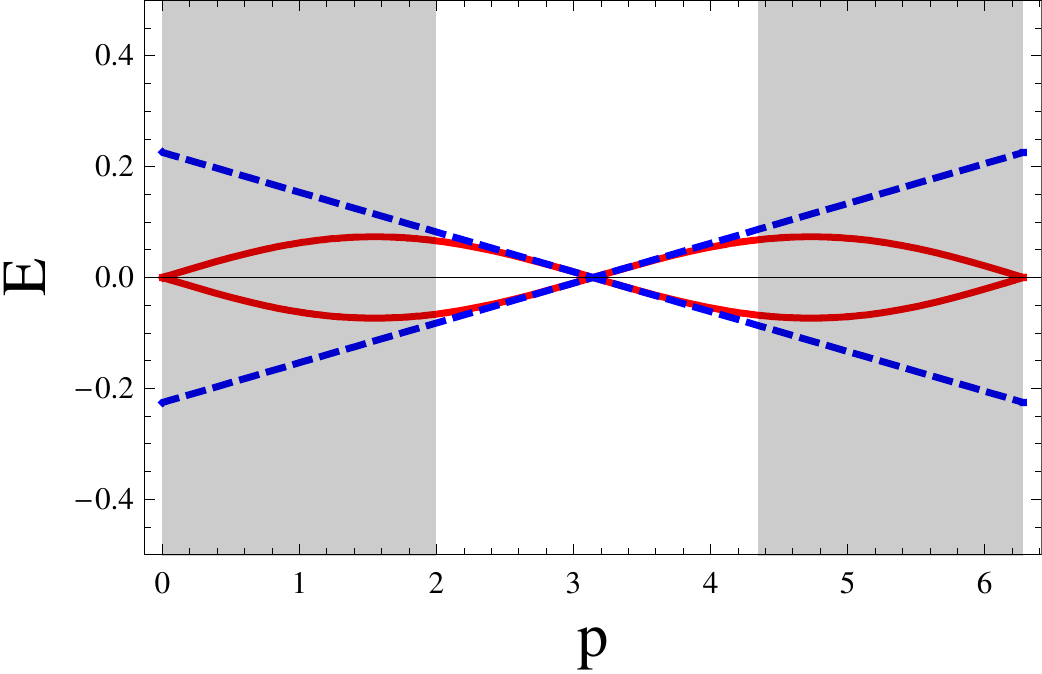}
\end{array}$
\end{center}
\caption{\label{fig:11} Existence of end modes and validity of the edge-mode expression in ( \ref{eq:EnergyBoundaryModes}). Top row:  zeros of the denominator ($|z|:z\in D(p)$, in blue) and the zeros of the numerator ($|z|:z\in N_\pm(p)$, in red) of $\tilde{G}_+$ as a function of $p$ for $J=0.6$ and $K=0.15$ (Left) and for $J=1$ and $K=0.15$ (Right). Bottom row: dispersion of the edge-modes (\ref{eq:EnergyBoundaryModes}) and their linear approximation (\ref{eq:EnergyBoundaryModesLinear}) each for the values of $J$ and $K$ corresponding to the upper row (same side). In dark grey we highlight the momentum space region where edge-modes are \emph{not} present so that the validity of (\ref{eq:EnergyBoundaryModes}) (bottom row) is limited to the white area. The dispersion with negative (positive) angular coefficient at $p=\pi$ corresponds to modes at the $j=1$ ($j=N$) boundary.
}
\end{figure}

We now restrict to the situation where end modes exist in a system with \emph{two} boundaries. In this regime, for each energy $E_\pm$, \emph{all} poles of the generating function have~\cite{Pershoguba} either norm bigger or smaller than $1$. Then,  the two facts just explained immediately imply that poles $(z^\pm_1,z^\pm_2)$ of $G$ associated with $E_\pm$ correspond to poles $\left(\frac{1}{z^\pm_1},\frac{1}{z^\pm_2}\right)$ of $G_N$  associated to the same energy $E_\pm$. In turn, this means that, if at the boundary $j=1$ we have an end mode with energy $E_+$ then, at the boundary $j=N$ the end mode cannot be associated with the same energy dispersion $E_+$ since $|\frac{1}{z^+_1}|,|\frac{1}{z^+_2}|<1$. In summary, if we have an end mode on one boundary defined by $E_+$ then the other boundary has an end mode with dispersion $E_-$ and viceversa (see Fig.~\ref{fig:11}). For each boundary, each dispersion $E_+$ or $E_-$ corresponds to a couple of poles of the generating function \emph{both} of which have norm \emph{either} bigger \emph{or} smaller than 1.

From the analysis given above we know that there are no zero-energy modes for $p=0$. So the analytical expression (\ref{eq:EnergyBoundaryModes}) for the edge modes is valid only in a window $\Delta p$ in momentum space. In order to determine the range in momentum space where such an expression is valid, we need to make sure that the zeros in the numerator actually simplify the unwanted (norm smaller than 1) poles in the denominator. Unfortunately, the expression for the poles of $G(z)$ does not take a simple analytical form but we can get good insights from numerical studies.

\begin{figure}[t]
\begin{center}
$\begin{array}{ll}
\includegraphics[scale=0.35]{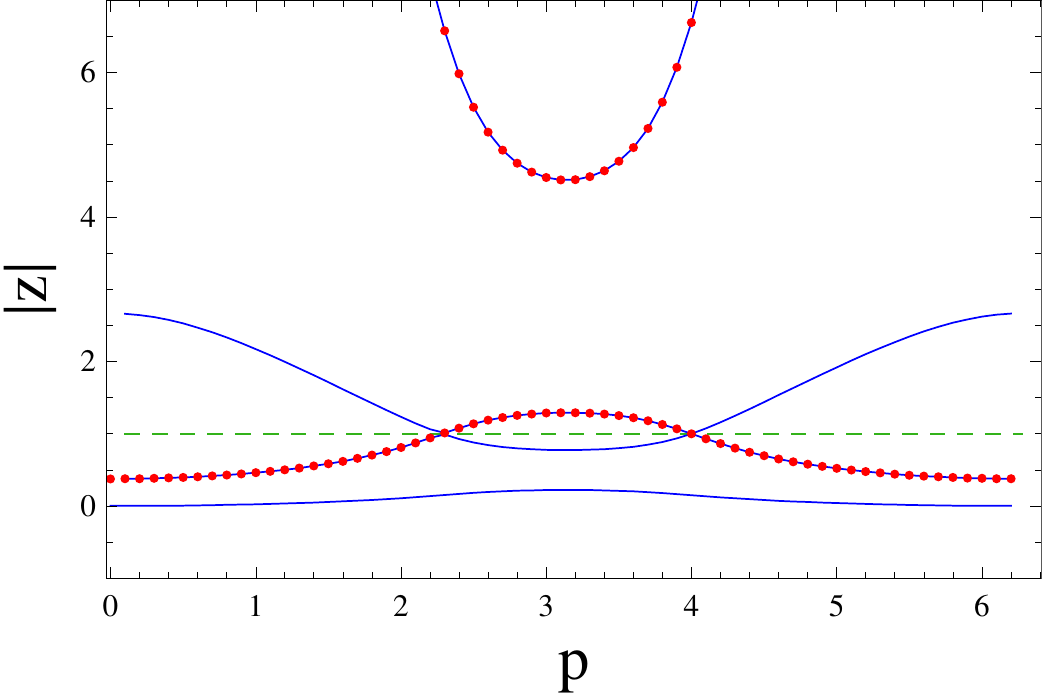}&\includegraphics[scale=0.35]{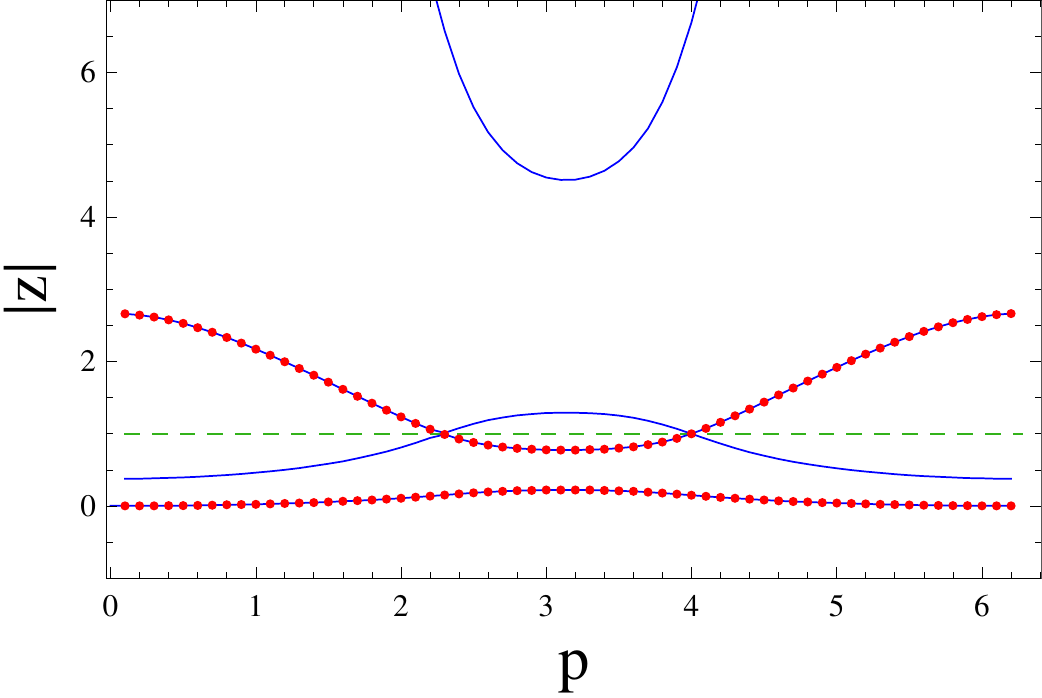}\\
\includegraphics[scale=0.35]{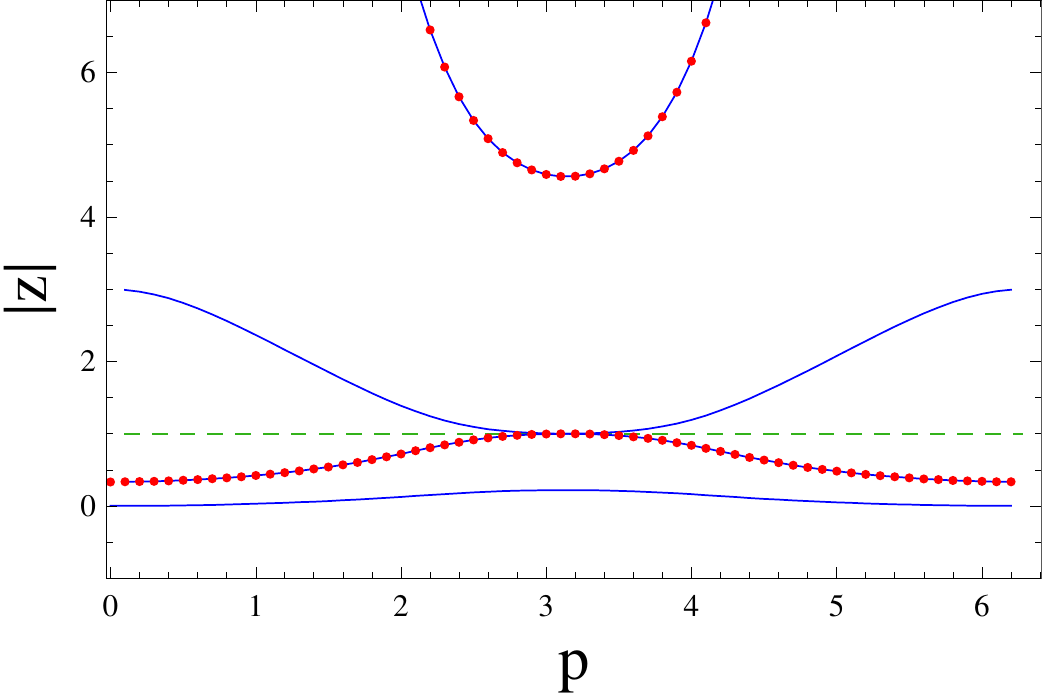}&\includegraphics[scale=0.35]{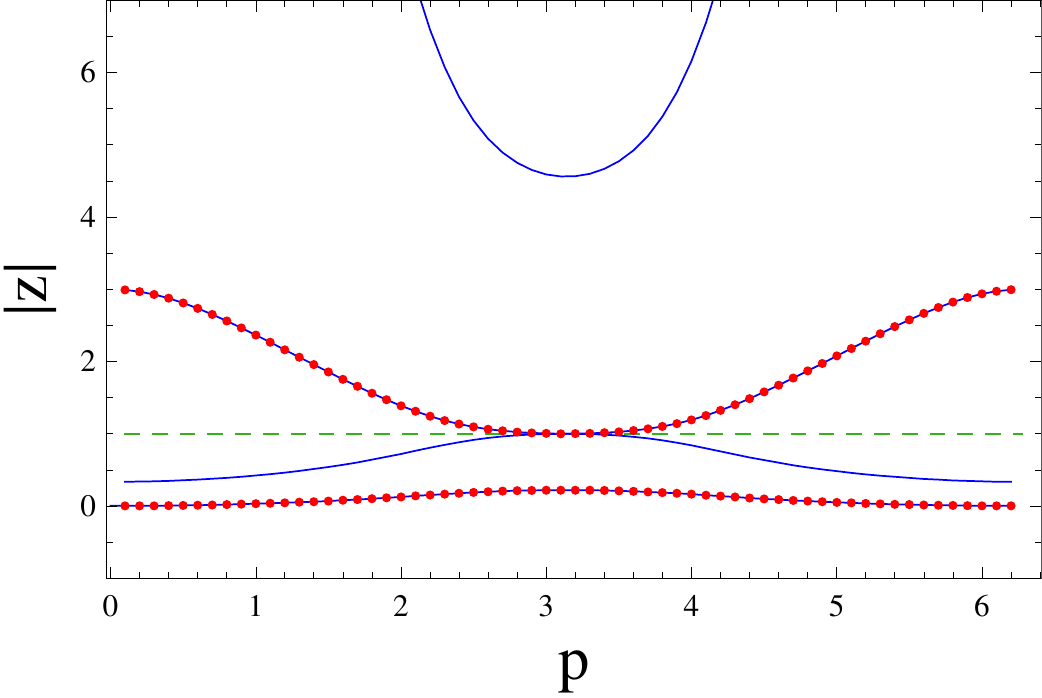}\\
\includegraphics[scale=0.35]{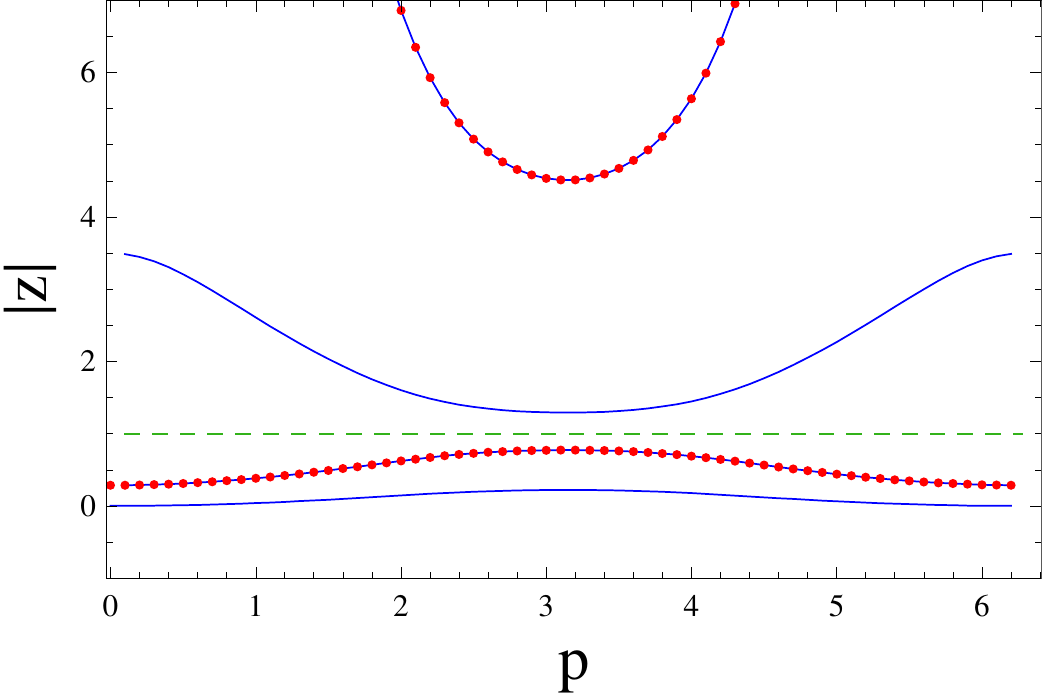}&\includegraphics[scale=0.35]{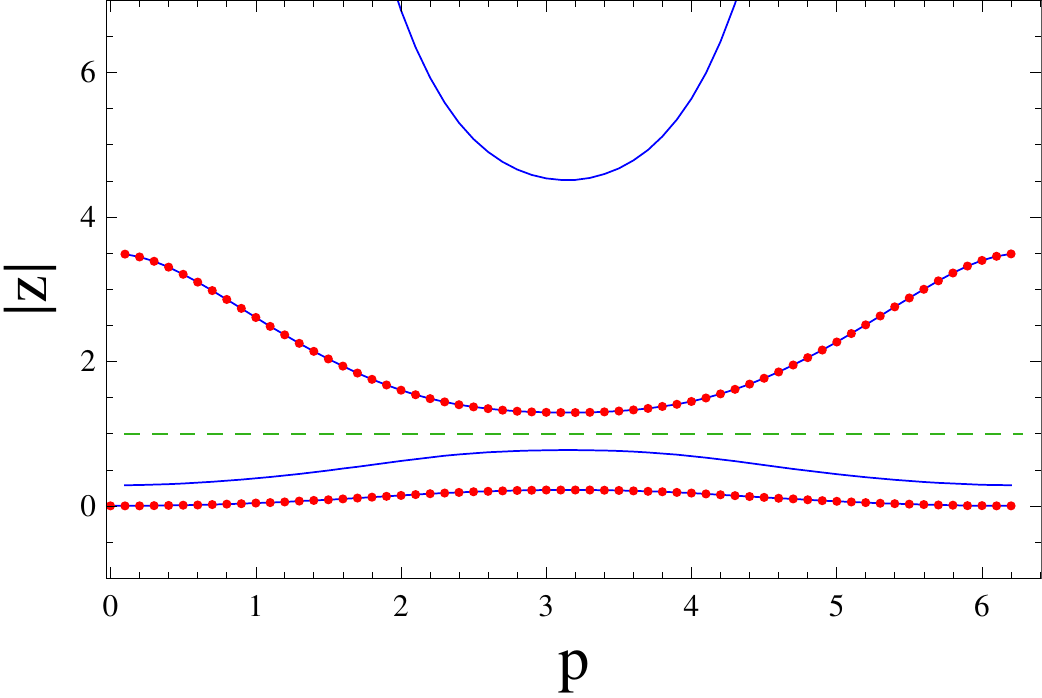}
\end{array}$
\end{center}
\caption{\label{fig:10} Comparison between the absolute value of the zeros of the denominator ($|z|:z\in D(p)$, in blue) and the zeros of the numerator ($|z|:z\in N_\pm(p)$, in red) of $\tilde{G}_\pm$ as a function of $p$ in different regimes. From top to bottom, each row considers fixed parameters $J$ and $K$. In particular $J=0.6$ and $K=0.15$ for the first row (non-Abelian regime), $J=1/2$ and $K=0.15$ for the second row (phase transition), and $J=0.4$ and $K=0.15$ for the third row (Abelian regime). Within each row, on the Left (Right) we plot the zeros of the numerator associated with $G_-$ ($G_+$). In green we plot the line at $|z|=1$. For each plot, the blue lines not covered by red dots represent the poles of $G_\pm$ (when the red dots cover the blue line the zeros of the numerator simplify the zeros of the denominator). In the non-Abelian regime, a range of $p$ exists where \emph{all} poles of $G_+$ ($G_-$) on the right (Left) plot  have absolute value bigger (smaller) than 1 (allowing for end modes at $j=1$ (Right) and $j=N$ (Left)), see also Fig.~\ref{fig:9} and \ref{fig:11}. In the Abelian case there is no $p$ where \emph{all} poles have absolute value bigger or smaller than one. In this case end modes do not arise~\cite{Pershoguba}.
}
\end{figure}

We want to study the poles of the following quantity
\be
\label{eq:GenFun}
\tilde{G}_\pm(z)=
\frac{z^2(-\gamma \xi_\pm-J )+z[(-\zeta-E_\pm)\xi_\pm-\omega]-{\gamma}^*\xi_\pm}{\Delta},
\ee
obtained by taking the first component  of (\ref{eq:G}) without loss of generality since the two components are proportional. The energy $E_\pm$ is given by (\ref{eq:EnergyBoundaryModes}) and $\Delta=\det{(z^2 \Gamma_2+z(\Gamma_1-E)+\Gamma_2^\dagger)}$. For each solution $E_\pm$ we then consider the set of zeros of the numerator
\be
\label{eq:Numerator}
N_\pm(p)=\{z:z^2(-\gamma \xi_\pm-J )+z[(-\zeta-E_\pm)\xi_\pm-\omega]-{\gamma}^*\xi_\pm=0\},
\ee
of size $|N_\pm(p)|=2$, $\forall p$, and the set of zeros of the denominator
\be
\label{eq:Denominator}
D(p)=\{z:\Delta (z)=0\},
\ee
of size $|D(p)|=4$, $\forall p$. Moreover, for each fixed momentum, the elements of $D(p)$ come in pairs $(z_{1,2},\frac{1}{z_{1,2}})$. The position of these two sets can be found in Fig.~\ref{fig:9}, while their norms can be found in Fig.~\ref{fig:10}. The poles of $\tilde{G}_\pm$ are collected in the set
\be
Z_\pm=\{z: z\in D(p)~ \land ~z\not\in N(p)\}.
\ee

\begin{figure}[t]
\begin{center}
$\begin{array}{ll}
\includegraphics[scale=0.35]{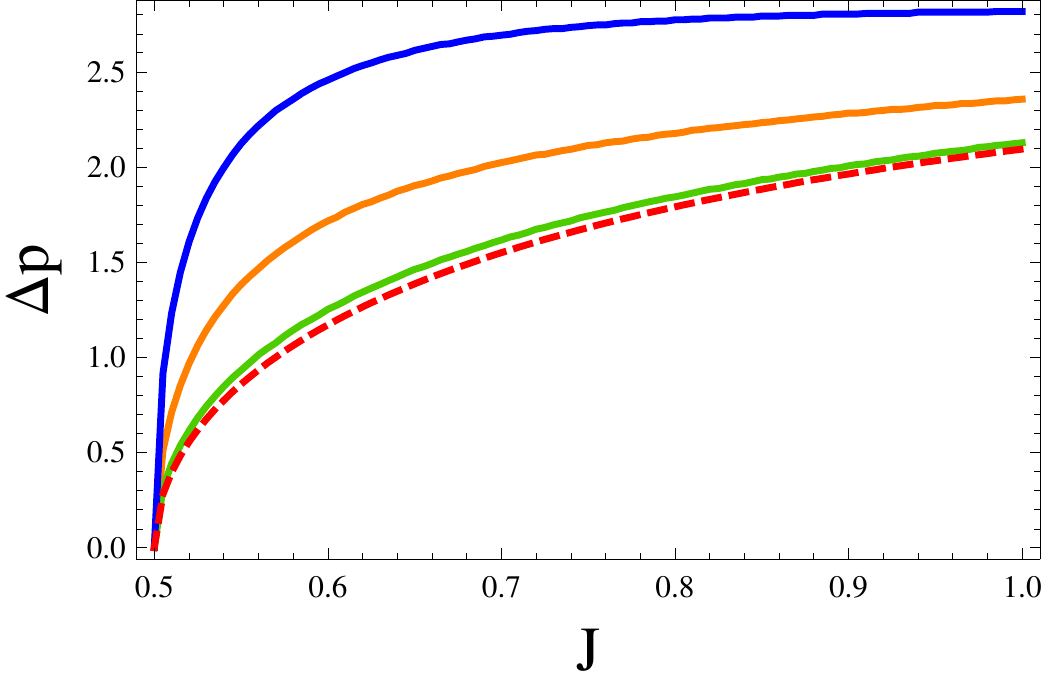}&\includegraphics[scale=0.35]{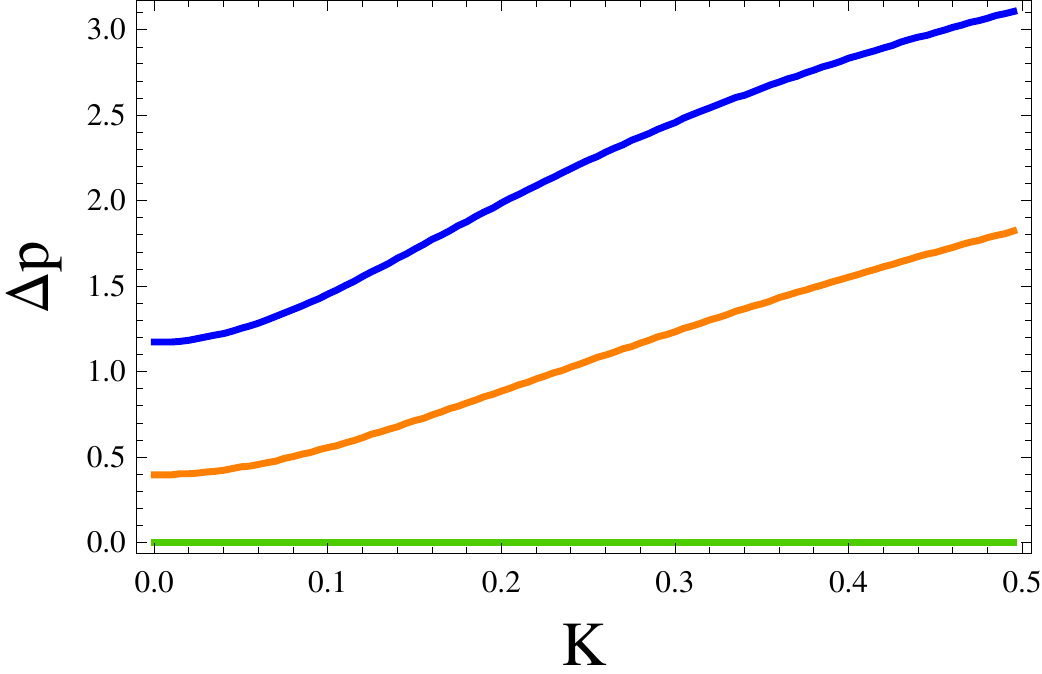}
\end{array}$
\end{center}
\caption{\label{fig:12} Momentum space range $\Delta p$ where edge-modes are present (\ref{eq:DeltaP}) as a function of $J$ (Left) and $K$ (Right). On the left different lines correspond to different values for $K$: dashed line ($K=0$, (\ref{eq:DeltaForK0})), green line ($K=0.05$), orange line ($K=0.15$), blue line ($K=0.3$). On the right different lines correspond to different values of $J$: green line ($J=1/2$), orange line ($J=0.51$), blue line ($J=0.6$). Both plots show that edge-modes exist only in the non-Abelian phase, as expected. In fact, on the left we can see that $\Delta p\rightarrow 0$ as $J\rightarrow {1/2}^{+}$ and on the right $\Delta p=0$ for the $J=1/2$ line consistently with the plots in Fig.~\ref{fig:10}. On the plot on the left we can see that, for $K\rightarrow 0$ the behaviour becomes that described in (\ref{eq:DeltaForK0}) (dashed line).}
\end{figure}

In  Fig.~\ref{fig:9} the elements of $D(p)$ (Blue) and $N(p)$ (Red) show that points which are blue but not red are elements of $Z_\pm$. Edge-modes exists only for those momentum $p$ such that all $z_p\in Z_\pm$ have absolute value bigger than one i.e. $|z_p|>1$. In Fig.~\ref{fig:9} and Fig.~\ref{fig:10} the elements of $Z_\pm$ are represented by the blue lines not covered with red dots. In light of the considerations made previously, end modes exist for $p\in P^\pm_I$ where
\be
P^\pm_I=\{p: |z_p|>1\lor |z_p|<1, \forall z_p\in Z_\pm\}.
\ee
This set encodes the validity region of length $\Delta p$ of (\ref{eq:EnergyBoundaryModes})
\be
\label{eq:DeltaP}
\Delta p =|P_I|,
\ee
so that edge-modes exist in the window $p=\pi\pm\frac{\Delta p}{2}$.
In Fig.~\ref{fig:11} the validity window $\Delta p$ of the linear approximation of the edge-mode dispersion is shaded for an example of the model in the topological phase. The dependence of $\Delta p$ with the model parameters $J,K$ is shown in Fig.~\ref{fig:12}.

In Fig.~\ref{fig:10} and Fig.~\ref{fig:11} we show the absolute value of the elements of $D(p)$ (in blue) and $N(p)$ (in red) as a function of p. In this case edge-modes exist for those $p$ where all the points which are in $D(p)$ (Blue) but not in $N(p)$ (Red) have norm bigger than 1.

From Fig.~\ref{fig:10} and Fig.~\ref{fig:12} we can see that, as expected, edge-modes exist only in the non-Abelian phase, $J>1/2$, while (\ref{eq:EnergyBoundaryModes}) and (\ref{eq:EnergyBoundaryModesLinear}) are valid only for $J>1/2$ inside an interval of length $\Delta p$ centred around $p=\pi$.
For example, in Fig.~\ref{fig:10} we can see that, in the non-Abelian phase, both energies $E_\pm$ from \rf{eq:EnergyBoundaryModes} have a range in momentum space where \emph{all} poles have norm bigger \emph{or} smaller than 1, corresponding to end-states. On the contrary, in the Abelian phase, for all $p$, we have a \emph{mixed} situation with  one pole having norm bigger than 1 and the other pole with norm smaller than 1 not allowing end modes to exist.

For $K\rightarrow 0$ the system becomes critical because the energy gap closes and the correlation length diverges and (\ref{eq:EnergyBoundaryModesLinear}) gives us 
\be
\label{eq:Euguale0}
E_\pm=0.
\ee

We want to compute the interval of validity $\Delta p$ around $\pi$ of such a result. The determinant $\Delta$ has, in this case a couple of solutions such that $(z_1\rightarrow 0,\frac{1}{z_1}\rightarrow\infty)$ and another couple $(z_2=\frac{1+J e^{ip}}{J},\frac{1}{z_2})$. The interval of validity of (\ref{eq:Euguale0}) is given by $\{p: |z_2|<1\}$ which gives the condition $|J e^{ip}+1|\le J$ which is satisfied in the interval
\be
\label{eq:DeltaForK0}
{\Delta p}_{K=0} =2\arccos{\frac{1}{2J}}.
\ee

\end{document}